\documentclass[a4paper,10pt]{article}

\usepackage{vmargin}
\usepackage{amsfonts}
\usepackage{amsmath}
\usepackage{graphics,graphicx,color}
\usepackage{psfrag}
\usepackage{verbatim}
\usepackage[bookmarks=false]{hyperref}

\newcommand{\beq}{\begin{equation}}
\newcommand{\eeq}{\end{equation}}
\newcommand{\beqa}{\begin{eqnarray}}
\newcommand{\eeqa}{\end{eqnarray}}
\newcommand{\nn}{\nonumber}

\newcommand{\R}{\mathbb{R}}
\newcommand{\V}{\mathbb{V}}
\newcommand{\C}{\mathbb{C}}

\newcommand{\G}{\mathcal{G}_{\D}}

\newcommand{\D}{\Delta}
\newcommand{\e}{\mathbf{e}}

\newcommand{\su}{\mathfrak{su}}
\newcommand{\SU}{\mathrm{SU}}
\newcommand{\SO}{\mathrm{SO}}
\newcommand{\Spin}{\mathrm{Spin}}

\newcommand{\tr}{\mathrm{tr}}
\newcommand{\Aut}{\mathrm{Aut}}
\newcommand{\Hom}{\mathrm{Hom}}
\newcommand{\End}{\mathrm{End} \,}

\setmarginsrb{39mm}{12mm}{28mm}{8mm}{12pt}{11mm}{0pt}{13mm}
\begin{document}

\title{\Large\bf Observables in 3d spinfoam quantum gravity with fermions}

\author{Richard J. Dowdall$^a$\footnote{richard.dowdall@maths.nottingham.ac.uk}\ , Winston J. Fairbairn$^{a,b}$\footnote{winston.fairbairn@uni-hamburg.de}\
\\ [1mm]
\itshape{\normalsize{$^a$School of Mathematical Sciences, University of Nottingham,}} \\
\itshape{\normalsize{University Park, Nottingham,}} \\
\itshape{\normalsize{NG7 2RD, UK.}}
\\
\\
\itshape{\normalsize{$^b$Department Mathematik, Universit\"at Hamburg,}} \\
\itshape{\normalsize{Bundesstra\ss e 55,  20146 Hamburg,}} \\
\itshape{\normalsize{Germany.    }}
}

\date{}
\maketitle

\begin{abstract}
We study expectation values of observables in three-dimensional spinfoam quantum gravity coupled to Dirac fermions.
We revisit the model introduced by one of the authors and extend it to the case of massless fermionic fields.
We introduce observables, analyse their symmetries and the corresponding proper gauge fixing. The Berezin integral over the fermionic fields is performed and the fermionic observables are expanded in open paths and closed loops associated to pure quantum gravity observables. We obtain the vertex amplitudes for gauge-invariant observables, while the expectation values of gauge-variant observables, such as the fermion propagator, are given by the evaluation of particular spin networks.
\end{abstract}

\section{Introduction}

In \cite{fairbairn-2007-39}, a method for the coupling of fermions to 3d spinfoam quantum gravity for Euclidean signature metrics was proposed.  The resulting model was expressed as a sum over fermionic loops throughout the spinfoam which were then coupled to the quantum geometry via grasping operators \cite{freidel-1999-2,hackett-2007-24}.  This yielded vertex amplitudes in the state sum given by modified $6j$ symbols incorporating the presence of matter fields.  While various other methods of coupling matter to spinfoam models have been proposed, for example \cite{Mikovic:2001xi, freidel:2004vi,livine-2004-7,barrett-2006-23,freidel-2006-23,Xu:2009bz,Livine:2007dx}, the simplicial  formalism used here, and in \cite{speziale-2007,oriti-2002-66,Mikovic:2002uq}, has very clear links to the classical action and closely resembles gauge theory on a random lattice \cite{randomLGT1}.

In this paper, we firstly extend the framework of the original proposal to massless fermions. This will turn out to clarify, and simplify the calculation. Motivated by recent interest in the graviton propagator \cite{Alesci:2007tx,Alesci:2007tg,Alesci:2008ff,Bianchi:2009ri}, we then show how to compute the fermion propagator and a few other simple observables.  This requires a careful analysis of the symmetries of the system and a proper gauge fixing. We then integrate out the fermions using a different, more straight forward technique than in the original proposal. This procedure expands the observables into Feynman diagrams which, we show, naturally couple to the quantum geometry.


The organisation of the paper proceeds as follows.  In section \ref{classical}, we recall the proposed coupling of fermions.  In section \ref{quantum theory} we perform the integrations in the path integral and arrive at spin foam models or spin network diagrams for the various observables. In section \ref{examples} we demonstrate the techniques by considering some simple examples.
The Appendix provides some useful facts on the representation theory of $\SU(2)$.

\section{Classical theory}
\label{classical}

Let $M$ be a  connected, oriented, compact, three-dimensional differential manifold endowed with an Euclidean metric $g$ with diagonal form $\eta = (+++)$. The relevant symmetry group is the spin group associated to the Euclidean metric $\eta$, that is, the Lie group $\Spin(3) \cong \SU(2)$. We will assume that the manifold $M$ is endowed with a given spin structure, i.e. a $\SU(2)$-principal bundle $\hat{\mathcal{P}}$ mapped with a two-to-one bundle homomorphism onto the $\SO(3)$ bundle of $g$-orthonormal frames $\mathcal{P}$ over $M$. Note that the principal bundle $\hat{\mathcal{P}}$ is necessarily trivial. Hence, we will choose once and for all a global trivialising section.

Since we are focusing on fermions, we will work in the spinor representation of 3d gravity. To this aim, we define the linear map $\gamma : \R^3 \rightarrow \End (\C^2)$, mapping the Clifford algebra $\mathcal{C}(3,0)$ (in which the (dual of the) vector space $\R^3$ is embedded) onto the Pauli algebra, i.e., the
endomorphism algebra of the two-dimensional complex vector space $\V \equiv \C^2$. In fact, we will think of the linear map $\gamma$ as taking value in the Lie algebra $\su(2)$ of traceless, anti-Hermitian matrices instead of the  true Pauli algebra (traceless, Hermitian matrices).

The dynamical fields of the bosonic sector of the theory are the image ${\bf e} = \gamma(e) = e^a \gamma_a$, $a=1,2,3$, of the soldering form $e$ on $\mathcal{P}$ under the map $\gamma$, and a principal connection on the spin bundle $\hat{\mathcal{P}}$, that is, a spin connection $\omega=\frac{i}{2} \omega^a \sigma_a$, where the symbols $\sigma_a$ denote the Pauli matrices and the symbols $\gamma_a= \frac{i}{2} \sigma_a$ generate the Lie algebra $\su(2)$.
The fields of the fermionic sector of the theory consist of a 2-component Dirac fermion $\psi$ which is a section of the spinor bundle, that is, the vector bundle $E = \hat{\mathcal{P}} \times_{\rho} \V$ associated to $\hat{\mathcal{P}}$ via the fundamental representation $\rho$ of the spin group $\Spin(3)$ on $\V$, and its Dirac conjugate $\bar{\psi}=\psi^{\dagger}$, where the dagger $^{\dagger}$ denotes Hermitian conjugation.

\subsection{Simplicial theory}
\label{simplicial theory}

In order to regularise the functional integrals that would be used in a formal path integral approach to quantisation, we now introduce a cut-off on the number of degrees of freedom of the theory by using lattice-like methods.

The so-called triangulation conjecture (or Hauptvermutung) states that all topological three-manifolds admit a piece-wise linear structure. In other words, all three-dimensional manifolds are triangulable. Accordingly, we can
choose a fixed triangulation $\Delta$ of the spacetime manifold $M$ and discretise the fields accordingly.
To this aim, we will also need to consider the two skeleton $\kappa=(v,e,f)$ of the topological dual $\Delta^*$ of the triangulation $\Delta$, that is, the complex formed by the vertices $v$, edges $e$ and faces $f$ of $\Delta^*$.
More precisely, we will work with the derived complex $\Delta^+$ obtained by refining the dual triangulation $\Delta^*$ by subdividing the dual faces $f$ into wedges $w$ (see \cite{Reisenberger:1996ib} and references therein).
The edges of the derived complex $\D^+$ (which include the half-edges of $\D^*$) will be denoted $\bar{e}$. 
 We will choose an orientation together with a distinguished vertex for each wedge. The orientation of the wedges naturally induces an orientation of the boundary edges. This prescription is necessary to have a well defined model, but the end results will be completely independent of these choices.

We finally assume that the global trivialising section of $\hat{\mathcal{P}}$ is piecewise constant in each tetrahedron of $\Delta$. Hence, the spinor and vector bundles associated to $\hat{\mathcal{P}}$ are also globally trivialised by a piecewise constant section and we have a copy $\SU(2)_v$ of the group $\SU(2)$, and copies $\V_v$, $\overline{\V}_v^*$ of the vector space $\V$, together with its dual complex conjugate vector space $\overline{\V}^*$, above each vertex $v$ of $\D^*$.
We therefore have a basis $(\gamma_a(v))_a$ of $\su(2)$, a basis $(e_{A}(v))_{A}$ of $\mathbb{V}$ and a basis $(e^{A}(v))_{A}$ of $\overline{\mathbb{V}}^*$ for each tetrahedron, or equivalently for each dual vertex $v$ of $\Delta^*$, inside which the vector and spinor fields are globally defined.

\paragraph{Simplicial fields.}

The fields are discretised in the standard way, namely by using their geometrical nature to assign values to the elements of the cellular complexes defined above.

The (spin space) frame field $\mathbf{e}$, as a one form, maps the one-simplices $s$ of $\Delta$ in $ \su(2)$. Each such one-simplex is topologically dual to a face $f$ of $\D^*$, and the wedges of $f$ are in one-to-one correspondence with the tetrahedra sharing the one-simplex $s$. Accordingly, it makes sense to assign the discretised co-frame to the wedges :
$
\e_w = e^a_w \gamma_a(v),
$
where $e^a_w:=e^a(s)$ denotes the components of the image in $\R^3$ of the tangent space vector $s$, measured with respect to the frame attached to the dual vertex $v$ associated to the wedge $w$.  Hence, for a given face $f$, the wedge variables represent the same vector measured in different frames. Accordingly, these variables are related \footnote{Note this is not related to the flatness of the connection ; the vector attached to the one-simplex is in the orthogonal complement of the dual face and is accordingly unaffected by the rotation associated to the holonomy around the face.} by a connection matrix in $\SU(2)$ (image in spin space of a $\SO(3)$ rotation).

This matrix $g_e$, the holonomy $P \, \exp \int_e \omega$ of the connection along the corresponding edge $e$, is the discretised connection and is assigned to all edges of the derived complex $\D^+$. Reversing the orientation of an edge maps the associated group element into its inverse : $g_{e^{-1}}=g^{\dagger}_e$. Finally, the curvature around a wedge $w$ is measured by the holonomy $G_w = \prod_{\bar{e} \in \partial w} g_{\bar{e}}$ of the connection around $w$, by virtue of the Ambrose-Singer theorem. The matrix product is taken starting from the distinguished vertex and in the ordering induced by the orientation of the wedge.

The fermionic fields, as sections of the spinor bundle, assign a spinor $\psi_v$ and a co-spinor $\bar{\psi}_v$ to each vertex $v$ of the dual triangulation. More precisely, to encompass the appropriate statistics, the components of the simplicial fermions $\psi_v^{A}$, $\bar{\psi}_{vA}$ are chosen to define a generating system of the Grassmann algebra $\mathcal{G}_{\D} = \bigwedge \left( E \bigoplus \overline{E}^* \right)$ associated to the triangulation $\Delta$, with $E=\oplus_{v} \V_v$ and $\overline{E}^* = \oplus_{v} \overline{\V}^*_v$.

\paragraph{Simplicial action.}

With the fields defined above, the discretised action for the fermion/gravity system is the the sum of  the Dirac action and an action for the gravitational field. To define the fermionic action, we first introduce the simplicial Dirac operator assigned to the edges of the dual triangulation:
\beq
\label{Dmatrix}
D_{e} = \Sigma_e \, U_{e} - U_{e} \, \Sigma_{e^{-1}},
\eeq
where $U_e:=U(g_e)$ is the holonomy of the connection along the edge $e$ in the spinor representation, and $\Sigma_e$ is the discretised version of the two-form $\Sigma = \e \wedge \e$. It is evaluated on the infinitesimal triangles of $\Delta$ dual to the edges of $\Delta^*$ and defined as
\beq
\label{area}
\Sigma_e = \frac{1}{3} \sum_{w_e,w'_e} {\bf e}_{w_e} {\bf e}_{w'_e} \, \mathrm{sgn}(w_e,w'_e),
\eeq
where the sum is taken over the three possible pairs (without counting the permutations) of wedges meeting on the edge $e$ and on the vertex $s(e)$, and the factor $\mathrm{sgn}(w_e,w_e')$ equals $\pm 1$ depending on the sign of the associated (coordinate) area bivector.
The spinor components are measured with respect to the spin frame associated to the source vertex $s(e)$ of the edge $e$.

The Dirac action yields
\beq
\label{fermionaction}
S_{\mbox{{\tiny D}}}[{\bf e}_w, g_e, \overline{\psi}_v, \psi_v]= \frac{1}{8} \sum_{or(e)} \sum_{e} S_e,
\eeq
where the sum is taken over all orientations $or(e)$ of all edges $e$ of $\Delta^*$, and
\beq
\label{edgeaction}
S_e = (\psi_{s(e)}, D_{e} \, \psi_{t(e)}),
\eeq
with $s(e)$ and  $t(e)$ respectively denoting label the source and target vertices of the edge $e$ in the corresponding orientation, and the inner product $(,)$ being the standard Hermitian inner product on $\C^2$., i.e., introducing a basis of $\C^2$, $(\xi, \chi) = \overline{\xi}_A \chi^B \delta_B^A$.

The gravitational action is given by
\beq
\label{GRaction}
S_{\mbox{{\tiny GR}}}[{\bf e}_w, g_{\bar{e}}]= \frac{1}{16 \pi G} \sum_w \tr \left( {\bf e}_w G_w \right),
\eeq
where the trace is in the spinor representation.

It was proven in \cite{fairbairn-2007-39}  that the action for the coupled system  $S_{\mbox{{\tiny GR-D}}}  = S_{\mbox{{\tiny GR}}} + S_{\mbox{{\tiny D}}}$ converges point-wise towards the continuum Einstein-Cartan-Dirac action when the lattice spacing goes to zero.


\paragraph{Discrete symmetries.}

The introduction of a fixed background simplicial structure on $M$ has broken the diffeomorphism invariance of the continuum theory (see \cite{Dittrich:2008pw} for a detailed discussion of these issues). Unlike the pure gravity case, the diffeomorphisms can not be recovered by a combination of gauge transformations because the introduction of fermions has broken the topological character of pure 3d gravity.  The expectation values of observables will therefore be diffeomorphism covariant as opposed to invariant.  The diffeomorphism invariance should be recovered in the continuum limit with some appropriate renormalisation as in lattice gauge theory.

However, the discretisation procedure has not broken the local $\SU(2)$ symmetry.
The arbitrariness in the choice of a basis of $\su(2)$ and of $\V$ at each dual vertex $v$ is reflected in the invariance of the simplicial action under the following discrete gauge transformations parameterized by an $\SU(2)$-valued sequence $k : \{v\}  \rightarrow \SU(2)$:
\beqa
\label{ge}
{\bf e}_w & \mapsto &  k_v^{-1} \, {\bf e}_w \, k_v \nn \\
\label{gG}
G_w & \mapsto &  k_v^{-1} \, G_w \, k_v,
\eeqa
for the bosonic part $S_{\mbox{{\tiny GR}}}$ of the action, and
\beqa
\label{gg}
g_e & \mapsto &  k_{s(e)}^{-1} \, g_e \, k_{t(e)}  \nn \\
\label{gS}
\Sigma_e & \mapsto &  k_{s(e)}^{-1} \, \Sigma_e \, k_{s(e)} \nn \\
\psi_v & \mapsto &  k_v^{-1} \psi_v \nn \\
\overline{\psi}_v & \mapsto &  \overline{\psi}_v \, k_v,
\eeqa
for the fermionic sector $S_{\mbox{{\tiny D}}}$.
In the first two lines, \eqref{ge}, the transformation acts on the variables assigned to all wedges $w$ containing the vertex $v$, while the transformation \eqref{gg} takes place on all edges $e$ of $\D^*$.

Finally, note that the Dirac matrix $D$ enjoys the following antisymmetry property:
\beq
\label{antisymetry}
D^{AB}_e = - D^{BA}_{e^{-1}},
\eeq
where the indices have been raised and lowered with the standard symplectic metric $\epsilon_{AB}$ on  $\C^2$. Our conventions are the following. The two-dimensional totally antisymmetric tensor is normalised such that $\epsilon_{01} = \epsilon^{01} = +1$, which implies that  $\epsilon_{AB} \epsilon^{BC} = - \delta_{A}^{C}$. The raising and lowering of indices occurs according to
\beq
\label{convention}
\chi^A = \epsilon^{AB} \chi_B, \;\; \mbox{and} \;\; \xi_A = \xi^B \epsilon_{BA} = - \epsilon_{AB} \xi^B.\eeq
This implies in particular the following property for multi-component spinors
 \beq
 \label{seesaw}
 M^{...A...}_{\;\;\;\;\;\;... A ...} = - M_{...A...}^{\;\;\;\;\;\;...A...}.
 \eeq

\paragraph{Simplicial observables.}
\label{Fermionic integration}

In this paper, we are interested in the study of simplicial fermionic\footnote{These observables are called fermionic, even if they obey bosonic statistics, because they are functionals of the spinor fields.} observables $\mathcal{O}_{f,\D}:=\mathcal{O}_{f}$, that is, polynomial functions of the simplicial fermion fields. In particular, we will study the following observables
\begin{itemize}
\item $\mathcal{O}^{\;\,\, A}_{f \;\, B}(x,y) = \psi_x^{A} \,\, \bar{\psi}_{y B}$ : the gauge-variant two-point function of the fermionic field.
\item $\mathcal{O}_f(e) = (\psi_{s(e)} , \, V_{e} \, \psi_{t(e)})$, with $e$ the path linking $s(e)$ to $t(e)$ and $V_e$ the corresponding holonomy. We will call this gauge-invariant observable the Polyakov line. Note that this observable is sometimes argued to be the correct physical observable for the fermion propagator (see \cite{Mitra:2005zq} and references therein).
\item $\mathcal{O}_f(e,e') = M(e) \, M(e')$, with $M(e) = (\psi_{s(e)} , \, V_{e} \, \psi_{t(e)})$; the two-point function of the Polyakov line.
\end{itemize}

With the discretisation outlined above, the naive definition of the expectation value of an arbitrary simplicial fermionic observable $\mathcal{O}_{f}$ yields
\beq
\label{qobservable}
\langle \mathcal{O}_{f} \rangle_{\mbox{{\tiny GR-D}}} = \, \frac{1}{\mathcal{Z}_{\mbox{{\tiny GR-D}}}}
\left( \prod_{w} \int_{\su(2)} d {\bf e}_w \right)
\left( \prod_{\bar{e}} \int_{\SU(2)} d g_{\bar{e}} \right)
\left( \int_{\G} d \mu (\overline{\psi}_v, \hspace{1mm} \psi_v) \right) \hspace{1mm}
\mathcal{O}_{f} \,\,
e^{iS_{\mbox{{\tiny GR-D}}}}.
\eeq
Here $d {\bf e}_w$ is the Lebesgue measure on $\su(2)$, $d g_e$ is the normalised Haar measure on $\SU(2)$, and
the symbol $d \mu (\overline{\psi}_v, \hspace{1mm} \psi_v)$ denotes the Berezin integral on $\G$. The latter measure is the element of $\G^*$ defined by
$$
d \mu (\overline{\psi}_v, \hspace{1mm} \psi_v) = \prod_{v=n}^1 d \overline{\psi}_v \prod_{v=n}^1 d \psi_v,
$$
where $n$ is the number of vertices of $\D^*$ and
$$
d \overline{\psi}_v = \prod_{A=2}^1 \frac{\partial}{\partial \overline{\psi}_{v A}}, \,\,\,  \,\,\,d \psi_v = \prod_{A=2}^1 \frac{\partial}{\partial \psi_v^{A}}.$$
Finally, the normalisation factor
\beq
\label{partition}
\mathcal{Z}_{\mbox{{\tiny GR-D}}} = \left( \prod_{w} \int_{\su(2)} d {\bf e}_w \right)
\left( \prod_{\bar{e}} \int_{\SU(2)} d g_{\bar{e}} \right)
\left( \int_{\G} d \mu (\overline{\psi}_v, \hspace{1mm} \psi_v) \right) \hspace{1mm} e^{iS_{\mbox{{\tiny GR-D}}}}
\eeq
is the discretised path integral for the coupled system.

\paragraph{Gauge fixing.} According to a theorem in lattice gauge theory by Elitzur \cite{PhysRevD.12.3978}, the expectation value of a locally gauge dependent observable, such as the fermion two-point function, is zero without gauge fixing.  This result is reproduced in the spinfoam formalism as the expectation value of a non-gauge fixed, locally gauge-dependent observable will result in spin network vertex amplitudes with a single open end which vanishes by Schur's Lemma (see equation \eqref{Schur} in the Appendix). It is therefore crucial to appropriately fix the above gauge freedom in the naive expression \eqref{qobservable}. This can be achieved following the next lines \cite{Freidel:2002xb,freidel:2004vi}.

The idea is to introduce a maximal tree of the triangulation's dual $T \subset \D^*$, that is, a sub-complex of the one-skeleton of $\D^*$ touching every vertex of $\D^*$ without ever forming a loop.
One can pick an arbitrary preferred vertex $r$ on $T$, called the root, and denote $u$ the endpoints of the tree (recall that $T$ is an open graph). We will note $\D_T$ the simplicial pair $(\D,T)$.
Using this tree and the gauge invariance, it is possible to reach the gauge $g_e=1\!\!1$ for all $e$ in $T$. Remarkably, the Faddeev-Popov determinant associated to this gauge-fixing procedure :
\beq
\Delta_{FP}^{-1} = \prod_v \int_{\SU(2)} dk_v \prod_{e \in T} \delta (k_{s(e)}^{-1} \, g_e \, k_{t(e)}),
\eeq
with $dk$ denoting the normalised Haar measure on $\SU(2)$, is equal to one \cite{freidel-2003-662,freidel:2004vi} because
\beq
\Delta_{FP}^{-1} = \int_{\SU(2)} d k_{r} \left( \prod_{u} \int_G dk_{u} \right) \prod_{u} \delta (k_r^{-1} \,  g_{\gamma(u)} \,
k_{u}) = 1,
\eeq
where $\gamma(u)$ is the unique path linking the endpoint $u$ to the root of the tree.

If the observables are gauge-invariant, one simply follows the standard Faddeev-Popov procedure associated to the above gauge-fixing prescription. If the observables are not gauge-invariant, the situation is more involved.
One procedure to compute a gauge-variant observables in lattice gauge theory involves averaging over the gauge group to define a gauge invariant function which can then be computed \cite{PhysRevD.44.2558}.
Alternatively, following Faddeev and Popov, one can take the definition of a gauge dependent operator to include the necessary gauge fixing \cite{Giusti:2001xf}
\beq
\label{FPobservable}
\langle \mathcal{O}_{f} \rangle_{\mbox{{\tiny GR-D}}} = \, \frac{1}{\mathcal{Z}_{\mbox{{\tiny GR-D}}}^{\Delta_T}}
\left( \prod_{w} \int_{\su(2)} d {\bf e}_w \right)
\left( \prod_{\bar{e}} \int_{\SU(2)} d g_{\bar{e}} \right)
\left( \int_{\G} d \mu (\overline{\psi}_v, \hspace{1mm} \psi_v) \right) \hspace{1mm}
\prod_{e \in T} \delta(g_e) \,\,
\mathcal{O}_{f} \,\,
e^{iS_{\mbox{{\tiny GR-D}}}},
\eeq
where the normalisation factor is the path integral \eqref{partition} gauge fixed by the insertion of the simplicial gauge fixing function $\prod_{e \in T} \delta(g_e)$.
For a gauge invariant operator, this reduces to equation \eqref{qobservable}, while it {\em defines} the expectation value for a gauge-variant observable.
One can show that the expectation values of gauge invariant observables are independent of the choice of gauge fixing tree $T$. The gauge variant observables necessarily depend on the choice of tree.

This describes the classical simplicial theory. We can now proceed to the quantum computations. Indeed, having reduced the number of degrees of freedom from infinite to finite by the introduction of a discretisation scheme, we can now compute the integrals and accordingly access the quantum aspect of the simplicial theory.

\section{Quantum theory}
\label{quantum theory}

To compute the integrals in \eqref{FPobservable}, the idea is
to first integrate out the fermionic degrees of freedom, in other words, perform
a Feynman diagram expansion of the matter sector, before computing the gravitational integrals.
As a result, $\langle \mathcal{O}_{f} \rangle_{\mbox{{\tiny GR-D}}}$ is expressed as the expectation value of a bosonic observable $\mathcal{O}_{b}$ of the {\itshape pure quantum gravity} theory :
\beq
\label{refobservable}
\langle \mathcal{O}_{f} \rangle_{\mbox{{\tiny GR-D}}} =
\frac{1}{\mathcal{Z}_{\mbox{{\tiny GR-D}}}}
\left( \prod_{w} \int_{\su(2)} d {\bf e}_w \right)
\left( \prod_{\bar{e}} \int_{\SU(2)} d g_{\bar{e}} \right)
\hspace{1mm}
\prod_{e \in T} \delta(g_e) \,\,
\mathcal{O}_{b} \,\,
e^{iS_{\mbox{{\tiny GR}}}}.
\eeq
Therefore, the expectation value of a fermionic observable can be written
\beq
\label{dude}
\langle \mathcal{O}_{f} \rangle_{\mbox{{\tiny GR-D}}} = \frac{\langle \mathcal{O}_{b} \rangle_{\mbox{{\tiny GR}}}}{\langle \det \, D \rangle_{\mbox{{\tiny GR}}}},
\eeq
where
\beq
\label{bosonic}
\mathcal{O}_{b} = \left( \int_{\G} d \mu (\overline{\psi}_v, \hspace{1mm} \psi_v) \right) \hspace{1mm}
\mathcal{O}_{f} \,\, e^{i S_{\mbox{{\tiny D}}}},
\eeq
and, using the fact that the Dirac action $S_{\mbox{{\tiny D}}}$ is an element of $\bigwedge^2 \left( E \bigoplus \overline{E}^* \right)$,
$$\left( \int_{\G} d \mu (\overline{\psi}_v, \hspace{1mm} \psi_v) \right)
\,\, e^{i S_{\mbox{{\tiny D}}}} = \det D.$$

The next step is to expand the bosonic observables $\mathcal{O}_{b}$ in fermionic paths.

\subsection{Feynman diagram expansion}

The advantage of first performing the matter integrals is that the Feynman diagrams define, as mentioned above, pure quantum gravity observables that naturally couple to the dynamical geometry. The essence of the expansion relies on the properties of the Berezin integral.
The techniques used here are different than the ones introduced in \cite{fairbairn-2007-39}, which relied on the introduction of symplectic Majorana fermions.
Here, we explicitly integrate over the Dirac fields which offers a more direct way to compute the integrals.

The first step consists in Taylor expanding the exponential of the Dirac
action. We use the fact that each term in the sum \eqref{fermionaction} is an even element of the Grassmann algebra, that is,  has overall bosonic statistics $S_e S_{e'} = S_{e'} S_e$, for all couples of edges $e,e'$, and that the expansion is finite because $(S_e)^N = 0$, for all edges $e$ and all $N>2$. This leads to the following expansion of \eqref{bosonic}
\beq
\label{expansion}
\mathcal{O}_{b} = \left( \int_{\G} d \mu (\overline{\psi}_v, \hspace{1mm} \psi_v) \right) \hspace{1mm}
\mathcal{O}_{f} \,\, \prod_{e} \left( 1 + i \alpha \, S_e - \frac{\alpha^2}{2} (S_e)^2  \right) \left( 1 + i \alpha \, S_{e^{-1}} - \frac{\alpha^2}{2} (S_{e^{-1}})^2  \right),
\eeq
where $\alpha = 8 (2 \pi G)^2$ (we have rescaled the triad such that $e \mapsto e / 16 \pi G$).
The above expansion produces a finite but possibly large number of contributions. However, the number of terms surviving the Berezin integral is drastically reduced.

Being a derivation, the Grassmann integral yields a non-zero result only if
the integrand is the top form on $\G$, or in other words, if for each vertex $v$ of $\D^*$, there is a product of two non-identical spinor and two non-identical co-spinor components.

Consider first the vertices where there are no field insertions.
The edge action $S_e$ appearing in the expansion has a co-spinor (resp. a spinor) siting at the source (resp. target) of the edge $e$.
Hence, the only contributions surviving the integration are such that, for each such vertex $v$, there is a product of four edge actions
$S_{e}$ such that $v$ is exactly a source vertex for two edges and a target vertex for the two other edges.
Note that the four edges need not be distinct.


To integrate such contributions, we firstly consider the Berezin integral for the Grassmann algebra associated to a single vertex. Dropping the vertex label for a moment, it is easy to prove the following identity
\beq
\label{int}
 \int \, d \mu ( \overline{\psi}, \hspace{1mm} \psi ) \;  \overline{\psi}_{A} \, \psi^{B} \, \overline{\psi}_{C} \, \psi^{D} =
- \epsilon^{BD} \epsilon_{AC}.
\eeq
If one implements this formula at a vertex $v$ satisfying the requirements discussed above, it is immediate to see that the integration connects, via the symplectic metric, the indices of the Dirac matrices $D_e$ (appearing in the edge actions $S_e$) associated to the edges meeting at $v$ pairwise : the two matrices associated to the two ingoing edges are paired together and likewise for the two matrices corresponding to the two outgoing edges. This implies that two matrices being connected are necessarily associated to edges  having opposite orientations.
The symplectic metric raises (resp. lowers) the second (resp. first) index of one of the two ingoing (resp. outgoing) matrices. This index then gets contracted with the first (resp. second) index of the other ingoing (resp. outgoing) matrix. The orientation of the edge associated to the matrix with raised (resp. lowered) indices is flipped using the symmetry \eqref{antisymetry} for the orientation of the edges to be consistent with the matrix product. Therefore, the Berezin integral at vertex $v$ yields two products of Dirac matrices (one for the ingoing and one for the outgoing edges) with consistent orientation of the associated edges.
Repeating the procedure for each vertex with no field insertion will accordingly produce products of closed sequences of Dirac matrices associated to edges forming loops in the dual triangulation $\Delta^*$. The number of Dirac matrices appearing in each such sequence will necessarily be even
because the matrices getting paired are associated to edges with opposite orientation.

Let now $v$ be a vertex with a fermionic field insertion. If a spinor (resp. a co-spinor) component is inserted on a vertex $v$, one only requires another spinor (resp. co-spinor) component and two co-spinor (resp. spinor) components to saturate the vertex. Therefore, the contributions surviving the integral are the following. For each vertex $v$ where a spinor (resp. a co-spinor) is inserted, there is a product of three edge actions $S_e$ where the corresponding edges, which again are not necessarily distinct, are such that $v$ is exactly a target for one edge (resp. two edges) and a source for the other two edges (resp. for the other edge).  This implies that for each spinor (resp. co-spinor) insertion, there must be also a co-spinor (resp. spinor) insertion on some other vertex to obtain a non-vanishing result. 

The integration at a vertex $v$ where a spinor (resp. co-spinor) is inserted then connects the two Dirac matrices associated to the two outgoing (resp. ingoing) edges as before, but the Dirac matrix corresponding to the ingoing (resp. outgoing) edge is left open with a free second (resp. first) index corresponding to the index assigned to the spinor (resp. co-spinor) insertion. The other index of the open Dirac matrix gets connected at one of the adjacent vertices with another matrix and, using the orientation flip \eqref{antisymetry}, we obtain a product of Dirac matrices along an open oriented sequence of edges which starts (resp. terminates) at $v$ and terminates (resp. starts) where another field insertion given by a co-spinor (resp. spinor) has occurred. This path is necessarily supported by an odd number of edges.

Therefore, the result of the Berezin integral with $p$ spinor and $p$ co-spinor insertions on the vertices $x_1, ..., x_{2p}$ of $\Delta^*$ can be stated as follows. The following applies for observables with at most a single field insertion on each vertex, but can be trivially extended to higher insertions.
Each non-vanishing contribution is labelled by a graph $\Gamma(x_1, ..., x_{2p})$. This graph is the union of 1) a union of oriented loops $\mathcal{L}$ with 2) a union of $p$ oriented open paths
$\mathcal{P}$ connecting the $p$ vertices with spinor insertions to the $p$ vertices with co-spinor insertions.

This graph is such that
\begin{enumerate}
\item It has overall length\footnote{The length of a graph is the number of edges of the dual triangulation supporting the graph.} $E_{\Gamma}$ given by $2n-p$
\item The  length $E_{\mathcal{L}}$ of each connected loop component $\mathcal{L}$ is even, while the length $E_{\mathcal{P}}$ of each path components  $\mathcal{P}$ is odd
\item Each vertex $v$ of the dual triangulation $\Delta^*$ is either traversed twice by the edges of $\Gamma$, either traversed once if $v$ admits a field insertion, i.e., if $v \in \{x_1, ..., x_{2p}\}$
\item Each open path $\mathcal{P}$ is oriented such that $s(\mathcal{P})$ (resp. $t(\mathcal{P})$) is a vertex with a spinor (resp. co-spinor) insertion.
\item Consecutive edges of the graph $\Gamma$ cannot go back and forth along the same dual edge $e \in \Delta^*$ unless that particular connected component of $\Gamma$ is of length two.
\end{enumerate}
A graph satisfying all these properties will be called {\em admissible}.

All possible, admissible graphs can be obtained by the following graphical method. Draw all possible sets of arrows of cardinality $2n$ connecting all the vertices of the dual triangulation such that each vertex without field insertion has exactly two ingoing and two outgoing arrows, while a vertex with a spinor (resp. co-spinor) insertion has two outgoing (resp. ingoing) and one ingoing (resp. outgoing) arrow. For each such set of arrows, the corresponding admissible graph is  obtained by connecting at each vertex the ingoing arrows with the ingoing arrows and likewise for the outgoing arrows. The vertices with field insertions will accordingly be sources or targets for open lines. The connection process flips the orientation of one of the two arrows being paired to obtain a consistent orientation.
For example, on a simple four point lattice with a field insertion $\psi_1 \overline{\psi}_{2}$, the connecting of the arrows occurs as follows
$$
\psfrag{1}{1}
\psfrag{2}{2}
\psfrag{3}{3}
\psfrag{4}{4}
\begin{array}{c}
\includegraphics[scale=0.5]{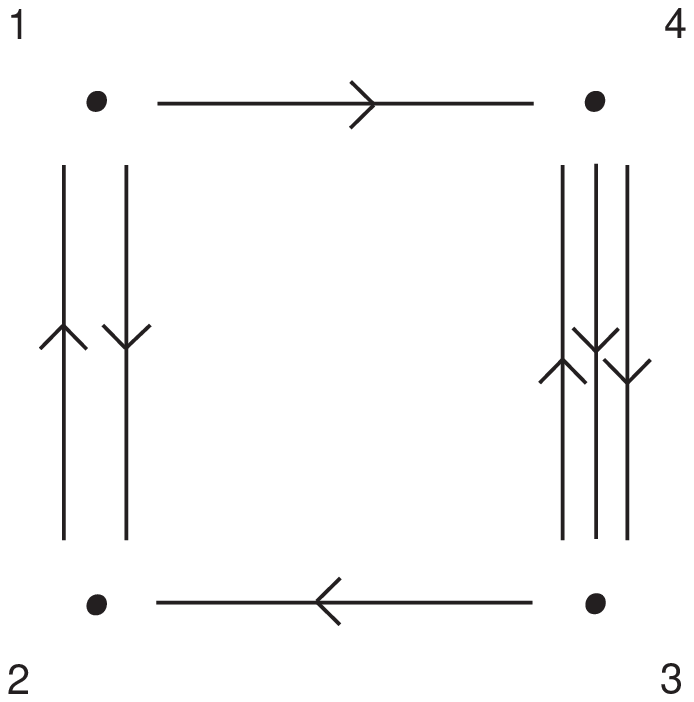}
\end{array}
 \ \ \ \ \ \ \ \ \Rightarrow \ \ \ \ \ \ \ \ \
 \begin{array}{c}
 \includegraphics[scale=0.5]{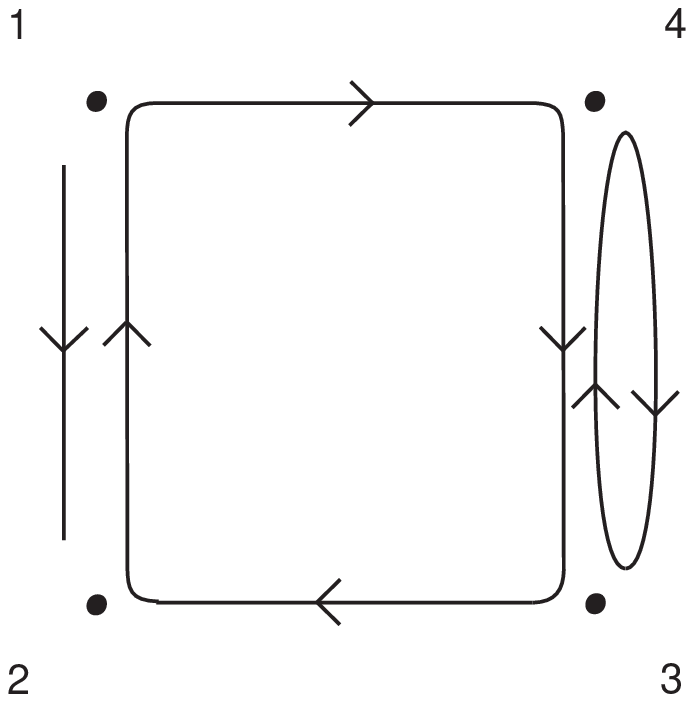}
 \end{array}
$$
It may seem that there is an ambiguity when choosing which arrow to flip; the two possible choices correspond to the two possible orientations for the loops and paths. However, the two contributions corresponding to the two possible orientations of a loop are in fact equal because of the orientation symmetry  \eqref{antisymetry}. For a path, the boundary data fixes the orientation such that the path always starts at a spinor insertion and terminates at a co-spinor insertion. Thus, there is no ambiguity.

We are now ready to establish a set of `Feynman rules' to obtain the bosonic observables $\mathcal{O}_{b}$ in terms of the fermionic ones $\mathcal{O}_{f}$. These rules are as follows.

\begin{itemize}
\item Consider all possible admissible graphs determined by $\mathcal{O}_{f}$.
\item To each admissible graph $\Gamma(x_1, ..., x_{2p})$, assign respectively to each open path $\mathcal{P}$ and to each closed loop $\mathcal{L}$ of $\Gamma(x_1, ..., x_{2p})$ the ordered products of Dirac matrices $D_e$
$$
D_{\mathcal{P}} =  (i \alpha)^{E_{\mathcal{P}}} \prod_{e \in \mathcal{P}} D_e, \;\;\;\;\; D_{\mathcal{L}} = (i \alpha)^{E_{\mathcal{L}}} \; \tr \prod_{e \in \mathcal{L}} D_{e},
$$
following the orientation of the path $\mathcal{P}$ or the loop $\mathcal{L}$ to which they are associated. 
\item Assign an global sign $\epsilon_{\Gamma(x_1, ..., x_{2p})} = \pm 1$ to each graph $\Gamma(x_1, ..., x_{2p})$ which controls the overall sign appearing as a result of  raising and lowering some of the indices with the symplectic metric (see equation \eqref{convention}), using the orientation antisymmetry \eqref{antisymetry}, the `see saw' property \eqref{seesaw}, and rearranging the fermionic fields to compute the full Berezin integral.
\item Sum over all such contributions:
\beq
\label{contributions}
\mathcal{O}_{b} = \sum_{\Gamma(x_1, ..., x_{2p})} \epsilon_{\Gamma(x_1, ..., x_{2p})} I_{\Gamma(x_1, ..., x_{2p})},
\eeq
where $I_{\Gamma(x_1, ..., x_{2p})}$ is the amplitude associated to the admissible graph $\Gamma(x_1, ..., x_{2p})$ constructed from the above assignments, and, depending on the obervables, the holonomies $V_e$ appearing in the definition of $\mathcal{O}_{f}$.
\end{itemize}

One may wander if there are no numerical factors supplementing the signs in the rules depicted above. The only factors to take into consideration are the $1/2$ factors coming from the quadratic terms in the Taylor expansion of the exponential. In fact, these factors cancel exactly the multiplicities of the contributions to which they are associated. Indeed, the $1/2$ factors are associated to loops going back and forth along a single edge because they correspond to terms of the form $(S_e)^2$ in the Taylor expansion. But for each such term there is also a contribution of the form $(S_{e^{-1}})^2$ giving rise to exactly the same loop contribution which hence appears twice.

Using this framework, we can study the expectation value of all polynomial observables in the fermion fields by computing the Berezin integrals according to the rules depicted above.
Firstly, we can consider the somewhat trivial case where there are no field insertions $p=0$, i.e., the computation of the functional determinant $\det D$.
In this case, the open paths $\mathcal{P}$ are simply reduced to the null graph and the admissible graphs are accordingly given by collections of closed loops which correspond to the terms of order $2n$ in the expansion parameter $\alpha$. An example is given by a single loop going through all the vertices of the triangulation twice. Another example is provided by a chain-type graph given by a collection of length two loops each going back and forth between two adjacent vertices. The amplitude $I_{\Gamma}$ associated to a such admissible graph $\Gamma$ is thus given by
\beq
I_{\Gamma} = \prod_{\mathcal{L}} D_{\mathcal{L}},
\eeq
where the product is over all the loops $\mathcal{L} \subset \Gamma$.
We therefore obtain that the functional determinant yields
\beq
\det D = \sum_{\Gamma} \epsilon_{\Gamma} \prod_{\mathcal{L} \subset \Gamma} D_{\mathcal{L}}.
\eeq

We can also compute the bosonic observable associated to the gauge-variant two-point function of the fermionic field
$$
\mathcal{O}^{\;\,\, A}_{f \;\, B}(x_1,x_2) = \psi_{x_1}^{A} \,\, \overline{\psi}_{x_2 B}.
$$
The corresponding bosonic observable is defined by specifying
the set of admissible graphs $\Gamma(x_1,x_2)$ and the corresponding amplitudes $I_{\Gamma(x_1,x_2)}$. The admissible graphs are characterised by the fact that they contain one open path $\mathcal{P}(x_1,x_2)$ going from $x_1$ to $x_2$. An example is given by an open path and a single closed loop satisfying the admissibility requirements.  The amplitude associated to an admissible graph $\Gamma(x_1,x_2)$ yields
\beq
\label{propagator}
I^{\,\,\;\;\;\;\;\;\;\;\;\;\;A}_{\Gamma(x_1,x_2) \ B} = D^{\;\;\;\;\;\;\;\;\;\;\,\,\, A}_{\mathcal{P}(x_1,x_2) \,\, B} \left( \prod_{\mathcal{L}} D_{\mathcal{L}} \right).
\eeq
Note that, for all admissible graphs $\Gamma$, the above amplitude is of order $2n - 1$ in the expansion parameter $\alpha = 8 / m_p^2$, where $m_p^2 = 1 / 2 \pi G$ is the Planck mass. Therefore, since the two-point function is given by the quotient \eqref{dude} of the expectation value of the above bosonic observable with the expectation value of the functional determinant, it will be of mass dimension two as expected for the three-dimensional fermion propagator in the spatial representation.

The bosonic observable corresponding to the gauge-invariant Polyakov line
$$
\mathcal{O}_f(e) = (\psi_{s(e)} , \, V_{e} \, \psi_{t(e)}),
$$
is associated to the same type of admissible graphs as the two-point function because the field insertions are of the same type. The corresponding amplitudes are given by the expression
\beq
\label{polyakov obs}
I_{\Gamma(x_1,x_2)} = \tr \,\, V_e D_{\mathcal{P}(x_2,x_1)} \; \left( \prod_{\mathcal{L}} D_{\mathcal{L}} \right)
\eeq
where $\mathcal{P}(x_2,x_1)$ is a path connecting the endpoints $x_2=t(e)$ to $x_1=s(e)$ of the Polyakov line.

Another interesting observable is the two point function of the Polyakov line
$$
\mathcal{O}_f(e,e') = (\psi_{s(e)} , \, V_{e} \, \psi_{t(e)}) \, (\psi_{s(e')} , \, V_{e'} \, \psi_{t(e')}).
$$
If $x_1$, $x_2$ and $x_3$, $x_4$ respectively label the source and target vertices of $e$ and $e'$,
the result of the integration over the fermion fields is labelled by two types of admissible graphs $\Gamma(x_1, ... , x_4)$.  The first type contains two open paths linking $x_2$ to $x_1$ and $x_4$ to $x_3$. The corresponding amplitude yields
\beq
\label{greson obs1}
I_{\Gamma(x_1, ... , x_4)} =  \tr \,\, V_e D_{\mathcal{P}(x_2,x_1)} \, \tr \, V_{e'} D_{\mathcal{P}(x_4,x_3)} \; \left( \prod_{\mathcal{L}} D_{\mathcal{L}} \right).
\eeq
The second type of admissible graphs admit two open paths respectively linking $x_2$ to $x_3$ and $x_4$ to $x_1$, and the corresponding amplitude follows
\beq
\label{greson obs2}
I_{\Gamma(x_1, ... , x_4)}  =  \tr \, V_e D_{\mathcal{P}(x_2,x_3)} V_{e'} D_{\mathcal{P}(x_4,x_1)} \; \left( \prod_{\mathcal{L}} D_{\mathcal{L}} \right).
\eeq

The above expression suggest that the amplitudes associated to both types of gauge invariant observables considered here can be written generically in the following form
\beq
\label{general}
I_{\Gamma} =
\left( \prod_{\gamma} \tr \prod_{\mathcal{P} \in \gamma} D_{\mathcal{P}}  V_{e(\mathcal{P})}  \right) \; \left( \prod_{\mathcal{L}} D_{\mathcal{L}} \right).
\eeq
Here, the first product  is over all possible closed loops $\gamma$ formed by composing the open paths $\mathcal{P}$ of the graph $\Gamma$ with the edges $e$ supporting the holonomies $V_e$ defining the observable $\mathcal{O}_f$. The edge $e(\mathcal{P})$ denotes the edge $e$ supporting the observable holonomy $V_e$ which is such that $t(e) = s(\mathcal{P})$.

The obtained amplitudes are not yet expressed as products of holonomies as one would expect.
In fact, all the above amplitudes are given by a sum over $2^{E_{\Gamma}}$ contributions since each Dirac matrix $D_e$ is a sum of two terms (see equation \eqref{Dmatrix}). All these contributions are associated to the same graph $\Gamma$ but vary by the pattern of the assignments of holonomies and simplical area two-forms to each edge of $\Gamma$.
We will call each such term a configuration associated to the graph $\Gamma$ and note it $I^{c}_{\Gamma}$, with the parameter $c \equiv c(\Gamma)$, ranging from $0$ to $2^{E_{\Gamma}}-1$, labeling the configuration. An example of a such term can be obtained for each observable given above by replacing the Dirac matrix $D_e$ with the quantity $\Sigma_e U_e$. Another configuration is obtained, for example, by assigning the factor $U_e \Sigma_{e^{-1}}$, instead of $\Sigma_e U_e$,  to one of the edges.
Hence, for each graph $\Gamma$, $I_{\Gamma} = \sum_{c} I_{\Gamma}^c$ and all bosonic observables will be written as a sum
\beq
\label{O}
\mathcal{O}_b = \sum_{\Gamma, c}  \epsilon_{\Gamma} I_{\Gamma}^c.
\eeq

We close this section on the fermionic integration with a remark. Interestingly, considering the fermionic observable given by the top form
$$
\mathcal{O}_{f} = \left( \frac{1}{2} \right)^n \prod_v (\psi_v , \psi_v)^2
$$
leads to a bosonic obervable simply given by
$$
\mathcal{O}_{b} = \pm 1.
$$
This implies that the expectation value of this observable is simply given by the ratio of the Ponzano-Regge amplitude and the expectation value of the fermionic determinant.
We now turn towards the integration over the quantum geometry fluctuations.

\subsection{Gravitational integrals}

We have shown that the bosonic observable $\mathcal{O}_b$ corresponding to an arbitrary fermionic observable $\mathcal{O}_f$ could be written as the finite sum \eqref{O}, where $I_{\Gamma}^c = I_{\Gamma}^c(\mathbf{e}_w,U_e)$
As a result, the expectation value of a fermionic observable $\mathcal{O}_f$ can be written as a {\em finite} sum over spinfoams :
\beq
\label{sumoverconfigurations}
\langle \mathcal{O}_f \rangle_{\mbox{{\tiny GR-D}}} = \frac{1}{\langle \det D \rangle_{\mbox{{\tiny GR}}}} \sum_{\Gamma, c} \epsilon_\Gamma  A_{\Gamma}^c,
\eeq
where the amplitude $A_{\Gamma}^c$ associated to the graph $\Gamma$ and to the configuration $c$ is given by
\beq
\label{I}
A_{\Gamma}^c = \left( \prod_{w} \int_{\su(2)} d {\bf e}_w \right)
\left( \prod_{\bar{e}} \int_{\SU(2)} d g_{\bar{e}} \right)
\hspace{1mm}
\prod_{e \in T} \delta(g_e) \,\,
I_{\Gamma}^c(\mathbf{e}_w,U_e)  \,\,
\, \exp \, \sum_w \tr \left( {\bf e}_w G_w \right).
\eeq

The calculation of the observable now boils down to evaluating each term \eqref{I} appearing in the sum over spinfoams \eqref{sumoverconfigurations} in a similar way to evaluating Feynman diagrams in ordinary quantum field theory. To this aim, we now need to perform the integration over the discretised frame field and holonomy variables.

Let us take a closer look at the functional $I_{\Gamma}^c(\mathbf{e}_w,U_e)$ associated to a particular term $\Gamma$ in the Feynman diagram expansion and to a particular configuration $c$. The amplitudes  $I_{\Gamma}^c$ for all observables considered here share the same structural properties : the functional $I_{\Gamma}^c(\mathbf{e}_w,U_e)$ is given by products of terms of the form $\Sigma_e U_e$ or $U_e \Sigma_{e^{-1}}$ along the edges of the graph $\Gamma$ which can close or not. There can also be holonomy matrices $V_e$ linking the open ends of the graph. We therefore need to understand how to integrate the Dirac matrices and holonomy variables against the complex exponent of the gravity action. We start by considering the case of gauge invariant observables \eqref{general}.

\subsubsection{Gauge invariant observables}

To calculate gauge invariant observables, we will use the fact that, because the Fadeev-Popov determinant is equal to one, the gauge group is compact and the Haar measures used in the path integral are normalised, the expectation value of a gauge invariant observable calculated by inserting the gauge fixing function $\prod_{e \in T} \delta(g_e)$ in the path integral is equal to the expectation value obtained without the gauge fixing. Therefore, for the computation of gauge invariant observables, we will simply omit the gauge fixing function.

To understand how to perform the gravity integrals, we will choose a particular configuration $I_{\Gamma}^c$ associated to an arbitrary graph $\Gamma$ in the sum over configurations appearing in \eqref{general}. It will be sufficient to consider only one type of configurations as the techniques developed for this particular case will equally apply to all other terms.
This particular prototypical contribution is the one obtained by assigning only terms of the form $\Sigma_e U_e$ to the edges of the graph $\Gamma$.
We arrange the labels $c$ such that this configuration corresponds to $c= 0$. It is then given by the expression
\beq
I_{\Gamma}^{0}(\mathbf{e}_w,U_e) =
(i \alpha)^{E_{\Gamma}} \left( \prod_{\gamma} \tr \prod_{\mathcal{P} \in \gamma} \prod_{e \in \mathcal{P}} \Sigma_e U_e V_{e(\mathcal{P})} \right) \; \left( \prod_{\mathcal{L} } \tr \prod_{e \in \mathcal{L}} \Sigma_e U_e \right).
\eeq

\paragraph{Integration over simplicial triads.}

The Dirac matrices \eqref{Dmatrix} involve holonomies and discretised area two-forms \eqref{area} which are quadratic in the simplicial triad field.
To integrate over the wedge variables ${\bf e}_w$, the idea, borrowed from conventional quantum field theory, is to replace polynomial combinations of the discretised triad appearing in the path integral by  source derivatives  \cite{freidel-1999-2,hackett-2007-24}
$$
\mathbf{e}_w \mapsto \frac{i}{2} \frac{\delta}{\delta J_w},
$$
acting on an appropriate generating functional \cite{fairbairn-2007-39}, where the sources $J_w = J_w^a \gamma_a$ represent a Lie algebra valued two-form $J$ discretised on the wedges $w$. The role of the free theory is played by the pure topological 3d gravity action. In this framework, the discretised area two-form
$\Sigma_e= \Sigma_e^a \gamma_a$ appearing in the Dirac matrices becomes the differential operator
\beq
\label{sigma}
\hat{\Sigma}_e^a = - \frac{1}{12} \sum_{w_e,w'_e} \epsilon^{abc} \frac{\delta}{\delta J^b_{w_e} }  \frac{\delta}{\delta J^c_{w'_e} } \,  \mathrm{sgn}(w_e,w'_e),
\eeq
and the amplitude \eqref{I} is re-writen as
\beq
\label{operator}
A_{\Gamma}^{0} = (i \alpha)^{E_{\Gamma}} \left[ \left( \prod_{\gamma} \prod_{\mathcal{P} \in \gamma}\prod_{e \in \mathcal{P}} \hat{\Sigma}_e^{a_e} \right) \; \left( \prod_{\mathcal{L} } \prod_{e \in \mathcal{L}} \hat{\Sigma}_e^{b_e} \right) \; \left( A_{0}^{\; (a,b)}(J) \right) \right]_{J=0},
\eeq
where the generating functional, after integration over the simplicial triad, is given by
\beq
\label{generating}
A_{\Gamma}^{0 \; (a,b)}(J) = \left( \prod_{\bar{e}} \int_{\SU(2)} d g_{\bar{e}} \right) \,\,
 \left( \prod_{\gamma} \tr \prod_{\mathcal{P} \in \gamma} \prod_{e \in \mathcal{P}} \gamma_{a_e} U_e V_{e(\mathcal{P})} \right) \; \left( \prod_{\mathcal{L} } \tr \prod_{e \in \mathcal{L}} \gamma_{b_e} U_e \right)
\, \prod_w    \delta   \left( e^{J_w} G_w \right)
\eeq
where $(a,b)$ is a pair of cumulative indices such that $a$ runs over all of the $a_e$ in each path $\mathcal{P}$ in $\gamma$, and $b$ runs over each $b_e$ associated to each edge $e$ of each loop configuration $\mathcal{L}$. The delta function, which appears because the action appearing in the generating functional is linear in the simplicial triad, is assumed\footnote{In fact, the naive procedure followed here yields \cite{freidel:2004vi} the delta function on $\SO(3)$ and not the one on $\SU(2)$. To obtain the delta function on $\SU(2)$, one needs to insert the observable $\prod_w \frac{1}{2} ( 1 + \frac{1}{2} \tr \, G_{w})$ in the path integral \cite{Livine:2008sw}. In what follows, we will assume that this trick has been implemented and that we are thus working with a delta function on the spin group.} to be on the Lie group $\SU(2)$.

The next step consists of expanding the delta functions in terms of characters. Let $j \in \mathbb{N}/2$ label the unitary, irreducible representations $\pi^j: \SU(2) \rightarrow \Aut \, V_j$ of the Lie group $\SU(2)$. The Peter-Weyl theorem states that
\beq
\label{delta}
\forall g \in \SU(2), \;\;\;\; \delta(g)= \sum_{j \in \mathbb{N}/2} \dim j \chi_{j}(g),
\eeq
where $\dim j=2j + 1$ is the dimension of the representation $j$ and $\chi_j = \tr \, \pi^j$ is the corresponding character.
This step transfers the calculations to the arena of the representation theory of $\SU(2)$. The idea is then to expand the characters appearing in the generating functional \eqref{generating} into products of matrix elements of the group variables associated to the boundary of the wedges. The same group element then appears into a number of different representation matrices and
implementing the integrals over the simplicial connections then boils down to understanding how to integrate the tensor product of representations. All the formulae given in the following section are proved in the Appendix.

\paragraph{Group integrals.}

The group integrals are easier to understand with the aid of the graphical calculus of spin networks.
The graphical methods offer the advantage of accurately performing representation theory calculations which would involve lengthy formulae if written algebraically.

A spin $j$ representation matrix $\pi^j$ evaluated on a group element $g$ is depicted as an oriented line coloured by the spin $j$ traversing a little circle corresponding to the group element $g$ :
\begin{center}
\psfrag{i}{$j$}
\psfrag{g}{$g$}
\includegraphics[scale=0.25]{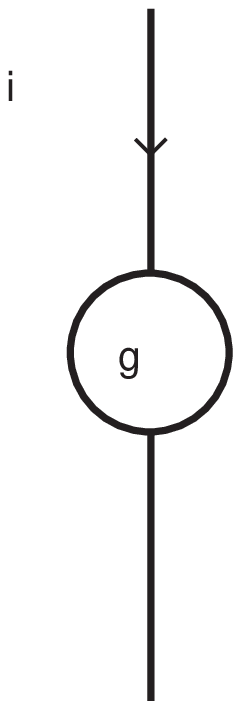}
\end{center}
The trace of a such matrix is simply obtained by joining the two ends of the open line to form a closed loop.
If the group element is the identity, the circle is simply omitted which implies that a spin $j$ line without a circle is the identity operator on $V_j$.  Each line comes with an arrow, which is inherited from the orientation of the derived complex $\Delta^+$, and reversing the arrow is equivalent to switching from a representation to its dual. Here it is assumed that, when no arrows are explicitly given, the orientation flows from the top to the bottom of the diagram.

The Haar integral over $g$ of a representation matrix $\pi^j(g)$ is understood when the circle is replaced by a rectangular box (a cable). Thus, the Haar integral of the tensor product of $n$ representation matrices with spins $j_1, ..., j_n$ evaluated on the same group element
\beq
\label{integral}
\int_{\SU(2)} d g \; \pi^{j_1}(g) \otimes ... \otimes \pi^{j_n}(g),
\eeq
is represented by a box in which the corresponding $n$ lines enter and exit
\beq
\begin{array}{c}
\psfrag{i}{$j_1$}
\psfrag{j}{$\cdots$}
\psfrag{k}{$j_n$}
\includegraphics[scale=0.25]{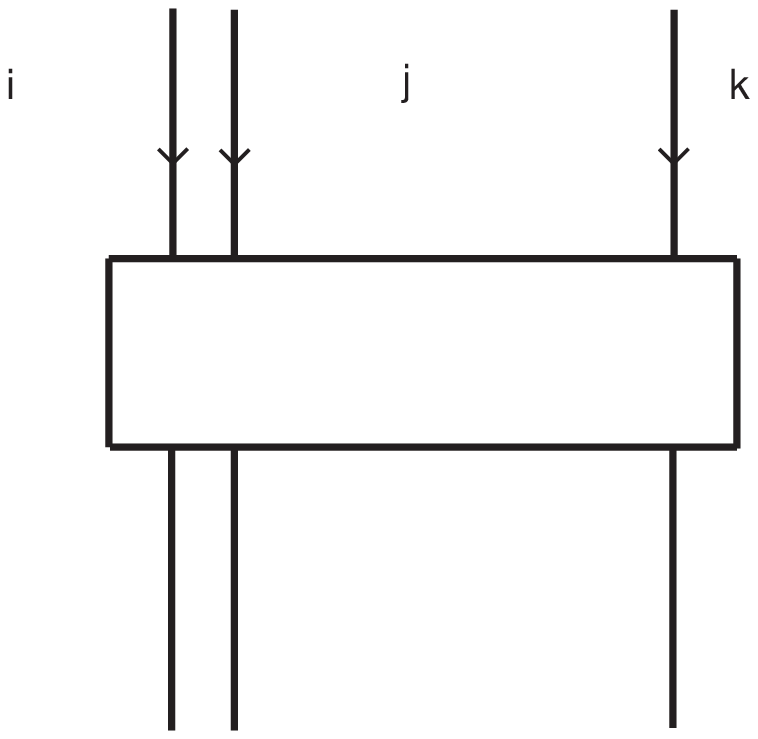}
\end{array}
\eeq
A box defines an endomorphism of the tensor product of the corresponding representations which projects onto the trivial subrepresentation. Accordingly, the result of the integration can be expressed in terms of intertwining operators. The diagrammatic representation of an intertwining operator $\iota$ of {\em unit norm} between the tensor product of $n$ representations $j_1, ..., j_n$ and the complex numbers
\beq
\label{iota}
\iota \, \in \, \Hom(V_{j_1} \otimes ... \otimes V_{j_n}, \C),
\eeq
is given by a vertex, or node (to avoid confusions with the vertices of the dual of the triangulation), with $n$ ingoing lines:
\beq
\label{inter}
\begin{array}{c}
\psfrag{i}{$j_1$}
\psfrag{j}{$\cdots$}
\psfrag{k}{$j_n$}
\psfrag{+}{$_+$}
\includegraphics[scale=0.25]{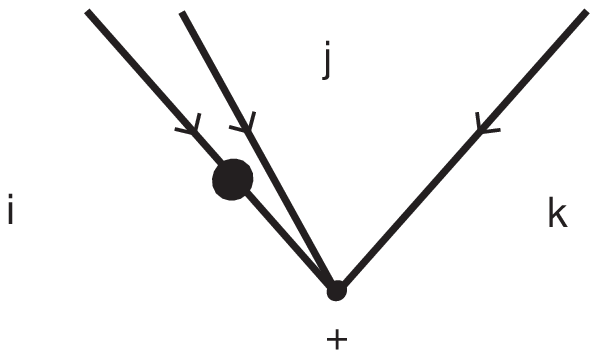}
\end{array}
\eeq
For this diagram to accurately describe an intertwining operator, one needs to specify an ordering of the lines to match the order of the tensor product in the corresponding algebraic expression.
Our conventions are such that the order of the tensor product is read clock-wise (resp. anti-clock-wise) starting from the line marked with a bullet point if the vertex carries a $+$ sign (resp. a $-$ sign). Therefore, the picture \eqref{inter} is precisely the diagrammatic form of the intertwiner $\iota$ in \eqref{iota}. The reversal of one or more arrows in the above diagram would represent an intertwining operator between the tensor product of the representation associated to the ingoing lines to the tensor product of the representation assigned to the outgoing lines.


We are now ready to perform the group integrals. There are three different types of Haar integrals appearing in the amplitude depending on what type of edge the integration is assigned to. The first type is when the edge $\bar{e}$ belongs to the interior of a face $f$ in which case it is shared by precisely two wedges and the corresponding group element thus appears in two characters. For every such interior edge $\bar{e}$ in a given face $f$, we use the orthogonality of the characters
\beq
\label{character}
\int_{\SU(2)} dg \;  \chi_j (g_1 g) \chi_k (g^{-1} g_2) = \frac{\delta_{jk}}{\dim j} \chi_j(g_1 g_2).
\eeq
Diagrammatically this is
\beq
\begin{array}{c}
 \psfrag{i}{$j$}
  \psfrag{j}{$k$}
  \psfrag{g1}{${g_1}$}
  \psfrag{g2}{${g_2}$}
 \hspace{-8mm}
\includegraphics[scale=0.45]{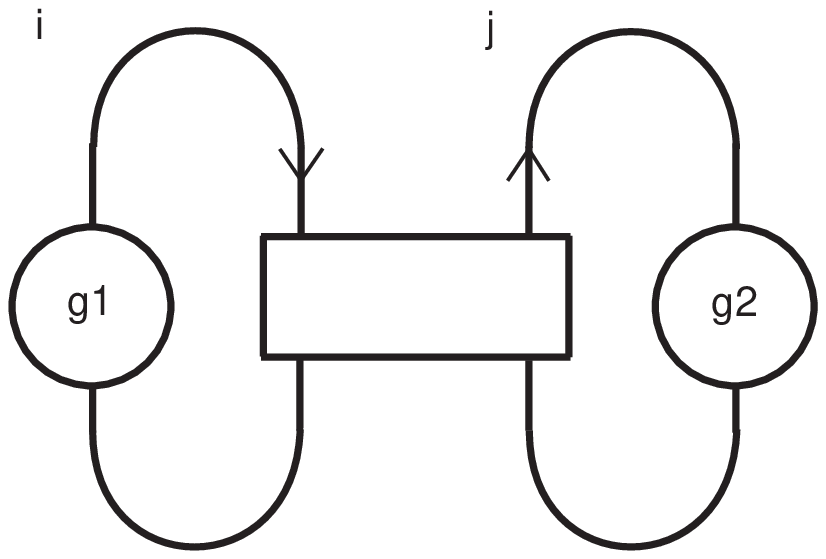}
\end{array}
=
\frac{1}{\dim j} \, \delta_{jk}
 \begin{array}{c}
 \psfrag{j}{$j$}
 \psfrag{d}{}
 \psfrag{g1}{${g_1}$}
  \psfrag{g2}{${g_2}$}
\includegraphics[scale=0.45]{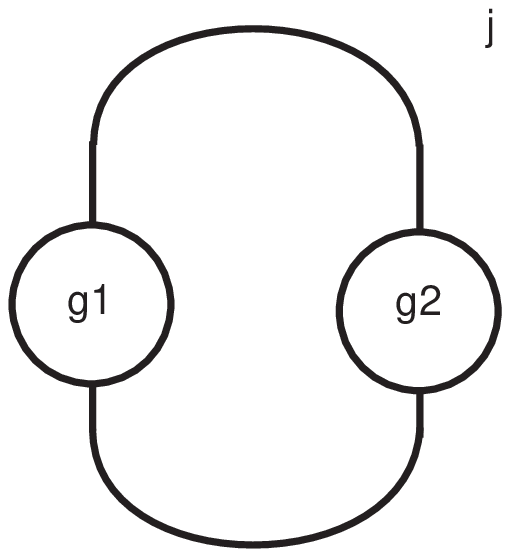}
\end{array}.
\eeq
Implementing the above formula repeatedly  forces all the spins $j_w$ associated to wedges belonging to the same face $f$ to be equal $j_w = j_f$ for all $w$ in $f$. The various powers of the dimension factors $\dim j=2j+1$ appearing  when performing the integrals combine with the dimension associated to each wedge by virtue of \eqref{delta} and produce, for each face $f$, a weight given by $(2j_f+1)^{\chi_f}$ where  $\chi_f$ is the topological invariant Euler characteristic of the face $f$. From here on, we will assume that each face is homeomorphic to a disk and thus that $\chi_f = 1$ for all $f$.
The generating functional \eqref{generating} thus reduces to
\beq
\label{I3}
A_{\Gamma}^{0 (a,b)}(J) = \sum_{j_f} \prod_f \dim j_f A_{\Gamma}^{0 (a,b)}(j_f,J),
\eeq
with
\beqa
\label{I5}
A_{\Gamma}^{0 (a,b)}(j_f,J) &=&
\left( \prod_{e} \int_{\SU(2)} d g_{e} \right) \left( \prod_{\gamma}  \tr \prod_{\mathcal{P} \in \gamma} \prod_{e \in \mathcal{P}} \gamma_{a_e} U_e V_{e(\mathcal{P})} \right) \; \left( \prod_{\mathcal{L} } \tr \prod_{e \in \mathcal{L}} \gamma_{b_e} U_e \right)
\\ \nn
&& \times \prod_f \chi_{j_f} \left( e^{J_f^1} g_{e_f}^1 \, ...  \, e^{J_f^m} g_{e_f}^m \right).
\eeqa
Here, the source insertions $J_f^v$ are along all the vertices $v$ of a given face $f $ ($m:=m_f$ denotes the number of
 vertices, i.e the number of wedges contained in $f$). The edges $\bar{e}$ of $\Delta^+$ that we have not integrated out, that is, the edges belonging to the boundaries of the faces $f$ (which are half-edges of $\Delta^*$), have been recombined into edges $e$ of $\Delta^*$.

To calculate \eqref{I5},  we need to perform the remaining integrals which, as we have seen, are associated to edges of the dual triangulation $\Delta^*$.
The edges of the dual triangulation are shared by three dual faces. If a such edge $e$ does not support an edge of the graph $\Gamma$, the corresponding group element appears in three characters and the associated integral involves the tensor product of three representations. For every such edge we use repeatedly the following formula
\beq
\label{3}
\int_{\SU(2)} d g \; \pi^i(g) \otimes \pi^j(g) \otimes \pi^k(g) = \iota \;  \iota^*,
\eeq
where $\iota : V_{i} \otimes V_j \otimes V_k \rightarrow \C$ is the Wigner $3j$ map, while  $\iota^*: \C \rightarrow V_{i} \otimes V_j \otimes V_k$ denotes the dual map. Introducing a basis $(e^j_{\alpha})_{\alpha=-j, .., j}$ of $V_j$,  the following evaluation
\beq
\iota(e_{\alpha}^i \otimes e_{\beta}^j \otimes e_{\gamma}^k) =
\left( \begin{array}{lll}
i & j & k \\
\alpha & \beta & \gamma
\end{array} \right),
\eeq
is the standard Wigner $3j$ symbol (see the Appendix for an account of our conventions). In the diagrammatic language, the formula \eqref{3} reads
 \beq
\label{3g}
\begin{array}{c}
\psfrag{i}{$i$}
\psfrag{j}{$j$}
\psfrag{k}{$k$}
\hspace{-8mm}
\includegraphics[scale=0.3]{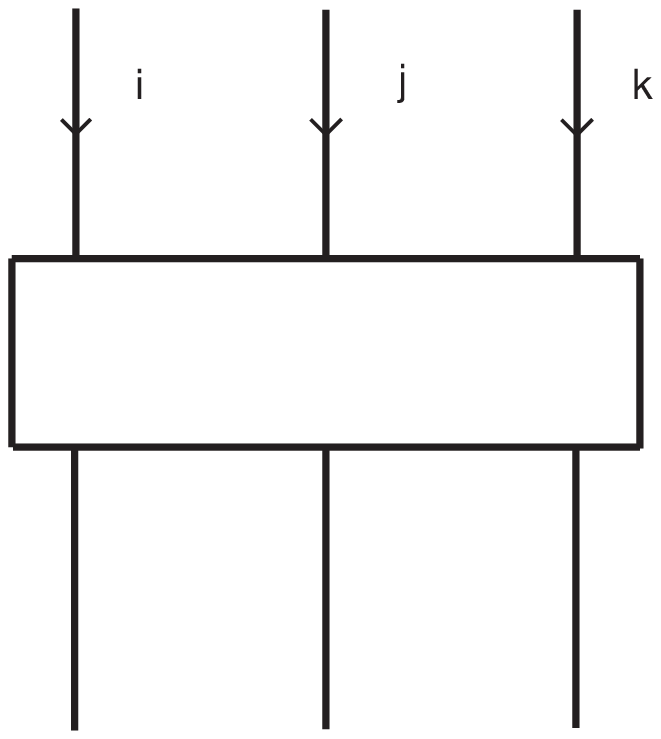}
\end{array}
= \begin{array}{c}
\psfrag{i}{$i$}
\psfrag{j}{$j$}
\psfrag{k}{$k$}
\psfrag{+}{$_+$}
\psfrag{-}{$_-$}
\includegraphics[scale=0.3]{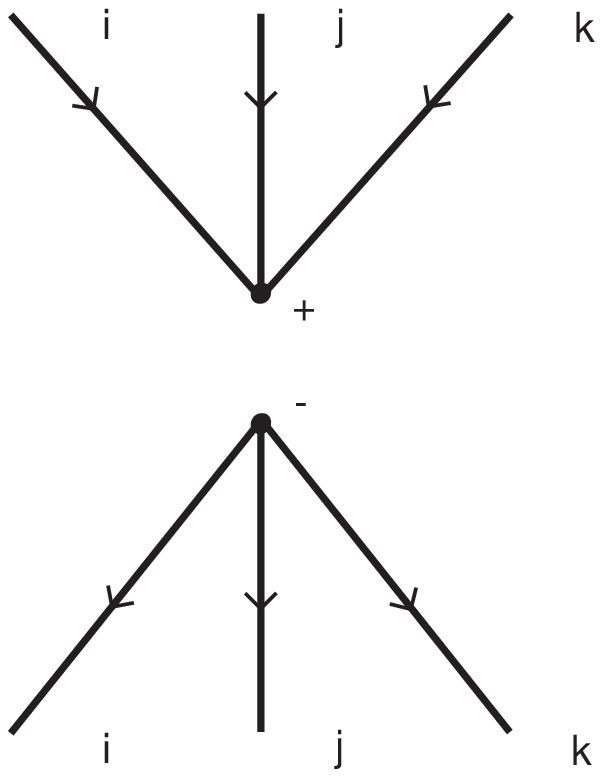}
\end{array}
\eeq
Note that no marking of lines is required for three-valent intertwiners, we only need to specify an orientation of the vertex. This is because of the cyclic symmetry of the $3j$ symbols (see equation \eqref{cyclicity} in the Appendix).

Note that we have only given the integration formula for the case where all three lines are pointing in the same direction, from top to bottom. If one line coloured by, say spin $k$, is pointing from bottom to top in equation \eqref{3valent integral}, the formula must be corrected by multiplying the right hand side by a phase given by $(-1)^{2k}$. This factor is the norm (squared) of the corresponding intertwiner. Since the Haar measure in invariant under the transformation $g \mapsto g^{-1}$, we now have covered all possible orientations configurations; if a line is pointing in a direction opposite to the two others, we pick up a phase given by $-1$ to the power twice the spin colouring the line.

Now, if an edge $e$ does support an edge of the graph $\Gamma$,
the corresponding integral involves the tensor product of the three representations assigned to the three wedges meeting on $e$ but also one (resp. two) fundamental $j=1/2$ representation associated to the holonomy matrix (matrices) $U_e$ appearing in the Dirac matrix (matrices) $D_e$ attached to the edge $e$. If there is only one spinor representation associated to $e$, the relevant formula to use is that of the tensor product of four representation matrices
\beq
\label{4}
\int_{\SU(2)} d g \; U(g) \otimes \pi^i(g) \otimes \pi^j(g) \otimes \pi^k(g) = \sum_s \iota_s \;  \iota^*_{s},
\eeq
where the spin $s$ labels a basis of the vector space of $4$-valent intertwiners obtained by choosing a recoupling scheme and decomposing the tensor product of two out of the four representations using the Clebsch-Gordan intertwining maps (see Appendix). For instance, if we choose the recoupling
$$
\Hom(V_i \otimes V_j \otimes V_k \otimes V_l, \C) = \bigoplus_s \left( \Hom(V_{i} \otimes V_j , V_s) \bigotimes \Hom(V_k \otimes V_l, V_s^*) \right),
$$
for arbitrary representations,
the four-valent intertwiner $ \iota_s$ is related to the 3j symbols by the following evaluation
\beq
\iota_s(e_{\alpha}^i \otimes e_{\beta}^j \otimes e_{\gamma}^k \otimes e_{\delta}^l) =
\left( \begin{array}{llll}
 i & j & k & l \\
\alpha & \beta & \gamma & \delta
\end{array} \right)_{s} = \sqrt{\dim s} \left( \begin{array}{lll}
i & j & \epsilon  \\
\alpha & \beta & s
\end{array} \right) \left( \begin{array}{lll}
s & k & l \\
\epsilon & \gamma & \delta
\end{array} \right),
\eeq
where the indices are raised and lowered, i.e., the lines inside a given column are interchanged, according to the convention displayed in the Appendix and summation over repeated indices is understood.  One could obviously choose another recoupling of representations. The coefficient appearing in the change of basis between the chosen decomposition and the one given by a different recoupling of representations is (proportional to) a six-j symbol (see equation \eqref{recoupling1} in the Appendix).

In the diagrammatic language, equation \eqref{4} yields
\beq
\label{4g}
\begin{array}{c}
\psfrag{i}{$i$}
\psfrag{j}{$j$}
\psfrag{k}{$k$}
\psfrag{h}{$\frac{1}{2}$}
\psfrag{I}{$ $}
\psfrag{1}{$1$}
\psfrag{s}{$s$}
\includegraphics[scale=0.3]{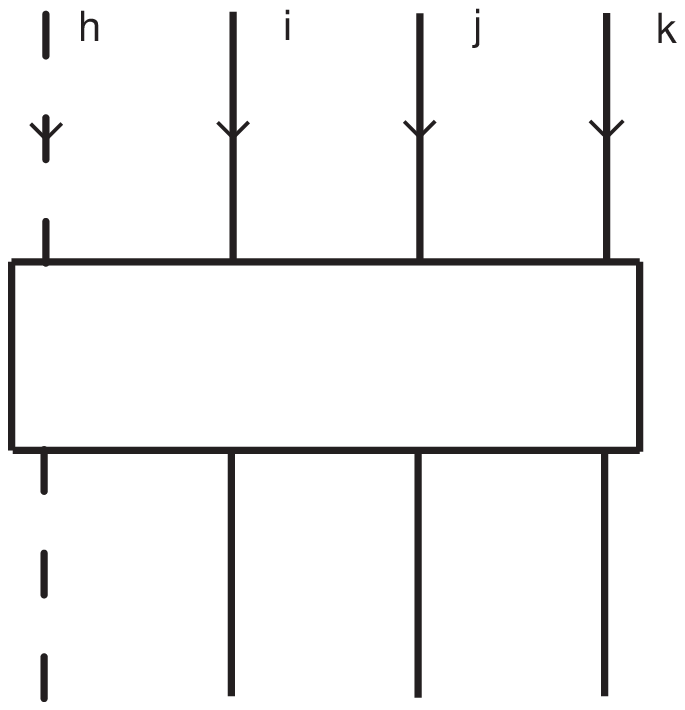}
\end{array}
=
\sum_s
\begin{array}{c}
\psfrag{i}{$i$}
\psfrag{j}{$j$}
\psfrag{k}{$k$}
\psfrag{h}{$\frac{1}{2}$}
\psfrag{I}{$ $}
\psfrag{1}{$1$}
\psfrag{s}{$s$}
\psfrag{+}{$_+$}
\psfrag{-}{$_-$}
\includegraphics[scale=0.3]{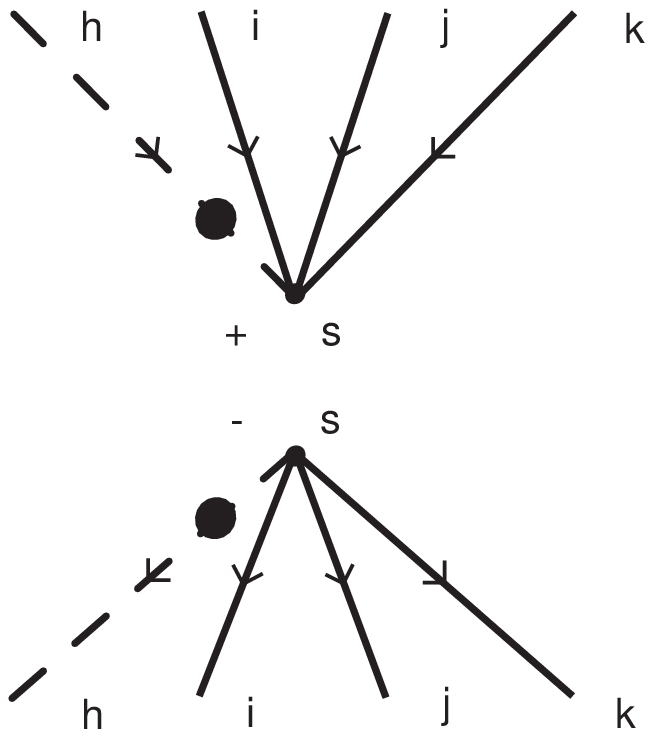}
\end{array}
=
\sum_s \dim s
\begin{array}{c}
\psfrag{i}{$i$}
\psfrag{j}{$j$}
\psfrag{k}{$k$}
\psfrag{h}{$\frac{1}{2}$}
\psfrag{I}{$ $}
\psfrag{1}{$1$}
\psfrag{s}{$s$}
\psfrag{+}{$_+$}
\psfrag{-}{$_-$}
\includegraphics[scale=0.3]{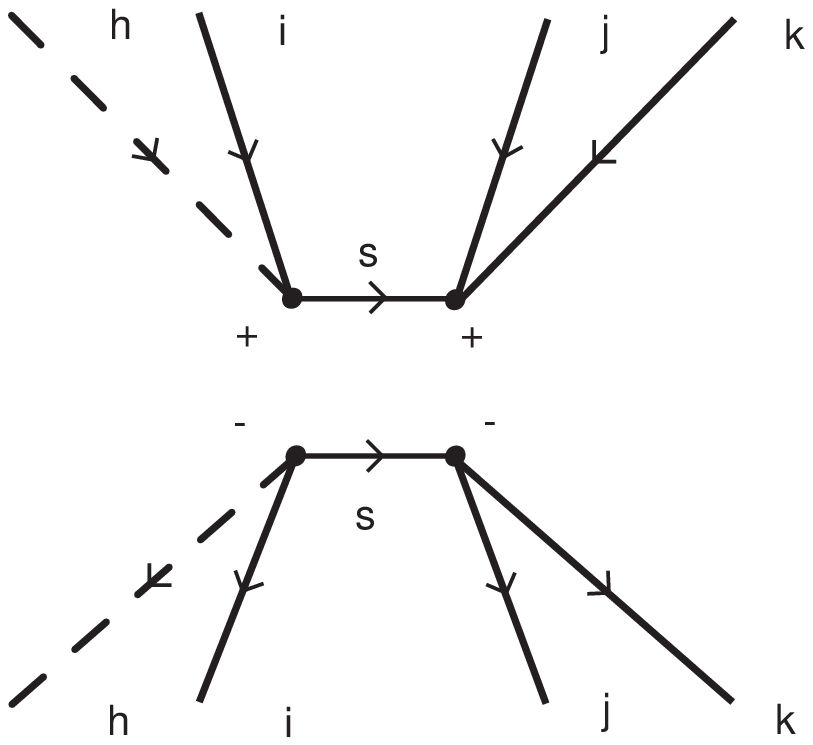}
\end{array}
\eeq
Note our convention in which the fundamental representations are represented by dashed lines.
As in the three-valent case, the formula would need to be corrected by a factor $(-1)^{2l}$ if the line of spin $l$ was pointing in a direction opposite to the three others, and by a factor $(-1)^{2(k+l)}$ if the lines $k$ and $l$ where going in opposite directions with respect to the two others.

There can also be two fundamental $j=1/2$ lines \footnote{Note that there could also be cases where a given edge $e$ is associated to three lines, two issued from the graph $\Gamma$ and one coming from a matrix $V_e$ defining the gauge-invariant observable. In such cases, one uses the obvious generalisation of the above formulae.} associated to the edge $e$. The corresponding integral involves the tensor product of five representations and the relevant formula is
\beq
\label{5}
\int_{\SU(2)} d g \; U(g) \otimes U(g) \otimes \pi^i(g) \otimes \pi^j(g) \otimes \pi^k(g) = \sum_{s,t} \iota_{s,t} \;  \iota^*_{s,t},
\eeq
where the spins $s,t$ now label two channels in a particular recoupling scheme. We will choose to couple the two spinor representations together and likewise for the $j$ and $k$ representations. The first coupling gives rise to $s$ while the second produces the label $t$.
For general representations, but under the same recoupling scheme, the five-valent intertwiner $\iota_{s,t}$ is also related to the 3j symbol as follows
\beq
\left( \begin{array}{ccccc}
 i & j & k & l & m \\ \beta & \delta & \zeta & \eta & \kappa \end{array} \right)_{s,t} = \sqrt{\dim s \dim t}\left( \begin{array}{ccc} i & j & \mu \\ \beta & \delta & s \end{array} \right)  \left( \begin{array}{ccc}  s & k & \sigma \\ \mu & \zeta & t \end{array} \right) \left( \begin{array}{ccc} t &  l & m \\ \sigma & \eta & \kappa \end{array} \right).
\eeq
Therefore, the diagrammatic representation of \eqref{5} yields
\beq
\label{5g}
\begin{array}{c}
\psfrag{i}{$\frac{1}{2}$}
\psfrag{j}{$i$}
\psfrag{k}{$j$}
\psfrag{l}{$k$}
\psfrag{h}{$\frac{1}{2}$}
\includegraphics[scale=0.35]{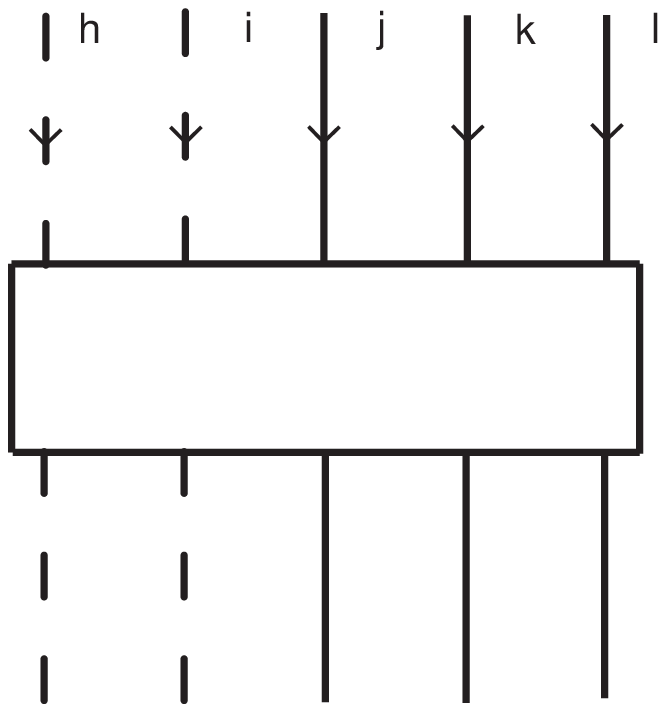}
\end{array}
 =
\sum_{s,t} \dim s \dim t
\begin{array}{c}
\psfrag{i}{$\frac{1}{2}$}
\psfrag{j}{$i$}
\psfrag{k}{$j$}
\psfrag{l}{$k$}
\psfrag{h}{$\frac{1}{2}$}
\psfrag{s}{$s$}
\psfrag{t}{$t$}
\psfrag{+}{$_+$}
\psfrag{-}{$_-$}
\includegraphics[scale=0.35]{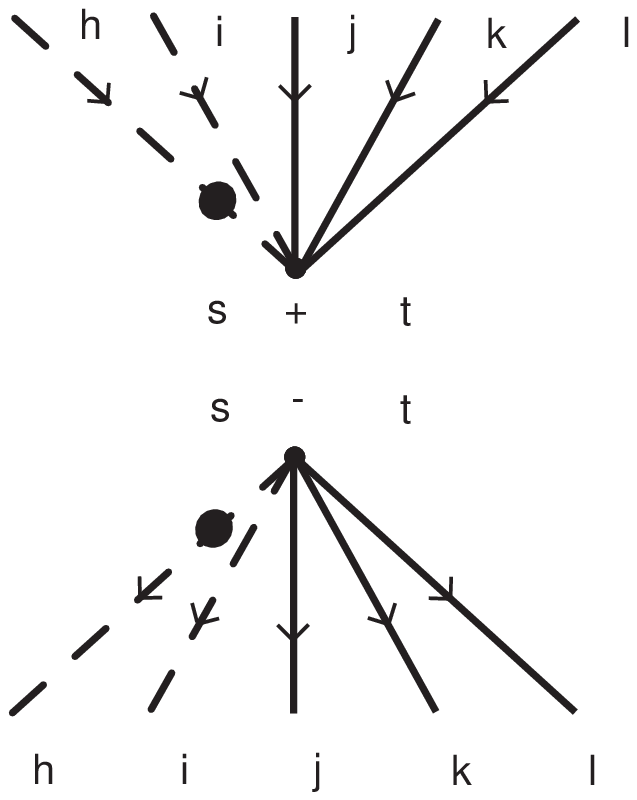}
\end{array}
=
\sum_{s,t} \dim s \dim t
\begin{array}{c}
\psfrag{i}{$\frac{1}{2}$}
\psfrag{j}{$i$}
\psfrag{k}{$j$}
\psfrag{l}{$k$}
\psfrag{h}{$\frac{1}{2}$}
\psfrag{s}{$s$}
\psfrag{t}{$t$}
\psfrag{+}{$_+$}
\psfrag{-}{$_-$}
\includegraphics[scale=0.35]{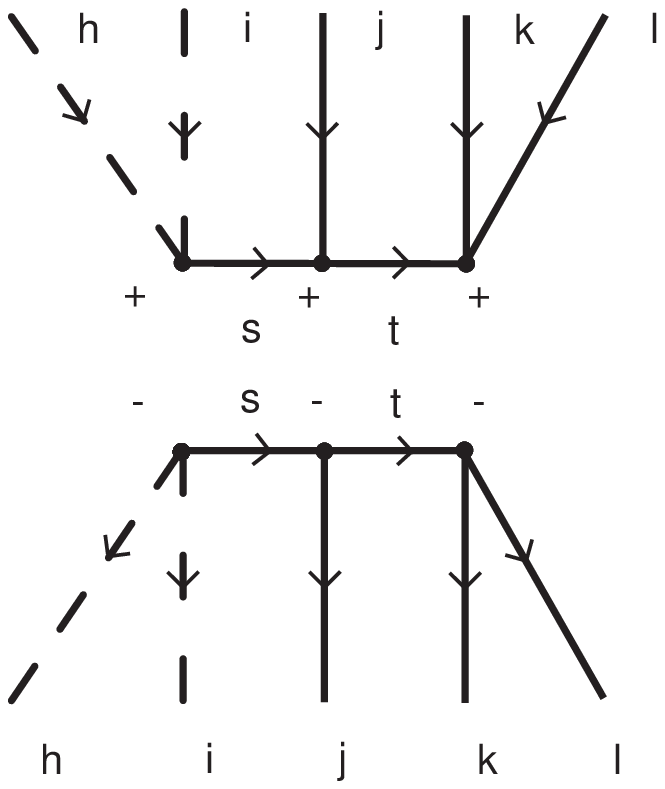}
\end{array}
\eeq
Again, to cover all possible orientation configurations, one simply needs to correct the formula by adding a phase $(-1)^{2k}$ if a line of spin $k$ is going in a direction opposite to the four others, and a phase $(-1)^{2(k+l)}$ is the two lines $k$ and $l$ are pointing in different directions than the three others.

Note that all the above integrals are invariant under permutation of the representations. Whilst a single intertwiner picks up a phase if two arguments are permuted, the whole integral is invariant because the phases appearing in each one of the two intertwiners appearing when performing the integrals cancel.

Putting everything together, we obtain that the generating functional \eqref{I5} for a fixed assignment of spins factorises into amplitudes for the vertices of the dual triangulation
\beq
\label{factorised}
A_{\Gamma}^{0 (a,b)}(j_f,J) = \prod_v A_v ^{(a,b)}(j_f,J).
\eeq
The vertex amplitudes $A_v$ are modified $6j$ symbols with source insertions \cite{freidel-1999-2} of the type discovered in  \cite{fairbairn-2007-39}. Their explicit expression depends on the observable, and thus on the graph $\Gamma$. Typically, the vertex amplitudes are functions of the six spins and of the six sources assigned to the six wedges sharing the corresponding vertex. They are calculated  by evaluating the representations on the sources. The indices of the resulting matrices $\pi^j(e^J)$ are contracted with the indices of the (three, four or five-valent) intertwiners issued from the Haar integrals. The tensor contraction pattern can be conveniently read out of the the diagrammatic representation of these amplitudes.

The spin network diagrams associated to the vertex amplitudes are all built out of four vertices, or nodes, and eight (possibly nine) lines. The nodes represent the intertwiners issued from the Haar integrations and six out of the eight lines correspond to the spins coloring the faces, or wedges, meeting on the vertex $v$ and contain a source insertion. The two (possible three) remaining lines are in the fundamental representation and are associated to the two edges of the graph $\Gamma$ traversing the vertex $v$, or to one (possibly two) edge of $\Gamma$ and one edge supporting the holonomy $V_e$ defining the observable if $v$ is a vertex with a fermionic field insertion. Note that a matter line can form a loop based at a vertex if it is associated to an edge of $\Gamma$.
In addition, the matter lines in the spinor representations are marked with a vertex with an open line in the adjoint representation colored by an adjoint index $a$ or $b$, weighted by the number $i\sqrt{6}/2$, if they are associated to an edge of $\Gamma$. This is because to the edges $e$ of $\Gamma$ are assigned the quantities $\gamma_{a_e} U_e$ (see e.g. \eqref{I5}). Thus the holonomies are composed with generators of $\su(2)$ in the spinor $j=1/2$ representation. As explained below \eqref{generator}, such a generator is an unnormalised intertwiner between the adjoint representation, a spinor representation and its dual and is thus represented diagrammatically by a vertex weighted by its norm, that is, the number $i\sqrt{6}/2$. This closes our state-sum representation of the generating functional. We are now ready to calculate the action of the source derivative in \eqref{operator}.

\paragraph{Grasping operators.}

Deriving an exponentiated current in the $j$ representation with respect to the $a$-th component of the source produces the evaluation of the $\su(2)$ generator $\gamma_a$ in the Lie algebra representation $j$ :
\beq
\label{grasp}
\left. \frac{\delta}{\delta J^a} \pi^j(e^J)\right|_{J=0} = \; \pi^j_*(\gamma_a),
\eeq
where $\pi_*^j$ is the Lie algebra representation induced by the group representation $\pi^j$.

Now, the linear map $\pi_*^j: \su(2) \rightarrow \End \, V_j$ commutes with the group action and is therefore an intertwining operator between the adjoint representation, a $j$ representation and its dual
\beq
\label{generator}
\pi_*^j \;\;\;\; \in \;\;\;\; \Hom(V_1,  V_j \otimes  V_j^*).
\eeq
This interwiner is not normalised and its norm (squared) is easily evaluated to be $\Theta(j)^2:=\Theta(1,j,j)^2= (-1)^{2j} j(j+1)(2j+1)$. Diagrammatically, a Lie algebra representation is thus represented as
\beq
\pi^j_* = \Theta(j)
\begin{array}{c}
\psfrag{i}{$j$}
\psfrag{j}{$1$}
\psfrag{k}{$j$}
\psfrag{+}{$_+$ }
\includegraphics[scale=0.25]{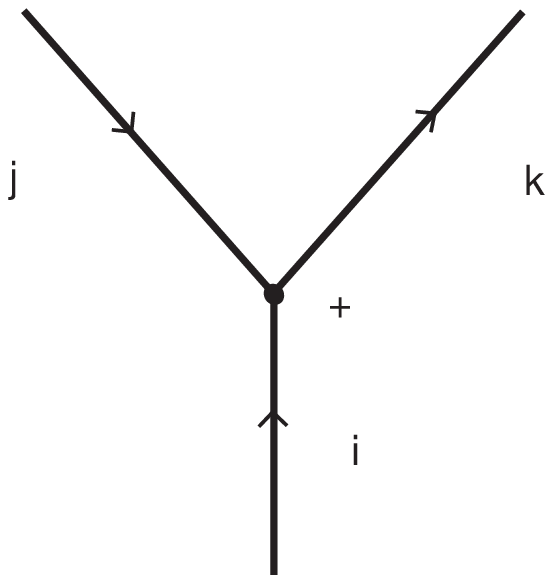}
\end{array}
\eeq
Therefore the action of source derivation with respect to the $a$-th component of the current, also
called grasping, modifies a spin network with source insertion by creating a new vertex attached to
an open line coloured by the index $a$ in the adjoint representation.

We can now readily see the action of the operator $\hat{\Sigma}_e^a$ associated to an edge $e$ of the graph $\Gamma$ in \eqref{operator} on the vertex amplitude associated to the vertex which is the source of $e$ taken at $J=0$. Each term in the sum over wedges in \eqref{sigma} marks each one of the two representation lines associated to the two wedges $w_e$ and $w'_e$ with a vertex attached to an open line. The two resulting open lines are coloured by the indices $b$ and $c$ in the adjoint representation. These two lines, together with the open line associated to the edge $e$ appearing in the vertex amplitude, connect through the totally antisymmetric three dimensional Levi-Civita tensor regarded as $3j$ symbol
\beq
\epsilon^{abc} = \sqrt{6}
\begin{array}{c}
\psfrag{i}{$1$}
\psfrag{j}{$1$}
\psfrag{k}{$1$}
\psfrag{a}{$a$}
\psfrag{b}{$b$}
\psfrag{c}{$c$}
\psfrag{+}{$_+$ }
\includegraphics[scale=0.25]{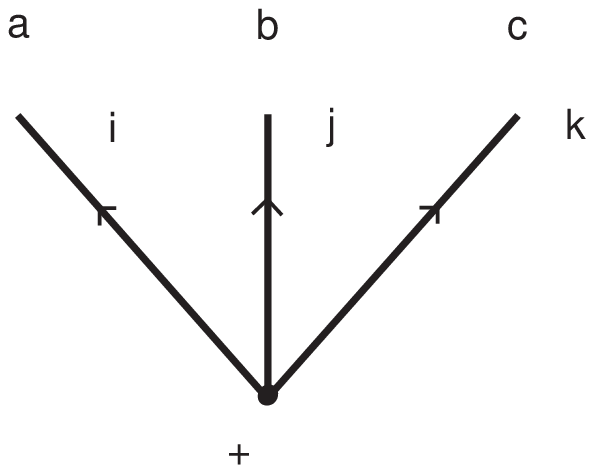}
\end{array}.
\eeq
This is the explicit picture of the soldering of the spin space to the quantum geometry
of spacetime. One must then sum over over all admissible wedges.

Finally, the amplitude \eqref{operator} associated to a graph $\Gamma$ and to the chosen configuration $c_{\Gamma} = 0$ is thus given by an expression of the type
\beq
A^0_{\Gamma} = (i \alpha)^{E_{\Gamma}} \prod_{f}  \sum_{j_f} d_{j_f} \prod_v \sum_{s_v} A_v(j_f,s_v).
\eeq
Here, the sum over $s_v$ labels each of the elements of the bases of intertwiners for the intertwiners of valence greater than three appearing in the amplitude $A_v$ associated to the vertex $v$. A sum over graspings is implicitly assumed in the definition of the amplitude $A_v$. Graphically, $A_v$ is given by a (sum over) spin network diagram(s) weighted by a function of the spins $N(j)$ obtained by assigning the quantity $(-i / 4) \Theta(j_{f_e}) \Theta(j_{f_{e'}})$ to each grasping occuring in the diagram, and a phase factor $(-1)^{\chi}$, with $\chi$ the sum of all spins associated to the lines oriented differently from the other lines at each vertex of the diagram which is issued from a Haar integration. All possible spin networks  are given in Figure $1$. Each one of these diagrams can be expressed in terms of $6j$ symbols using the recoupling identities given in the Appendix.
For a particular triangulation and a particular graph/configuration, the obtained vertex amplitudes are recoupled in the following section.

The extension to the other configurations $c \neq 0$ associated to the graph $\Gamma$ is immediately obtained from the above derivation.

\begin{figure}
\begin{center}
\begin{tabular}{cccc}
  \psfrag{j1}{$j_1$}
\psfrag{j2}{$j_2$}
\psfrag{j3}{$j_3$}
\psfrag{j4}{$j_4$}
\psfrag{j5}{$j_5$}
\psfrag{j6}{$j_6$}
  \includegraphics[scale=0.40]{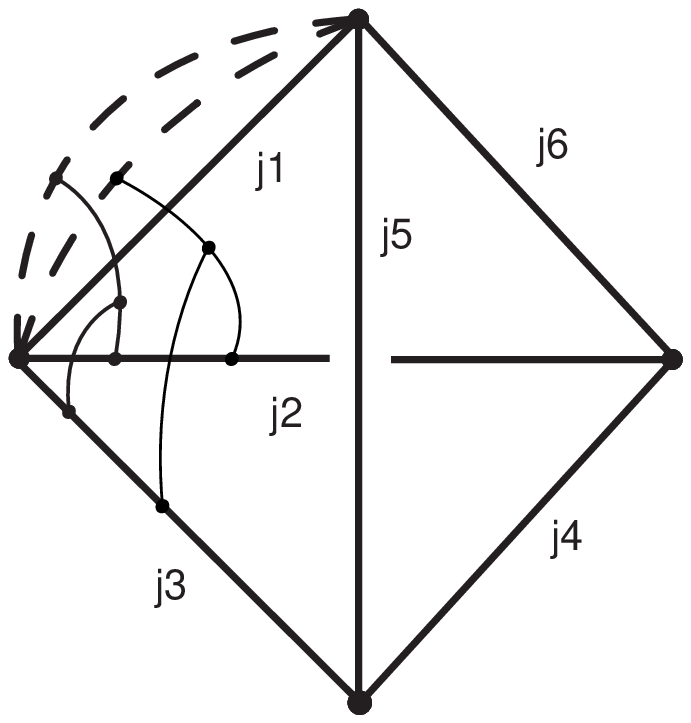} &
  \psfrag{j1}{$j_1$}
\psfrag{j2}{$j_2$}
\psfrag{j3}{$j_3$}
\psfrag{j4}{$j_4$}
\psfrag{j5}{$j_5$}
\psfrag{j6}{$j_6$}
\includegraphics[scale=0.40]{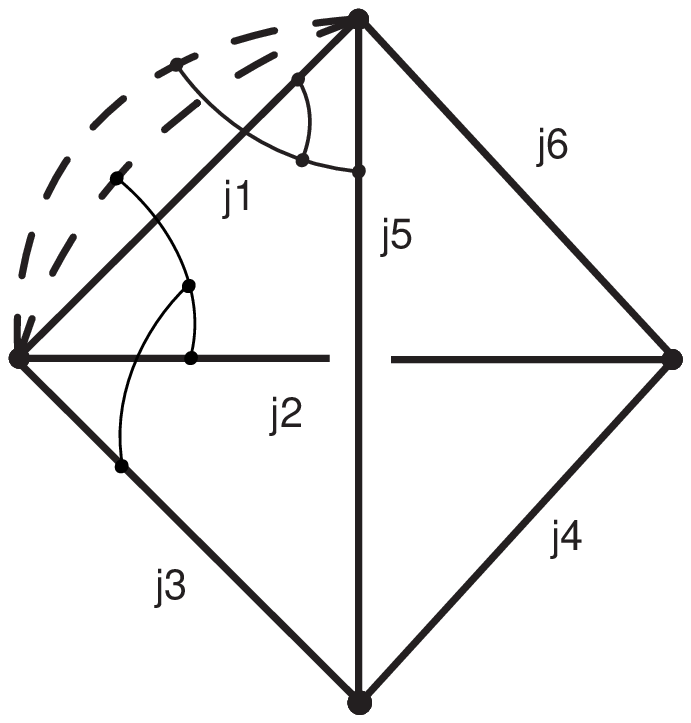} &
\psfrag{j1}{$j_1$}
\psfrag{j2}{$j_2$}
\psfrag{j3}{$j_3$}
\psfrag{j4}{$j_4$}
\psfrag{j5}{$j_5$}
\psfrag{j6}{$j_6$}
\includegraphics[scale=0.40]{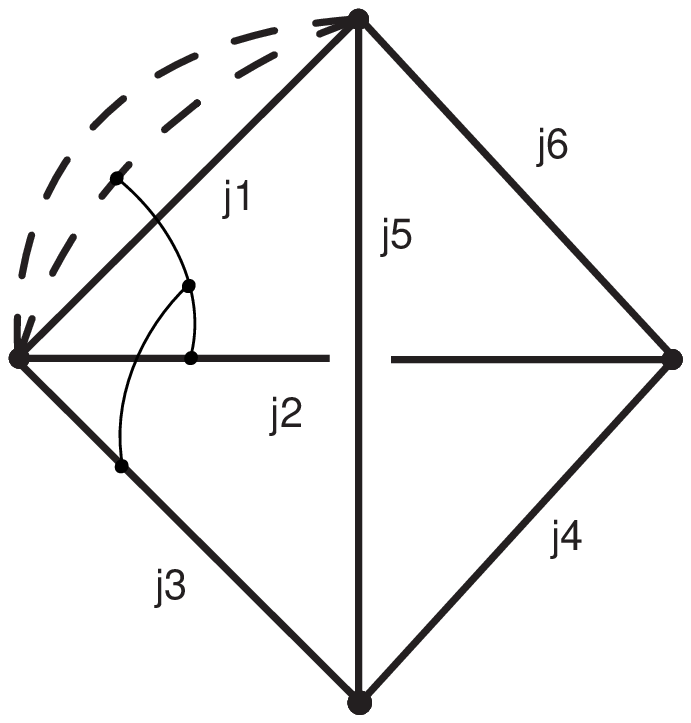} &
\psfrag{j1}{$j_1$}
\psfrag{j2}{$j_2$}
\psfrag{j3}{$j_3$}
\psfrag{j4}{$j_4$}
\psfrag{j5}{$j_5$}
\psfrag{j6}{$j_6$}
\includegraphics[scale=0.40]{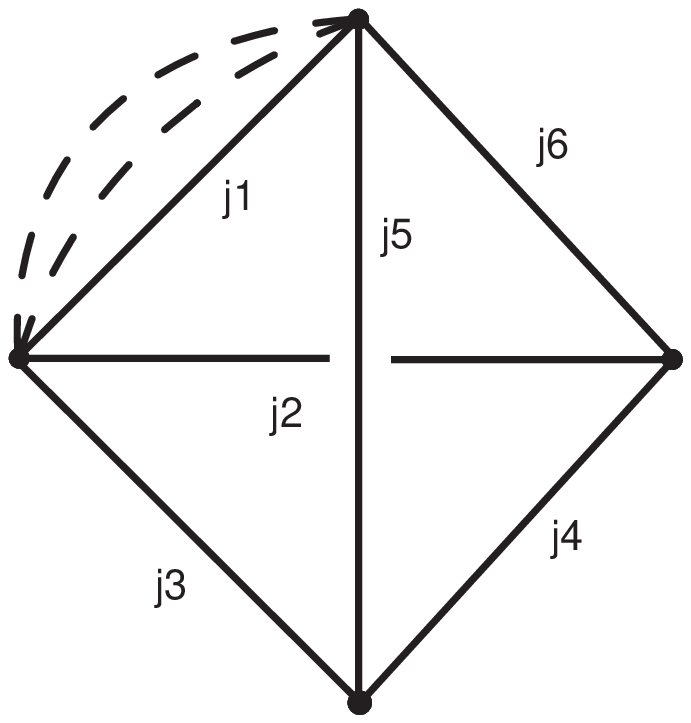} \\
  \psfrag{j1}{$j_1$}
\psfrag{j2}{$j_2$}
\psfrag{j3}{$j_3$}
\psfrag{j4}{$j_4$}
\psfrag{j5}{$j_5$}
\psfrag{j6}{$j_6$}
  \includegraphics[scale=0.40]{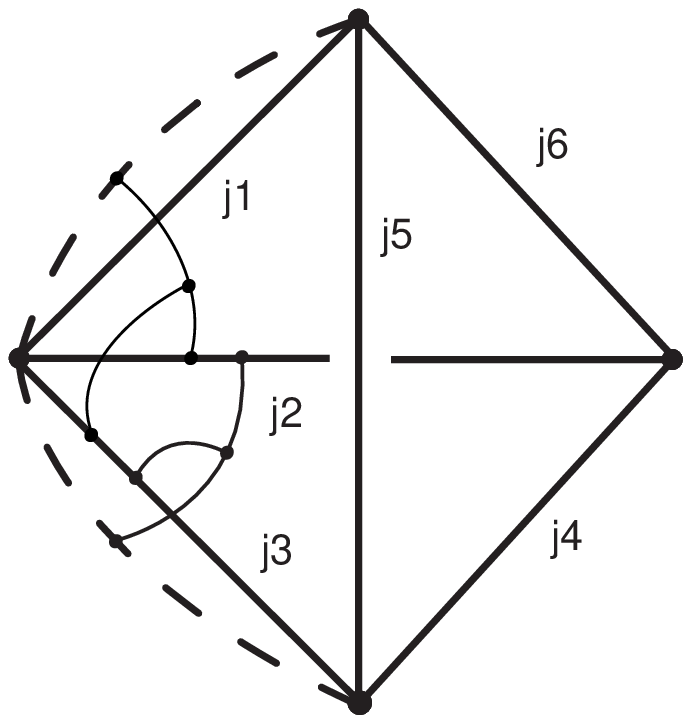} &
  \psfrag{j1}{$j_1$}
\psfrag{j2}{$j_2$}
\psfrag{j3}{$j_3$}
\psfrag{j4}{$j_4$}
\psfrag{j5}{$j_5$}
\psfrag{j6}{$j_6$}
\includegraphics[scale=0.40]{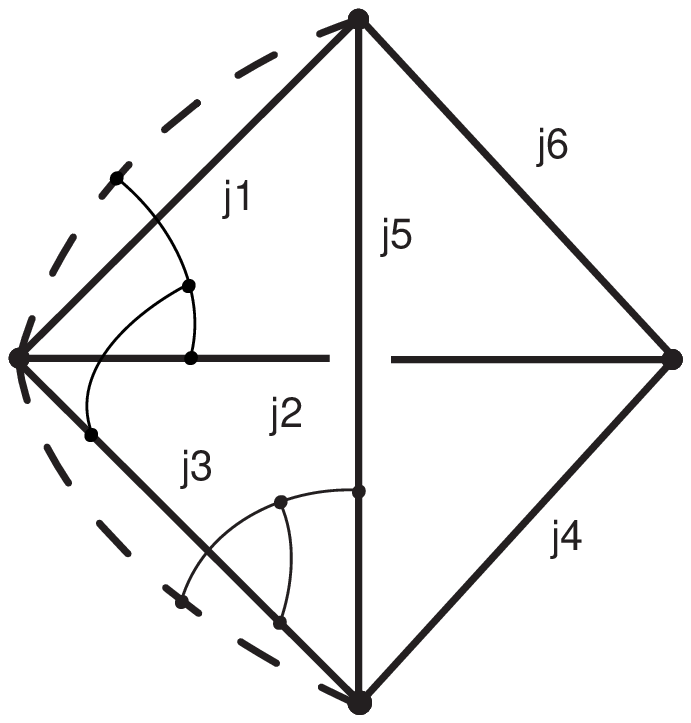} &
\psfrag{j1}{$j_1$}
\psfrag{j2}{$j_2$}
\psfrag{j3}{$j_3$}
\psfrag{j4}{$j_4$}
\psfrag{j5}{$j_5$}
\psfrag{j6}{$j_6$}
\includegraphics[scale=0.40]{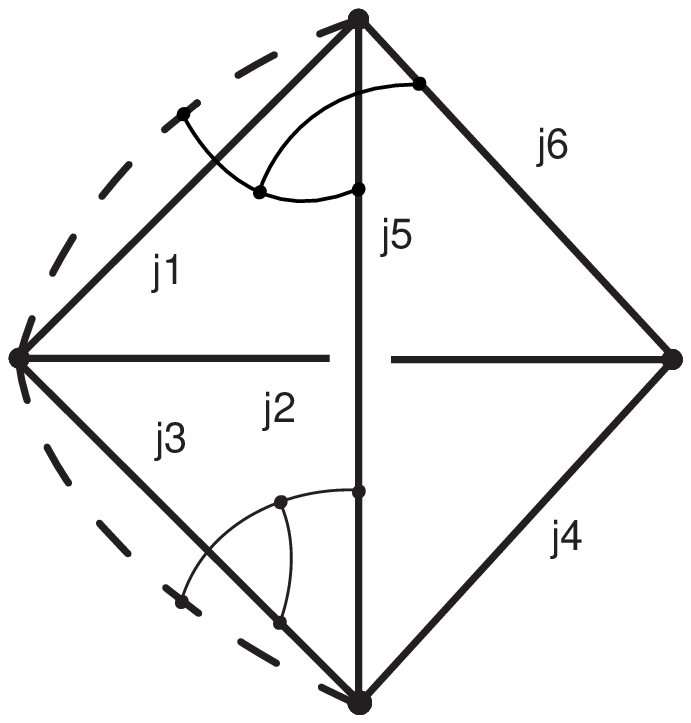} &
\psfrag{j1}{$j_1$}
\psfrag{j2}{$j_2$}
\psfrag{j3}{$j_3$}
\psfrag{j4}{$j_4$}
\psfrag{j5}{$j_5$}
\psfrag{j6}{$j_6$}
\includegraphics[scale=0.40]{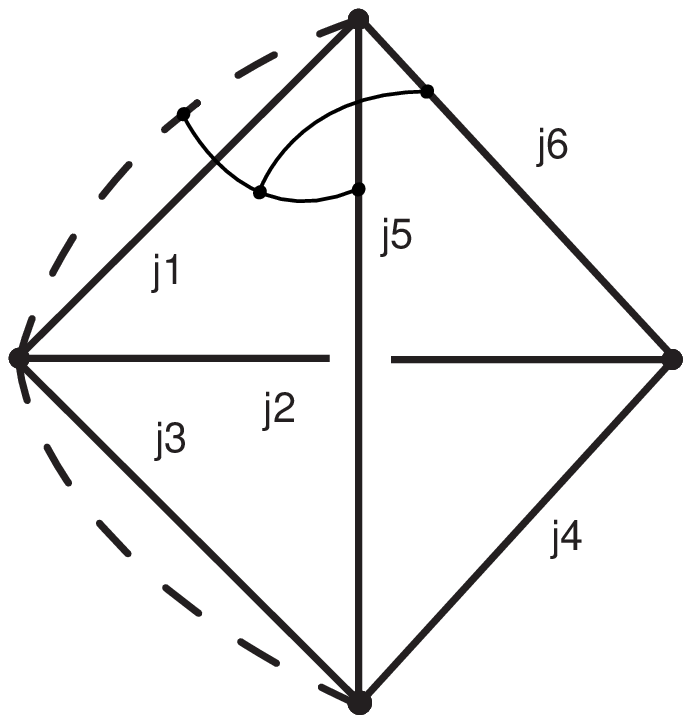} \\
  \psfrag{j1}{$j_1$}
\psfrag{j2}{$j_2$}
\psfrag{j3}{$j_3$}
\psfrag{j4}{$j_4$}
\psfrag{j5}{$j_5$}
\psfrag{j6}{$j_6$}
  \includegraphics[scale=0.40]{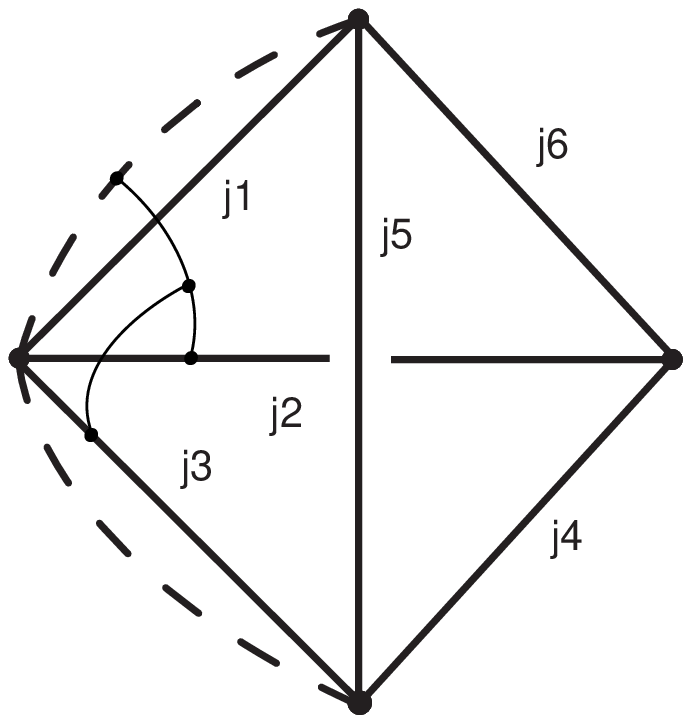} &
  \psfrag{j1}{$j_1$}
\psfrag{j2}{$j_2$}
\psfrag{j3}{$j_3$}
\psfrag{j4}{$j_4$}
\psfrag{j5}{$j_5$}
\psfrag{j6}{$j_6$}
\includegraphics[scale=0.40]{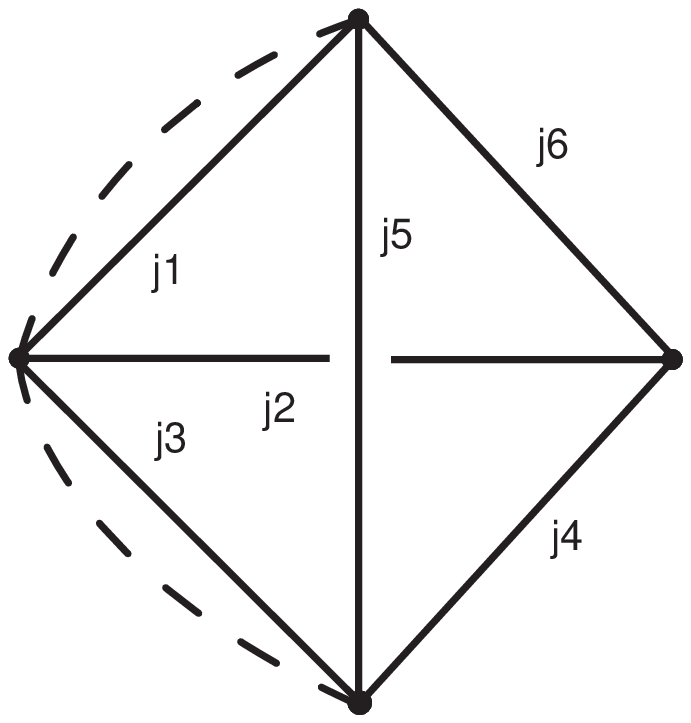} &
\psfrag{j1}{$j_1$}
\psfrag{j2}{$j_2$}
\psfrag{j3}{$j_3$}
\psfrag{j4}{$j_4$}
\psfrag{j5}{$j_5$}
\psfrag{j6}{$j_6$}
\includegraphics[scale=0.40]{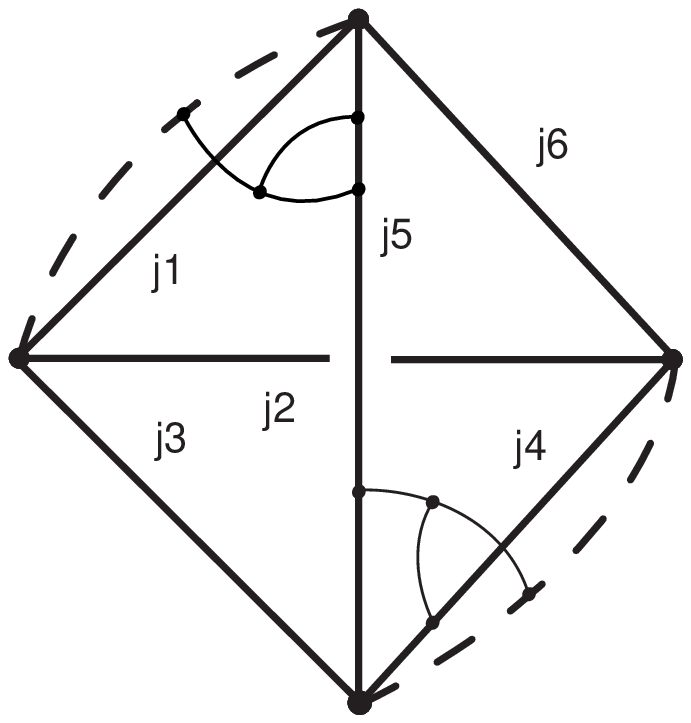} &
\psfrag{j1}{$j_1$}
\psfrag{j2}{$j_2$}
\psfrag{j3}{$j_3$}
\psfrag{j4}{$j_4$}
\psfrag{j5}{$j_5$}
\psfrag{j6}{$j_6$}
\includegraphics[scale=0.40]{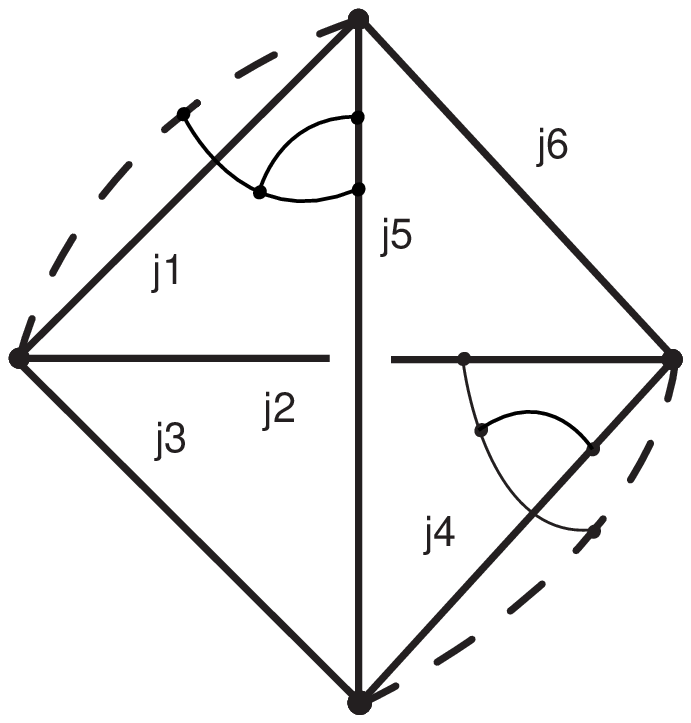} \\
\psfrag{j1}{$j_1$}
\psfrag{j2}{$j_2$}
\psfrag{j3}{$j_3$}
\psfrag{j4}{$j_4$}
\psfrag{j5}{$j_5$}
\psfrag{j6}{$j_6$}
  \includegraphics[scale=0.40]{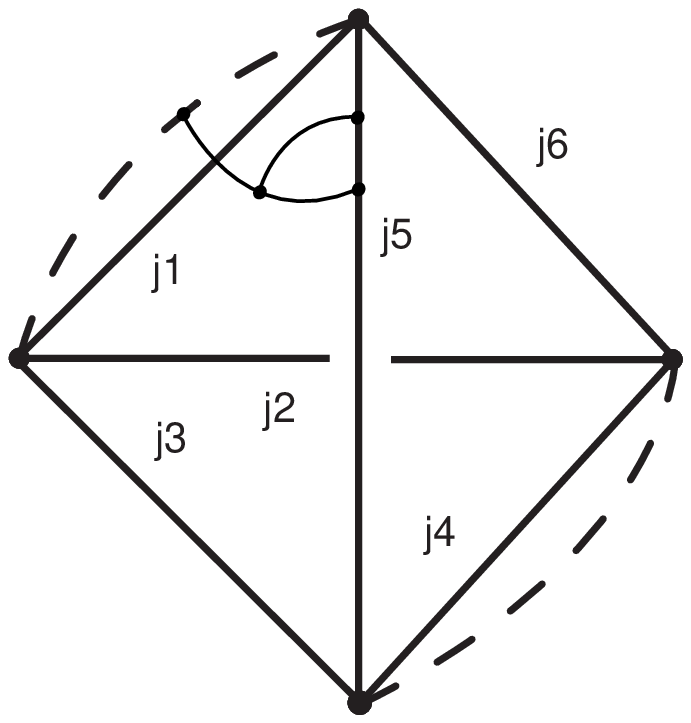} &
  \psfrag{j1}{$j_1$}
\psfrag{j2}{$j_2$}
\psfrag{j3}{$j_3$}
\psfrag{j4}{$j_4$}
\psfrag{j5}{$j_5$}
\psfrag{j6}{$j_6$}
\includegraphics[scale=0.40]{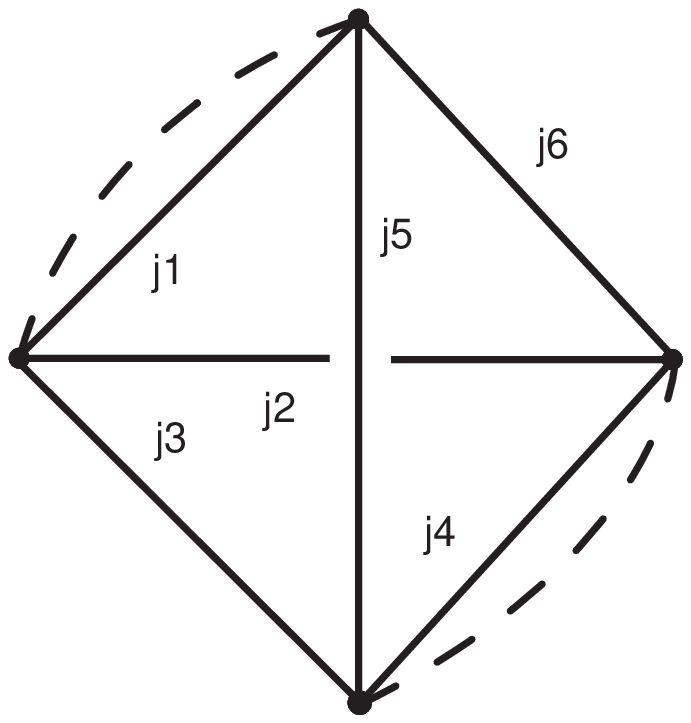} &
\psfrag{j1}{$j_1$}
\psfrag{j2}{$j_2$}
\psfrag{j3}{$j_3$}
\psfrag{j4}{$j_4$}
\psfrag{j5}{$j_5$}
\psfrag{j6}{$j_6$}
\includegraphics[scale=0.40]{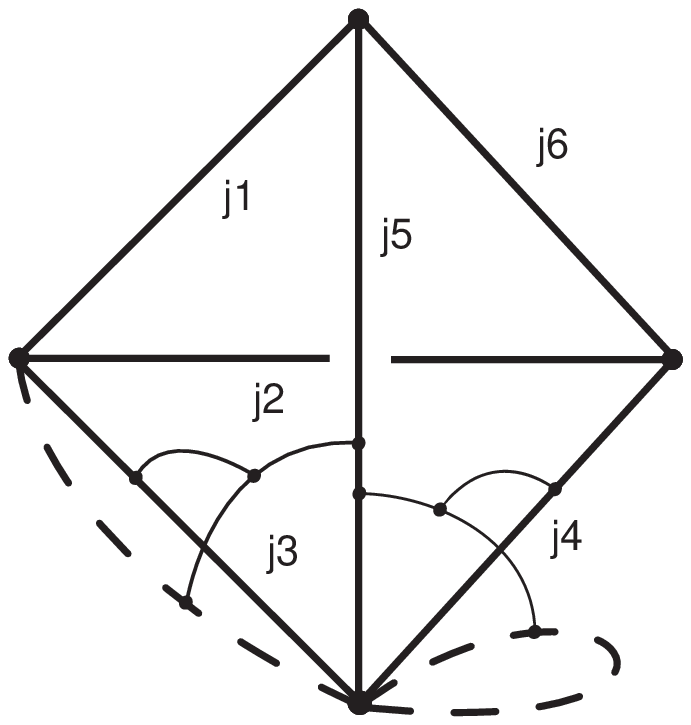} &
\psfrag{j1}{$j_1$}
\psfrag{j2}{$j_2$}
\psfrag{j3}{$j_3$}
\psfrag{j4}{$j_4$}
\psfrag{j5}{$j_5$}
\psfrag{j6}{$j_6$}
\includegraphics[scale=0.40]{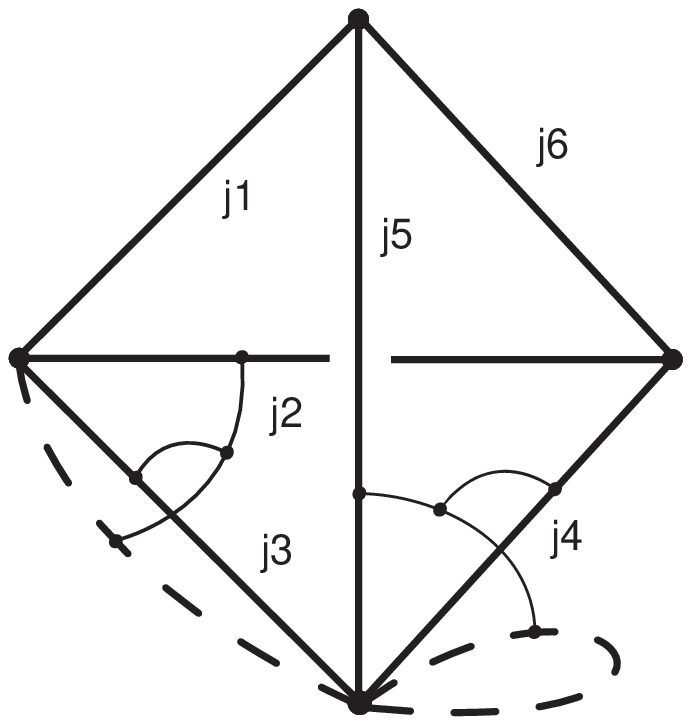} \\
\psfrag{j1}{$j_1$}
\psfrag{j2}{$j_2$}
\psfrag{j3}{$j_3$}
\psfrag{j4}{$j_4$}
\psfrag{j5}{$j_5$}
\psfrag{j6}{$j_6$}
  \includegraphics[scale=0.40]{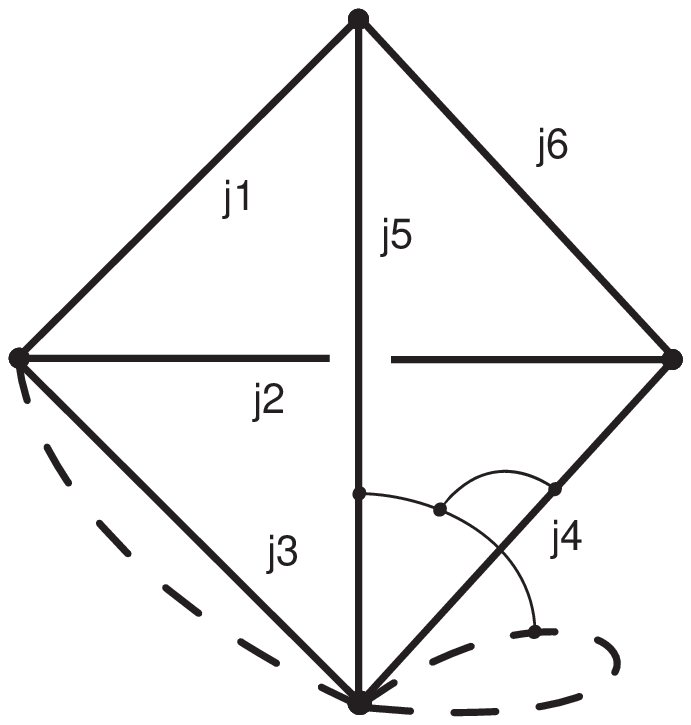} &
  \psfrag{j1}{$j_1$}
\psfrag{j2}{$j_2$}
\psfrag{j3}{$j_3$}
\psfrag{j4}{$j_4$}
\psfrag{j5}{$j_5$}
\psfrag{j6}{$j_6$}
\includegraphics[scale=0.40]{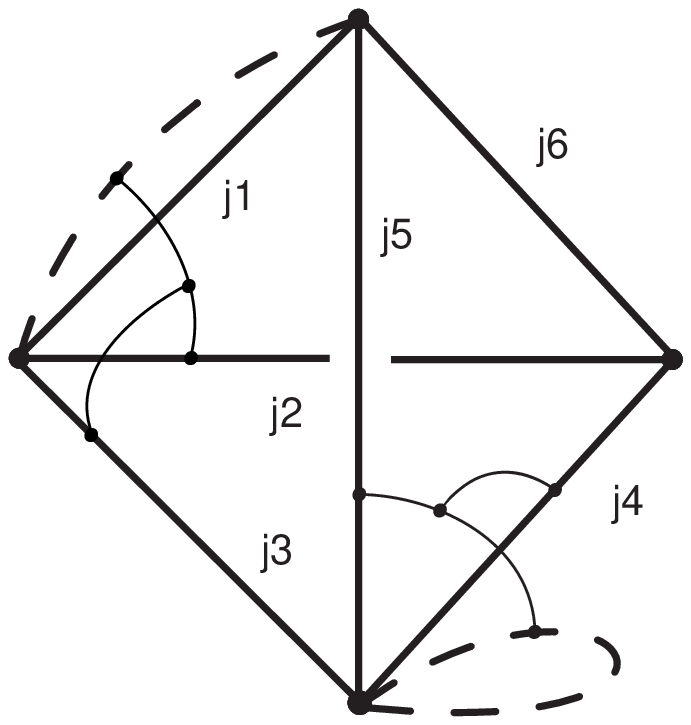} &
\psfrag{j1}{$j_1$}
\psfrag{j2}{$j_2$}
\psfrag{j3}{$j_3$}
\psfrag{j4}{$j_4$}
\psfrag{j5}{$j_5$}
\psfrag{j6}{$j_6$}
\includegraphics[scale=0.40]{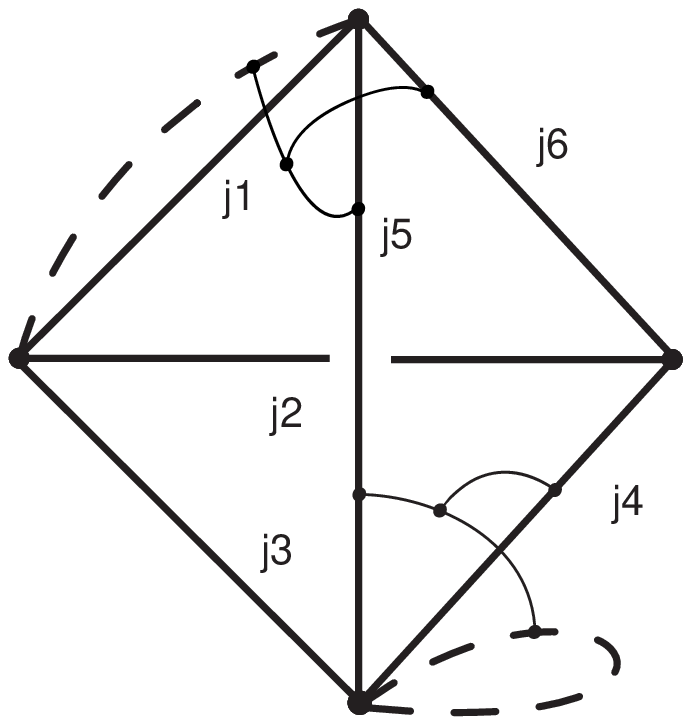} &
\psfrag{j1}{$j_1$}
\psfrag{j2}{$j_2$}
\psfrag{j3}{$j_3$}
\psfrag{j4}{$j_4$}
\psfrag{j5}{$j_5$}
\psfrag{j6}{$j_6$}
\includegraphics[scale=0.40]{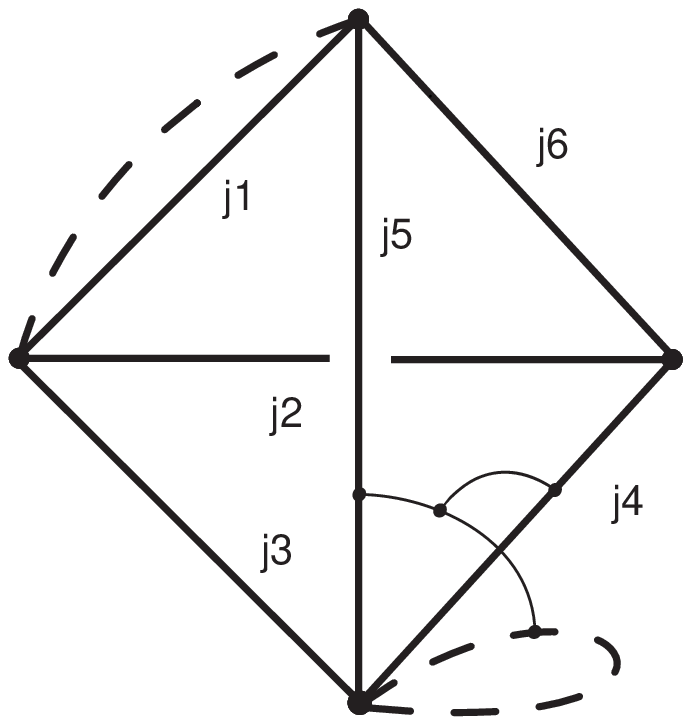} \\
&
\psfrag{j1}{$j_1$}
\psfrag{j2}{$j_2$}
\psfrag{j3}{$j_3$}
\psfrag{j4}{$j_4$}
\psfrag{j5}{$j_5$}
\psfrag{j6}{$j_6$}
  \includegraphics[scale=0.40]{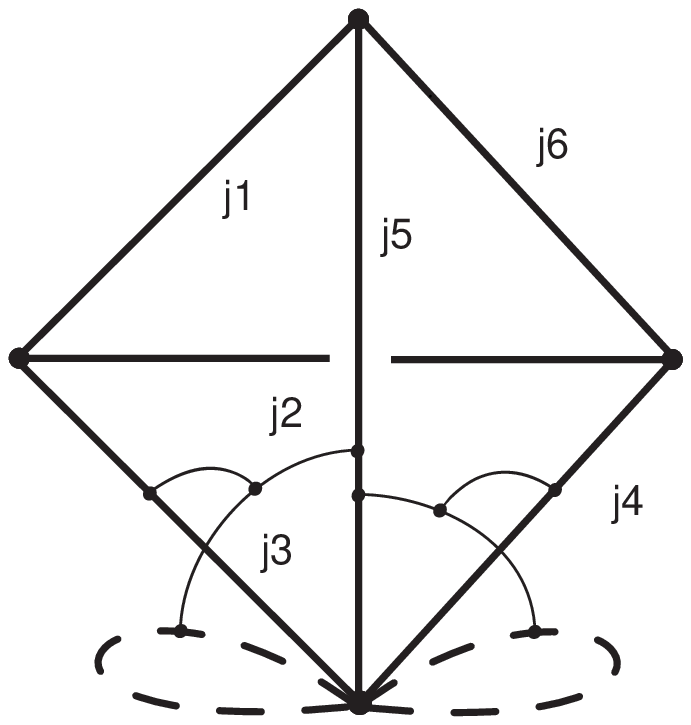} &
  \psfrag{j1}{$j_1$}
\psfrag{j2}{$j_2$}
\psfrag{j3}{$j_3$}
\psfrag{j4}{$j_4$}
\psfrag{j5}{$j_5$}
\psfrag{j6}{$j_6$}
\includegraphics[scale=0.40]{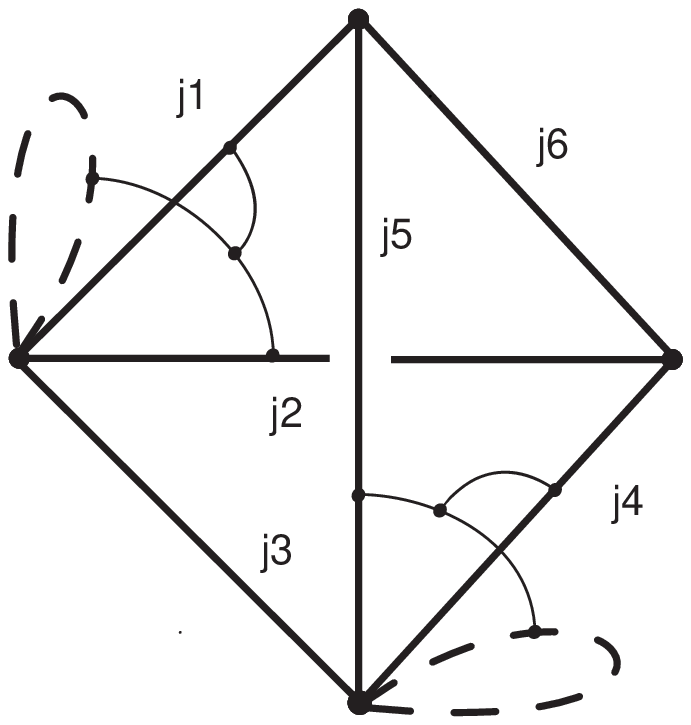}
 &

\end{tabular}

\caption{Spin network diagrams for the different possible vertex amplitudes $A_{v}$ (up to symmetry) of the gauge invariant observables.  The dashed lines are in the spin half representation and the curved lines denote the grasping operators in the spin one representation.
To each depicted diagram, one must add the two diagrams obtained by summing over the possible grapsings. Note that due to the $V_e$ terms there can be additional spin half lines that have not been included in these diagrams. We have also omitted for simplicity the ordering labels at each vertex and the arrows on the lines.}\label{fermion fig}
\end{center}
\end{figure}

\subsubsection{Gauge dependent observables}

When computing gauge dependent observables, the essential difference with the gauge invariant case comes from the fact that we can no longer suppress the gauge fixing function $\prod_{e \in T} \delta(g_e)$ from the computation if we do not want to obtain a vanishing expectation value. The resulting calculation will therefore depend on the gauge, and thus on the tree $T$.

As in the gauge invariant case, we choose a particular configuration $c = 0$ associated to an arbitrary graph $\Gamma$. The corresponding general form of the amplitude to be integrated yields
\beq
I^{0}_{\Gamma}(\mathbf{e}_w,U_e) =
(i \alpha)^{E_{\Gamma}} \prod_{\mathcal{P}} \prod_{e \in \mathcal{P}} \Sigma_e U_e  \; \left( \prod_{\mathcal{L} } \tr \prod_{e \in \mathcal{L}} \Sigma_e U_e \right).
\eeq

All the steps performed in the gauge invariant case for the calculation of $A_0$ transfer unchanged to the gauge dependent context up to equation \eqref{I5}, which is now modified with the inclusion of the gauge fixing function. After that, the derivation slightly differs from the gauge invariant case. The main consequence of the presence of the gauge fixing function is that not all group variables are integrated over. Therefore, although all formulae for the Haar integrals are implemented as in the gauge invariant case, the resulting generating functional at fixed spins \eqref{I5} can not put into factorised form as in \eqref{factorised}.

Typically, \eqref{factorised} is replaced by a function of the spins and of the currents given by a collection of invariant tensors the indices of which are contracted with the indices of representation matrices evaluated on the currents. The pattern of contraction depends on the triangulation, the tree, the graph and can be potentially complicated. Thanks to the diagrammatic methods, one can read the contraction patterns out of the corresponding (open) spin network. The evaluation of this diagram is a function of the currents and yields the value of the generating functional
\beq
A^{0 (a,b)}_{\Gamma}(j_f,J) = \langle \bigotimes_l \pi^{j_l}(e^{J_l}) ,  \bigotimes_n \iota_n \rangle,
\eeq
where $l$ and $n$ stand for the lines and nodes, or vertices, of the diagram, and the pairing $\langle, \rangle$ denotes the contraction of the indices determined by the combinatorics of the diagram.

The implementation of the grasping operator $\hat{\Sigma}_e^a$ proceeds in the same way than in the gauge invariant case and the amplitude associated to a graph $\Gamma$, to the chosen configuration $c_{\Gamma} = 0$ and to a particular tree $T$ is thus given by an expression of the form
\beq
A^0_{\Gamma} = (i \alpha)^{E_{\Gamma}} \prod_f \sum_{j_f} d_{j_f} \prod_v \sum_{s_v} A(j_f,s_v),
\eeq
where $A(j_f,s_v)$ is given by the evaluation of a particular spin network diagram determined by the triangulation $\Delta$, the tree $T$ and the graph $\Gamma$. A sum over all possible graspings of each operator $\Sigma_e$ for each edge $e$ is assumed in the notation so that $A(j_f,s_v)$ is in fact a sum over spin network diagrams.

The exact form of $A(j_f,s_v)$ can be obtained by the following graphical method based on the cable and wire technology \cite{oecklbook,Girelli:2001wr}.
\begin{itemize}
\item To each $e$ of the graph $\Gamma$, associate an oriented open $j=1/2$ line which is marked by a vertex from which emerges an open line in the adjoint $j=1$ representation.
\item To each face $f$ of the dual triangulation $\Delta^*$ which does not touch any edge $e$ of $\Gamma$ associate a closed loop coloured by a spin $j_w$ and orientated according to the orientation of the face.
\item For each edge $e$ of the graph $\Gamma$, select a couple of faces $f_e, f_{e'}$ (or of wedges $w_e, w_{e'}$) among the three faces (wedges) touching $e$ and $s(e)$, and assign an oriented closed loop marked with a vertex from which emerges an open line coloured by an adjoint representation $j=1$ to $f_e$ and $f_{e'}$.
\item For each edge $e$ of $\Gamma$, connect the two open lines emerging from the couple of faces $f_e, f_{e'}$ to a three-valent vertex  corresponding to the totally antisymmetric tensor $\epsilon^{a_eb_ec_e}$.
\item Connect, for each edge $e$ of $\Gamma$, the third line emerging from the three-valent vertex to the open line emerging from the vertex marking the $j=1/2$ line assigned to the edge $e$.
\item To each edge $e$ of the complex $\Delta^*$ not belonging to the tree $T$ is assigned a box traversed by all the lines associated to $e$, which correspond to the irreducible representations labelling the wedges adjacent to the edge $e$ possibly supplemented by spin $1/2$ representations \footnote{The order with which the lines enter the boxes is irrelevant because of the permutation symmetry discussed above.}.
\item Perform the integrals by replacing the boxes by the corresponding intertwiners using equations \eqref{3g}, \eqref{4g}, \eqref{5g}.
\item Assign the factor $(-i/4)\Theta(j_{f_e}) \Theta(j_{f_{e'}})$ to each grasping
\item Sum over the diagrams obtained by summing over all possible couples $f_e, f_{e'}$ for each edge $e$ of $\Gamma$.
\end{itemize}
The result of a such procedure for a given set of spins $\{j_f,s_v\}$ yields a number $A(j_f,s_v) \in \C$. From this spin network evaluation, one can calculate $A^0_{\Gamma}$. The same prescription then applies, modified appropriately, to all configurations $c$ of a given graph $\Gamma$. Note also that this graphical method applies to the gauge-invariant context by simply suppressing the tree.

This closes the general prescription to compute observables in 3d spinfoam quantum gravity with fermions. To sharpen the details and techniques of the procedure, we now fix a particular triangulated $3$-manifold and illustrate the techniques derived above to calculate the expectation value of the Polyakov line and of the fermion two-point function.

\section{Explicit calculations: observables on $\mathcal{S}^3$}
\label{examples}

In this section, we give examples of the calculations described above for a space-time manifold homeomorphic to the three-sphere, i.e.,  $M \cong \mathcal{S}^3$.
We consider the triangulation, denoted $\Delta$, of $\mathcal{S}^3$ with the boundary of a 4-simplex\footnote{The (pseudo-)triangulation constructed from only two tetrahedra can not be used as it is not possible to properly define the action. This is because there is more than one edge connecting two vertices. While this is not problematic when working with theories of flat connections, it becomes highly ambiguous when dealing with non-trivial curvature.} and label each vertex $v$ of $\Delta^*$ (the pentachoron graph) by the labels $I,J= 1, ... ,5$ which implies that the edges $e$ of $\Delta^*$ are labelled by couples $IJ$, with $I \neq J$, and the faces $f$ of $\D^*$ are labelled by triples $IJK$, with $I \neq J \neq K$ corresponding to the three vertices belonging to the face.
The Dirac part of the action \eqref{fermionaction} for $\Delta$, with rescaled triad, is
\beq
S_{\mbox{\tiny D}} =  \alpha \sum_{I \neq J} S_{IJ}.
\eeq
The examples we will consider are the Polyakov line $\mathcal{O}_{f }(e) = (\psi_{s(e)} , \, V_{e} \, \psi_{t(e)}) $, and the propagator $\mathcal{O}^{\;\,\, A}_{f  \;\, B}(x_1,x_2) = \psi_{x_1}^{A} \,\, \bar{\psi}_{x_2 B}$. We will now calculate the Berezin integrals explicitly to illustrate the Feynman rules for both observables.

\subsection{Gauge invariant observable: Polyakov line}

For this observable we consider $\mathcal{O}_{f}(23) = (\psi_{2} , \, V_{253} \, \psi_{3})  $, with $V_{253} = V_{25}V_{53}$. The bosonic observable $\mathcal{O}_{b}$ formed by integrating the Grassmann variables is
\beqa
\mathcal{O}_{b} = \left( \int_{\G} d \mu (\overline{\psi}_I, \hspace{1mm} \psi_I) \right) \hspace{1mm}
(\psi_{2} , \, V_{253} \, \psi_{3}) \,\, e^{i S_{\mbox{{\tiny D}}}}   ,
\eeqa
We will consider one relevant non-vanishing term in the Feynman expansion \eqref{expansion} in order to compute the amplitude $I_{\Gamma}$ of a single admissible graph  in equation \eqref{contributions}.
\beqa
\mathcal{O}_{b} &=& ...+  (i\alpha)^{9} \left( \int_{\G} d \mu (\overline{\psi}_I, \hspace{1mm} \psi_I) \right) \hspace{1mm}
\overline{\psi}_{2A}   V_{253 \ B}^{\ \, A}  \psi_{3}^B \, \nn \\
&& \times
\overline{\psi}_{1C}   D_{12 \, D}^{\ \  C} \psi_{2}^D \
\overline{\psi}_{2E}   D_{23 \, F}^{\ \  E} \psi_{3}^F \
\overline{\psi}_{3G}   D_{32 \, H}^{\ \  G} \psi_{2}^H \
\overline{\psi}_{3I}   D_{34 \, J}^{\ \  I} \psi_{4}^J \
\overline{\psi}_{4K}   D_{41 \, L}^{\ \  K} \psi_{1}^L \
\nn \\
&& \times
\overline{\psi}_{4M}   D_{45 \, N}^{\ \  M} \psi_{3}^N \
\overline{\psi}_{5P}   D_{54 \, Q}^{\ \  P} \psi_{4}^Q \
\overline{\psi}_{5R}   D_{51 \, S}^{\ \  R} \psi_{1}^S \
\overline{\psi}_{1T}   D_{15 \, U}^{\ \  T} \psi_{5}^U \
 + ...,
\eeqa
including the factor of 2 coming from the identical contribution of the $S_{IJ}^2$ and $S_{JI}^2$ terms.
This corresponds to a particular Feynman graph which we denote $\Gamma(2,3)$.
Taking into account the order of the Grassmann variables, performing the integrals and contracting the indices leaves us with a global sign of $\epsilon_{\Gamma(2,3)}= -1$. This agrees with the usual QFT Feynman rules where one associates a minus sign to each closed loop.

Hence
\beq
\label{s3 example traces of D}
\epsilon_{\Gamma(2,3)} I_{\Gamma(2,3)} = - (i \alpha)^9 \tr \left( D_{32} V_{253}\right) \   \tr \left( D_{21}D_{15}D_{54}D_{41}D_{15}D_{54}D_{43}D_{32} \right).
\eeq

Applying the graphical method described above gives us a number of admissible graphs, the graph $\Gamma(2,3)$ corresponding to the term in the Feynman expansion above is  given by the following
$$
\psfrag{1}{$1$}
\psfrag{2}{$2$}
\psfrag{3}{$3$}
\psfrag{4}{$4$}
\psfrag{5}{$5$}
\begin{array}{c}
\includegraphics[scale=0.35]{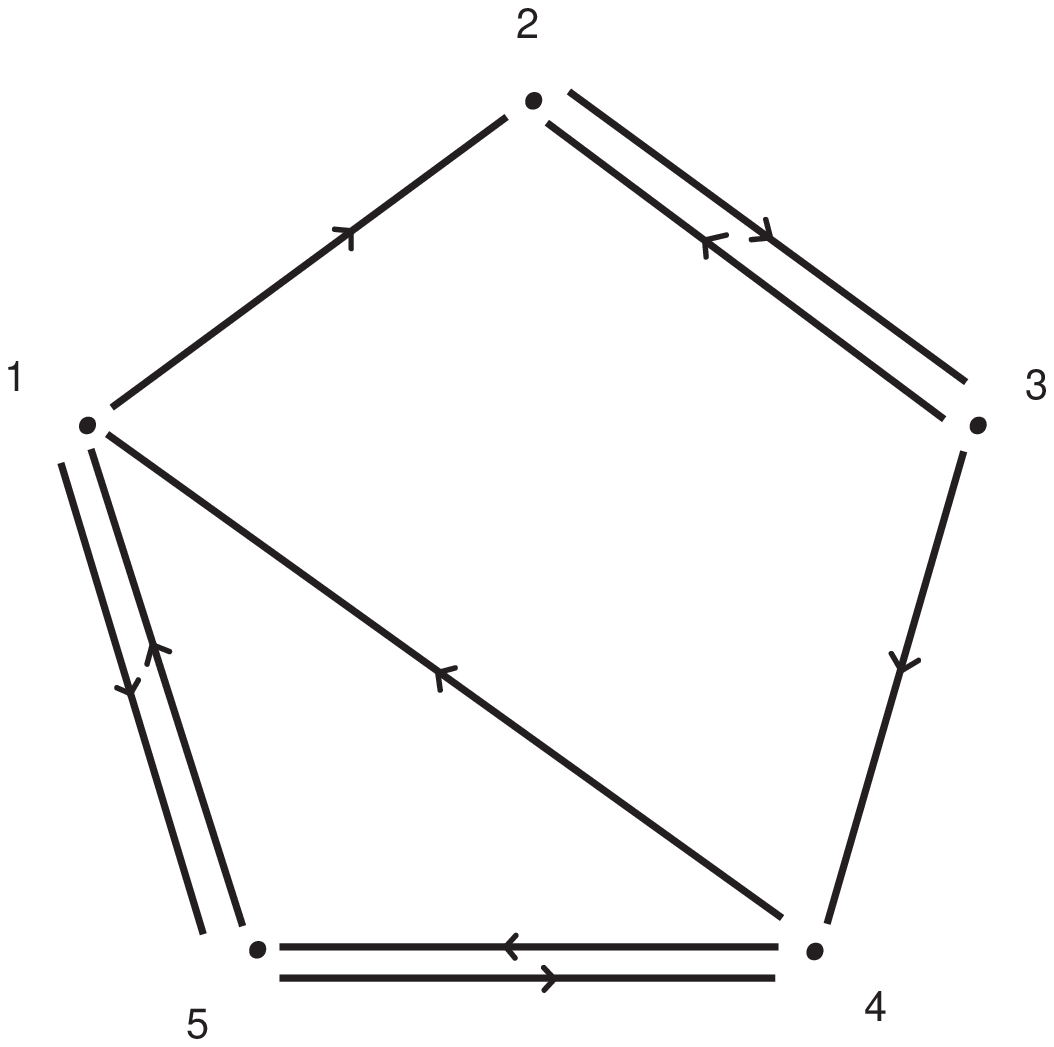}
\end{array}
 \ \ \ \ \ \ \ \ \Rightarrow \ \ \ \ \ \ \ \ \
\psfrag{1}{$1$}
\psfrag{2}{$2$}
\psfrag{3}{$3$}
\psfrag{4}{$4$}
\psfrag{5}{$5$}
\psfrag{p}{$\mathcal{P}(3,2)$}
\psfrag{l}{$\mathcal{L}$}
\begin{array}{c}
\includegraphics[scale=0.35]{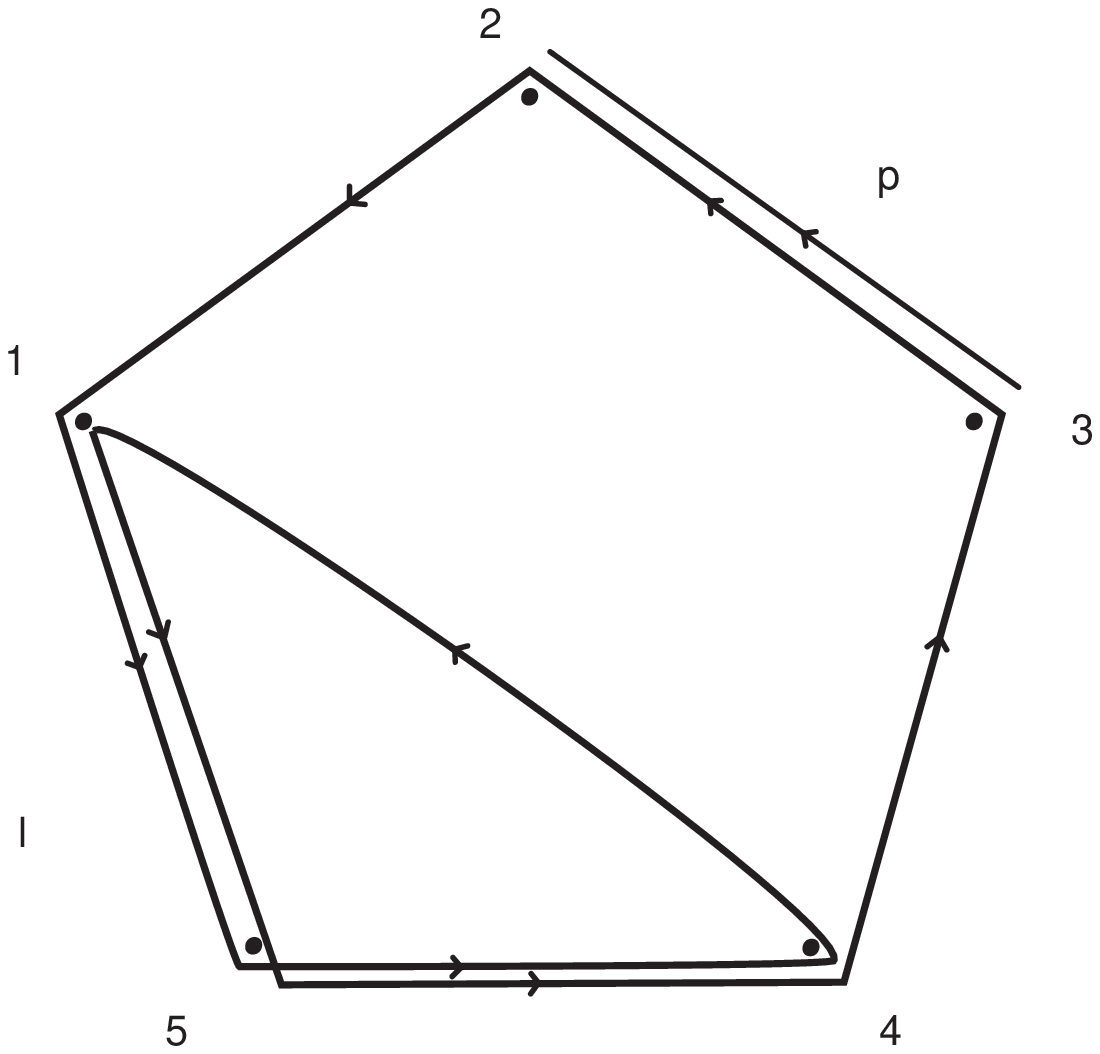}
\end{array}
$$
The path $\mathcal{P}(3,2)$ runs from vertex 3 to vertex 2 and has length $E_{\mathcal{P}} = 1$, and the loop term $\mathcal{L}$ has $E_\mathcal{L}=8$ and passes through the following chain of vertices: $154154321$. The overall length is therefore $E_{\Gamma(2,3)} = 2n-p = 9$.
The Feynman rules in section \ref{quantum theory} now associate a product of Dirac matrices $D_e$ to the edges of $\Gamma(2,3)$ which gives equation \eqref{s3 example traces of D}.

This completes the fermionic integration, we can now consider the gravitational integration. As above, we will consider the $c=0$ configuration $I^{0}_{\Gamma(2,3)}$, which is given by
\beqa
 I^{0}_{\Gamma(2,3)} &=& (i \alpha)^9  \tr \left( \Sigma_{32}U_{32}   V_{253}\right) \
\nn \\
&& \times
  \tr \left(
 \Sigma_{21}U_{21}
 \Sigma_{15}U_{15}
 \Sigma_{54}U_{54} \Sigma_{41}U_{41} \Sigma_{15}U_{15} \Sigma_{54}U_{54} \Sigma_{43}U_{43}
 \right).
\eeqa
Using this, one can apply the graphical methods of section \ref{quantum theory} to obtain the generating functional and then the amplitude $A^0_{\Gamma(2,3)}$ of the graph $\Gamma(2,3)$.
\beqa
\label{s3 gauge inv amplitude}
A^0_{\Gamma(2,3)}
&=&
 \left(\frac{\alpha}{4}\right)^{9}
  \prod_{I<J<K }
   \sum_{j_{IJK}}
 \dim j_{IJK} \, (-1)^{2(j_{145} + j_{123} + j_{245} + j_{234} + j_{135})}
  \\
& & \times \sum_{s_{15},t_{15}} \sum_{s_{23},t_{23}} \sum_{s_{45},t_{45}} \sum_{s_{14}} \sum_{s_{12}} \sum_{s_{25}} \sum_{s_{34}} \sum_{s_{35}}
\nn \\
&& \times \dim s_{\mbox{\tiny 15}} \dim t_{\mbox{\tiny 15}} \dim s_{\mbox{\tiny 14}} \dim s_{\mbox{\tiny 12}} \dim s_{\mbox{\tiny 23}}\dim t_{\mbox{\tiny 23}}
\dim s_{\mbox{\tiny 25}}\dim s_{\mbox{\tiny 34}}\dim s_{\mbox{\tiny 35}} \dim s_{\mbox{\tiny 45}}\dim t_{\mbox{\tiny 45}}
\nn \\
& & \times
  \Theta^2(j_{\mbox{\tiny 123}})
  \Theta^2(j_{\mbox{\tiny 124}})
  \Theta(j_{\mbox{\tiny 125}})
  \Theta^2(j_{\mbox{\tiny 134}})
  \Theta^2(j_{\mbox{\tiny 135}})
  \Theta^4(j_{\mbox{\tiny 145}})
  \Theta(j_{\mbox{\tiny 234}})
  \Theta^2(j_{\mbox{\tiny 235}})
  \Theta(j_{\mbox{\tiny 245}})
  \Theta(j_{\mbox{\tiny 345}})
 \nn \\
& & \times
 A_1(j_{\mbox{\tiny 1..}}, s_{\mbox{\tiny 1.}}    )
A_2(j_{\mbox{\tiny 2..}}, s_{\mbox{\tiny 2.}}    )
A_3(j_{\mbox{\tiny 3..}}, s_{\mbox{\tiny 3.}}    )
A_4(j_{\mbox{\tiny 4..}}, s_{\mbox{\tiny 4.}}    )
A_5(j_{\mbox{\tiny 5..}}, s_{\mbox{\tiny 5.}}    )
\nn
\eeqa
Here, the global factor is due to the multiplication of the $i\alpha$ factors coming from the Feynman rules with the factor $- i / 4$ coming from the global contribution attached to each grasping ($i\sqrt{6}/2$ arising from the $\gamma^a$, $\sqrt{6}$ from the $\epsilon^{abc}$ tensor and $-1/12$ from the $\hat{\Sigma}_e^a$).

The signs are due to the Haar integrals, and the $\Theta(j_{IJK})$ terms come from the normalisation of the source derivatives. The vertex amplitudes $A_I$, $I=1,...,5$, are given by evaluating the following spin networks.

\beq
A_1 =
\begin{array}{c}
\psfrag{j123}{$j_{\mbox{\tiny 123}}$}
\psfrag{j124}{$j_{\mbox{\tiny 124}}$}
\psfrag{j125}{$j_{\mbox{\tiny 125}}$}
\psfrag{j135}{$j_{\mbox{\tiny 135}}$}
\psfrag{j134}{$j_{\mbox{\tiny 134}}$}
\psfrag{j145}{$j_{\mbox{\tiny 145}}$}
\psfrag{s12}{$s_{\mbox{\tiny 12}}$}
\psfrag{s14}{$s_{\mbox{\tiny 14}}$}
\psfrag{s15}{$s_{\mbox{\tiny 15}}$}
\psfrag{t15}{$t_{\mbox{\tiny 15}}$}
\psfrag{-}{$_-$}
\psfrag{+}{$_+$}
\includegraphics[scale=0.45]{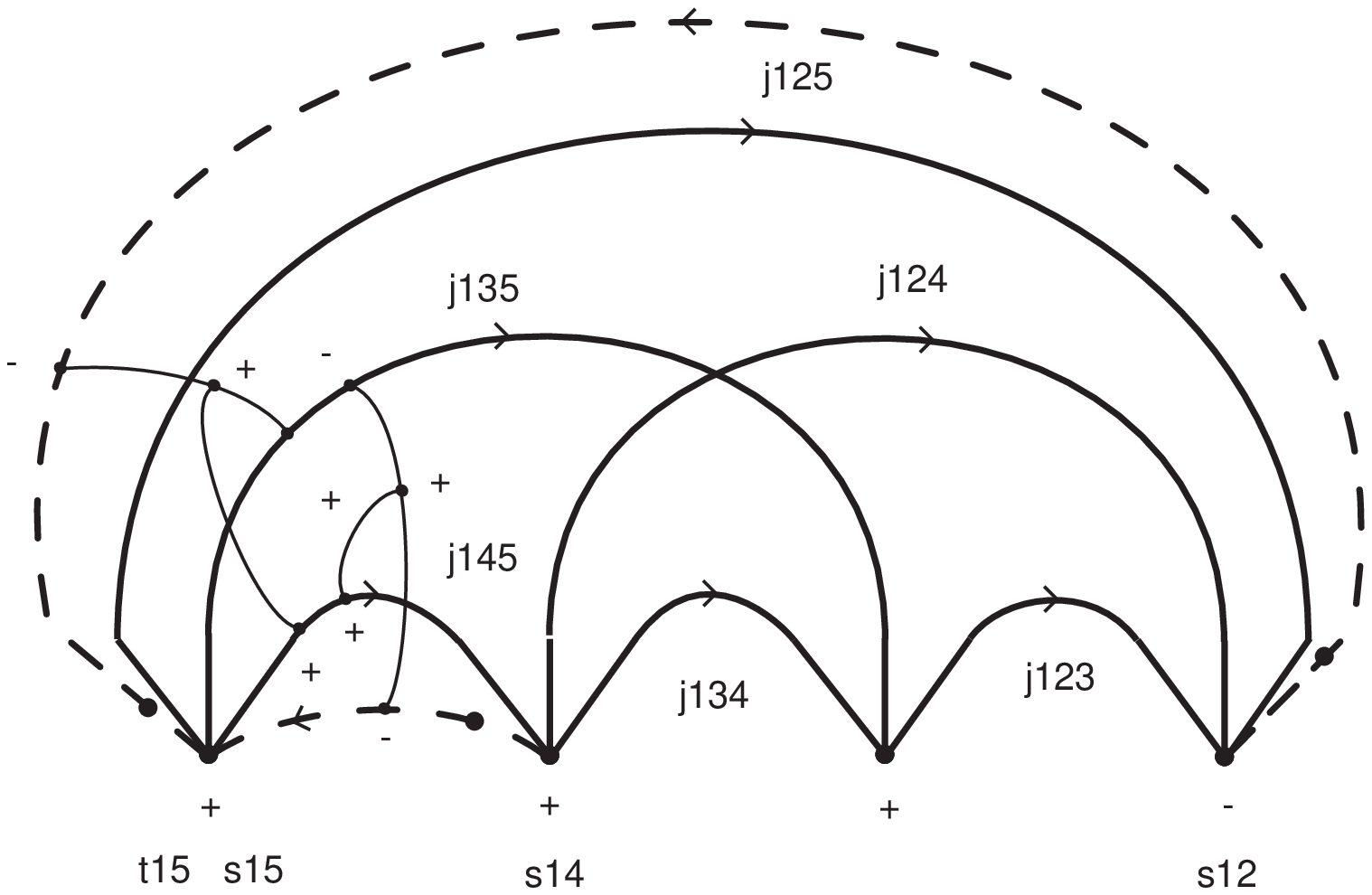}
\end{array}
\eeq
\beq
A_2 =
\begin{array}{c}
 \psfrag{j235}{$j_{\mbox{\tiny 235}}$}
\psfrag{j234}{$j_{\mbox{\tiny 234}}$}
\psfrag{j123}{$j_{\mbox{\tiny 123}}$}
\psfrag{j124}{$j_{\mbox{\tiny 124}}$}
\psfrag{j125}{$j_{\mbox{\tiny 125}}$}
\psfrag{j245}{$j_{\mbox{\tiny 245}}$}
\psfrag{s25}{$s_{\mbox{\tiny 25}}$}
\psfrag{s23}{$s_{\mbox{\tiny 23}}$}
\psfrag{t23}{$t_{\mbox{\tiny 23}}$}
\psfrag{s12}{$s_{\mbox{\tiny 12}}$}
\psfrag{-}{$_-$}
\psfrag{+}{$_+$}
\includegraphics[scale=0.45]{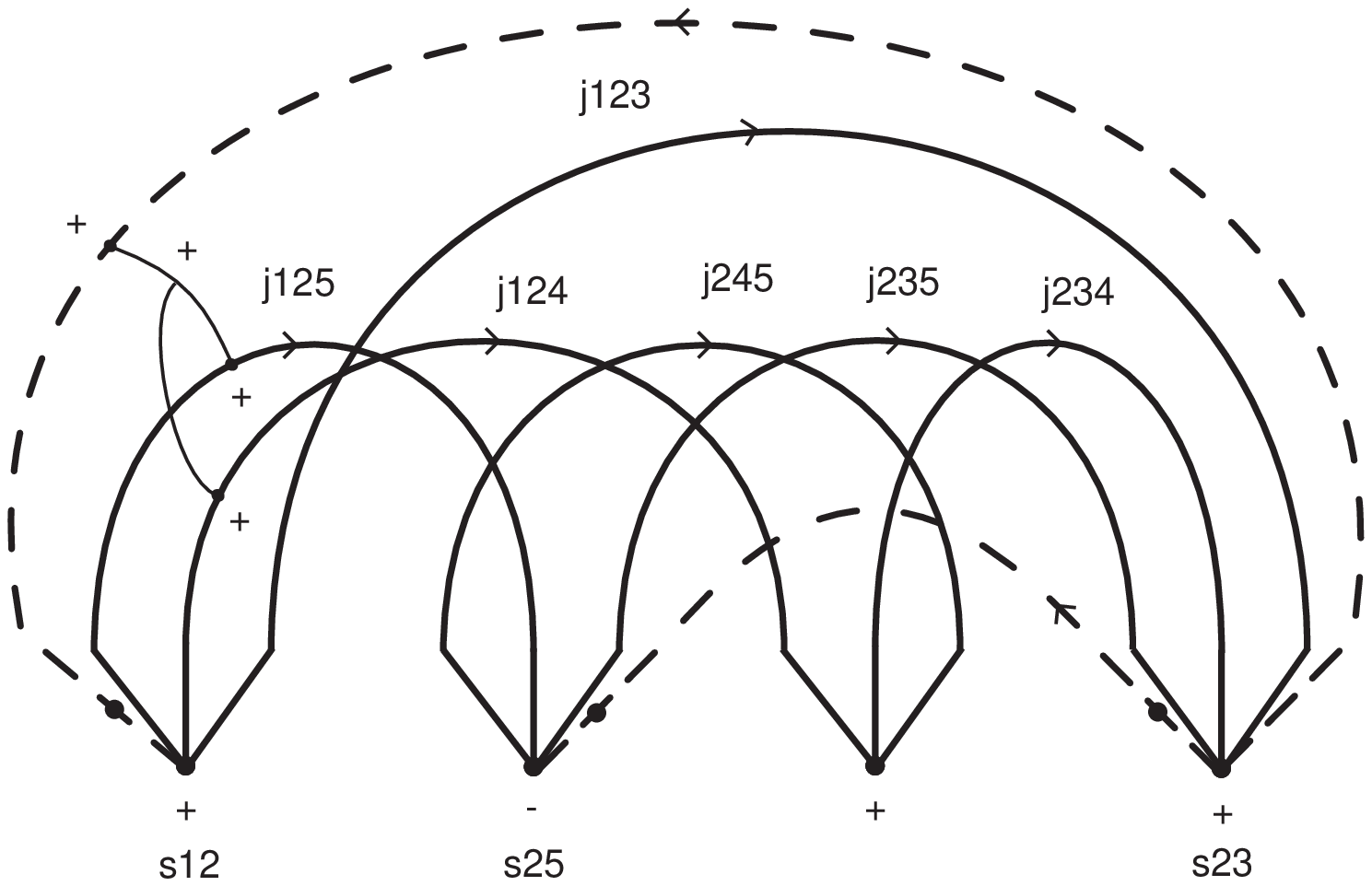}
\end{array}
\eeq
\beq
A_3 =
\begin{array}{c}
\psfrag{j123}{$j_{\mbox{\tiny 123}}$}
\psfrag{j134}{$j_{\mbox{\tiny 134}}$}
\psfrag{j135}{$j_{\mbox{\tiny 135}}$}
\psfrag{j345}{$j_{\mbox{\tiny 345}}$}
\psfrag{j235}{$j_{\mbox{\tiny 235}}$}
\psfrag{j234}{$j_{\mbox{\tiny 234}}$}
\psfrag{s23}{$s_{\mbox{\tiny 23}}$}
\psfrag{t23}{$t_{\mbox{\tiny 23}}$}
\psfrag{s34}{$s_{\mbox{\tiny 34}}$}
\psfrag{s35}{$s_{\mbox{\tiny 35}}$}
\psfrag{-}{$_-$}
\psfrag{+}{$_+$}
\includegraphics[scale=0.45]{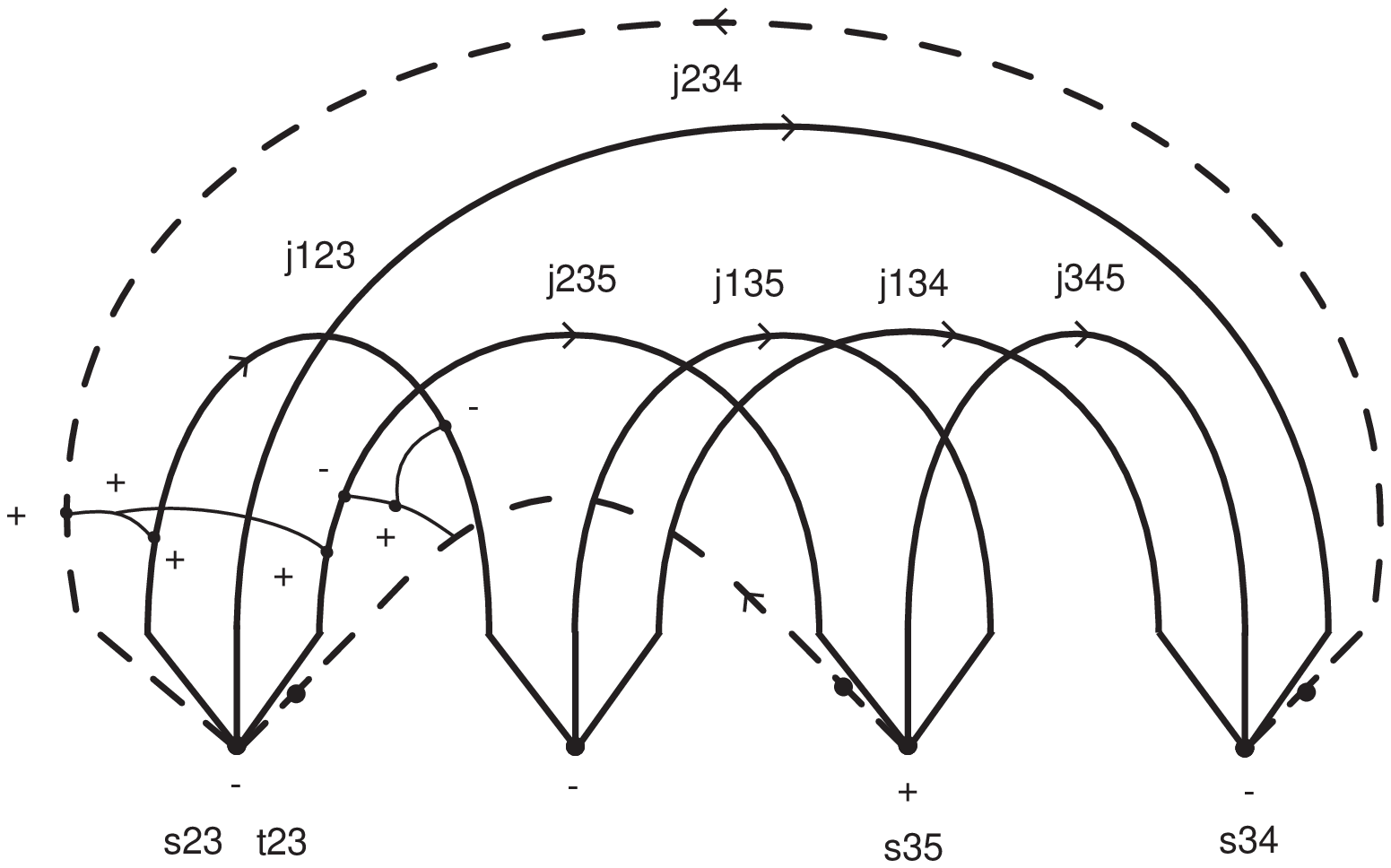}
\end{array}
\eeq
\beq
A_4 =
\begin{array}{c}
 \psfrag{j145}{$j_{\mbox{\tiny 145}}$}
\psfrag{j245}{$j_{\mbox{\tiny 245}}$}
\psfrag{j345}{$j_{\mbox{\tiny 345}}$}
\psfrag{j234}{$j_{\mbox{\tiny 234}}$}
\psfrag{j134}{$j_{\mbox{\tiny 134}}$}
\psfrag{j124}{$j_{\mbox{\tiny 124}}$}
\psfrag{s14}{$s_{\mbox{\tiny 14}}$}
\psfrag{s45}{$s_{\mbox{\tiny 45}}$}
\psfrag{t45}{$t_{\mbox{\tiny 45}}$}
\psfrag{s34}{$s_{\mbox{\tiny 34}}$}
\psfrag{-}{$_-$}
\psfrag{+}{$_+$}
\includegraphics[scale=0.45]{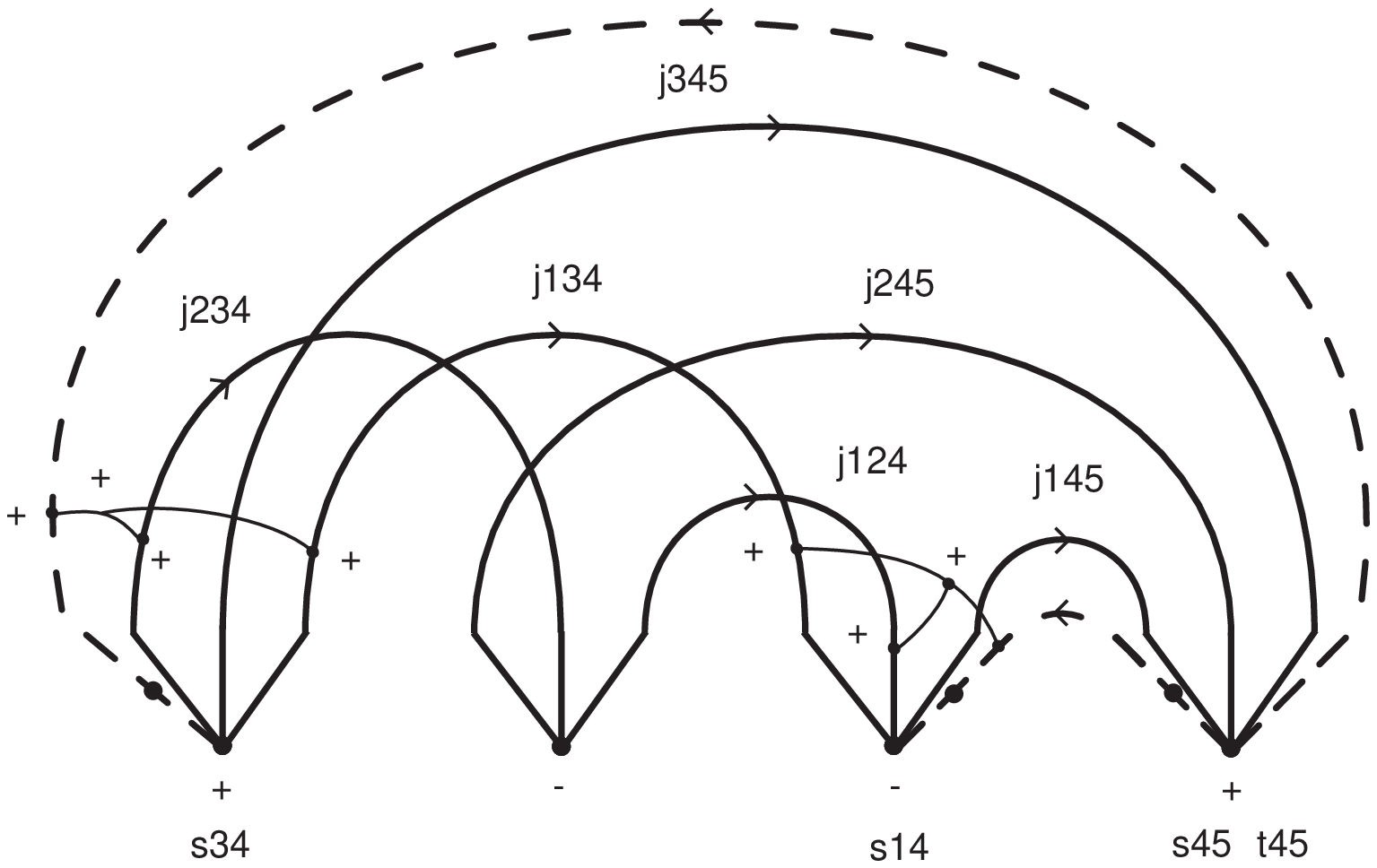}
\end{array}
\eeq
\beq
A_5 =
\begin{array}{c}
\psfrag{j145}{$j_{\mbox{\tiny 145}}$}
\psfrag{j345}{$j_{\mbox{\tiny 345}}$}
\psfrag{j245}{$j_{\mbox{\tiny 245}}$}
\psfrag{j235}{$j_{\mbox{\tiny 235}}$}
\psfrag{j125}{$j_{\mbox{\tiny 125}}$}
\psfrag{j135}{$j_{\mbox{\tiny 135}}$}
\psfrag{s15}{$s_{\mbox{\tiny 15}}$}
\psfrag{t15}{$t_{\mbox{\tiny 15}}$}
\psfrag{s45}{$s_{\mbox{\tiny 45}}$}
\psfrag{t45}{$t_{\mbox{\tiny 45}}$}
\psfrag{s25}{$s_{\mbox{\tiny 25}}$}
\psfrag{s35}{$s_{\mbox{\tiny 35}}$}
\psfrag{-}{$_-$}
\psfrag{+}{$_+$}
\includegraphics[scale=0.45]{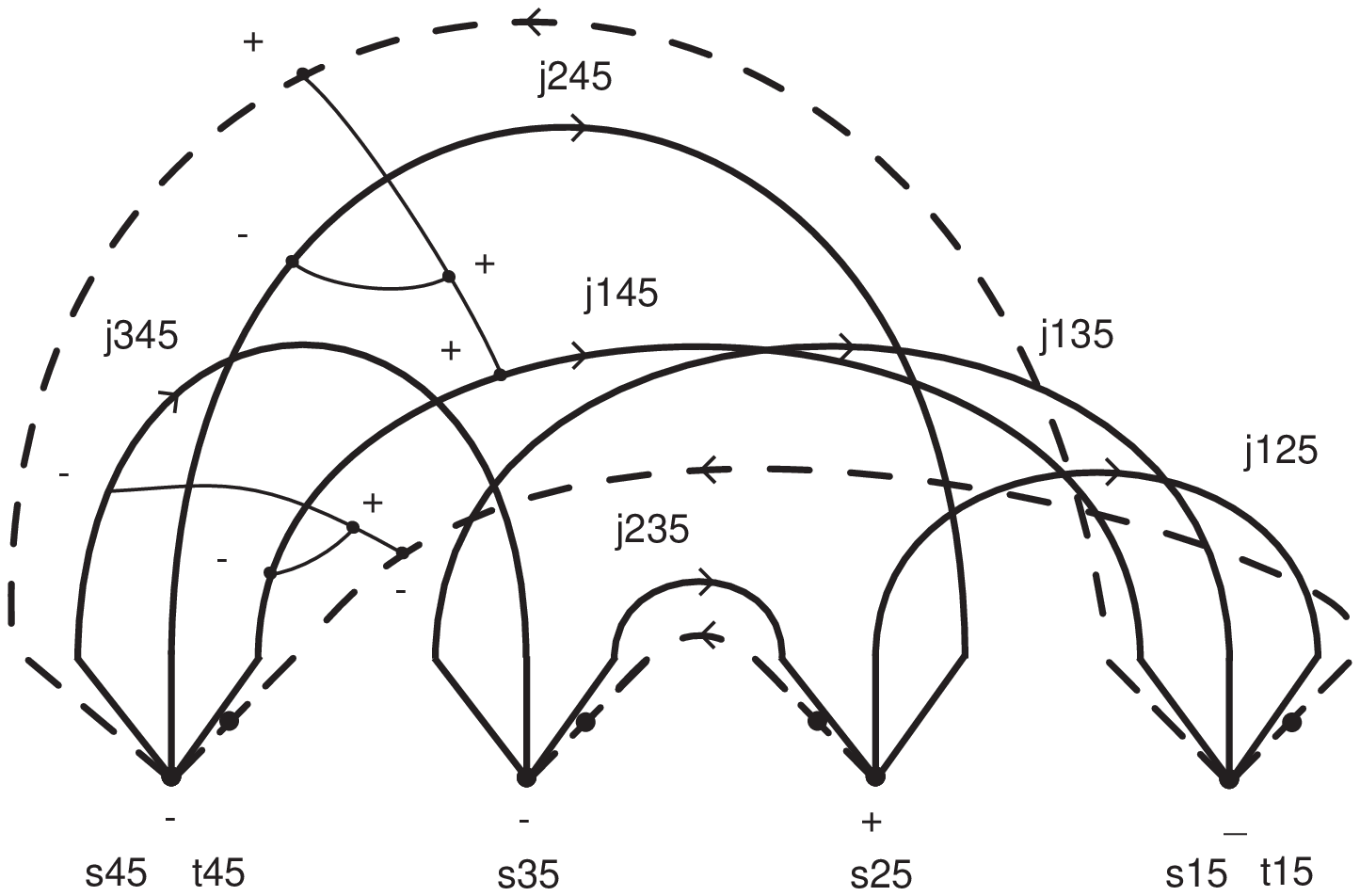}
\end{array}
\eeq

In order to write $A_{\Gamma(2,3)}^0$ explicitly, we will now recouple the vertex amplitudes appearing in the state sum in order to express them in terms of sums and products $6j$ symbols.
We start by decomposing the vertices of valence four and five into three-valent vertices. Then, we repeatedly use the recoupling identities \eqref{Schur}, \eqref{recoupling1}, \eqref{fusion move} displayed in the Appendix. As a result, we obtain the following decomposition of the vertex amplitudes $A_I(j_{IJK},s_{IJ})$ appearing in equation \eqref{s3 gauge inv amplitude}.
\beqa
A_1 =
\begin{array}{c}
\psfrag{j123}{$j_{\mbox{\tiny 123}}$}
\psfrag{j124}{$j_{\mbox{\tiny 124}}$}
\psfrag{j125}{$j_{\mbox{\tiny 125}}$}
\psfrag{j135}{$j_{\mbox{\tiny 135}}$}
\psfrag{j134}{$j_{\mbox{\tiny 134}}$}
\psfrag{j145}{$j_{\mbox{\tiny 145}}$}
\psfrag{s12}{$s_{\mbox{\tiny 12}}$}
\psfrag{s14}{$s_{\mbox{\tiny 14}}$}
\psfrag{s15}{$s_{\mbox{\tiny 15}}$}
\psfrag{t15}{$t_{\mbox{\tiny 15}}$}
\psfrag{-}{$_-$}
\psfrag{+}{$_+$}
\includegraphics[scale=0.45]{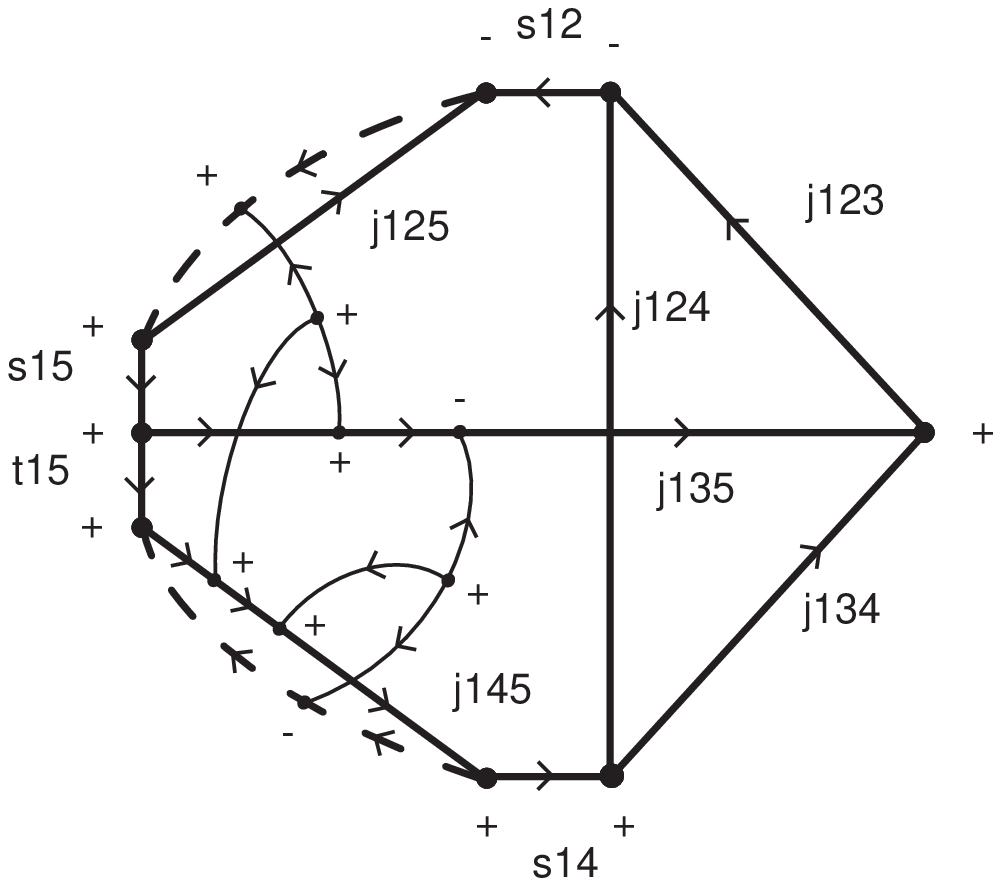}
\end{array}
 \hspace{-4mm}
 &=&
  \hspace{-4mm}
\begin{array}{l}
\sum_{p,n,q} \dim n \ \dim p \  \dim q
\nn \\ \times
(-1)^{3n+q + j_{\mbox{\tiny 135}} + s_{\mbox{\tiny 14}} + 2j_{\mbox{\tiny 123}}  + j_{\mbox{\tiny 134}} + j_{\mbox{\tiny 124}} +s_{\mbox{\tiny 12}} + t_{\mbox{\tiny 15}} + t_{\mbox{\tiny 15}}      }
\nn \\ \times
\left\{ \begin{array}{ccc}
 j_{\mbox{\tiny 124}} & \hspace{-2mm}j_{\mbox{\tiny 134}} & \hspace{-2mm} s_{\mbox{\tiny 14}} \\
 j_{\mbox{\tiny 135}} & \hspace{-2mm} s_{\mbox{\tiny 12}} & \hspace{-2mm} j_{\mbox{\tiny 123}}
\end{array} \right\}
 \left\{ \begin{array}{ccc}
 s_{\mbox{\tiny 14}} & \hspace{-2mm} j_{\mbox{\tiny 135}} & \hspace{-2mm} s_{\mbox{\tiny 12}} \\
 j_{\mbox{\tiny 135}} & \hspace{-2mm} n & \hspace{-2mm} 1
\end{array} \right\}
 \left\{ \begin{array}{ccc}
 1 & \hspace{-2mm} s_{\mbox{\tiny 14}} & \hspace{-2mm} n \\
 j_{\mbox{\tiny 145}} & \hspace{-2mm} \frac{1}{2} & \hspace{-2mm} \frac{1}{2}
\end{array} \right\}
\nn \\  \times
 \left\{ \begin{array}{ccc}
 1 & \hspace{-2mm} s_{\mbox{\tiny 14}} &  q \\
 j_{\mbox{\tiny 145}} & \hspace{-2mm} \frac{1}{2} &  \frac{1}{2}
\end{array} \right\}
 \left\{ \begin{array}{ccc}
 s_{\mbox{\tiny 12}} & \hspace{-2mm} j_{\mbox{\tiny 135}} &  n\\
 j_{\mbox{\tiny 135}} & \hspace{-2mm} p &  1
\end{array} \right\}
 \left\{ \begin{array}{ccc}
  \frac{1}{2} &  n & \hspace{-2mm} j_{\mbox{\tiny 145}} \\
1 &  j_{\mbox{\tiny 145}} & \hspace{-2mm} q
\end{array} \right\}
\nn \\  \times
 \left\{ \begin{array}{ccc}
 j_{\mbox{\tiny 135}} & \hspace{-2mm} p & \hspace{-2mm} n \\
 1 & \hspace{-2mm} t_{\mbox{\tiny 15}} & \hspace{-2mm} s_{\mbox{\tiny 15}}
\end{array} \right\}
 \left\{ \begin{array}{ccc}
 1 & \hspace{-2mm} 1 &  1 \\
 s_{\mbox{\tiny 15}} & \hspace{-2mm} s_{\mbox{\tiny 12}} &  p
\end{array} \right\}
 \left\{ \begin{array}{ccc}
 \frac{1}{2} &  t_{\mbox{\tiny 15}} & \hspace{-2mm} j_{\mbox{\tiny 145}} \\
 1 &  j_{\mbox{\tiny 145}} & \hspace{-2mm} n
\end{array} \right\}
\nn \\  \times
 \left\{ \begin{array}{ccc}
 1& \hspace{-2mm}  \frac{1}{2} & \hspace{-2mm}  \frac{1}{2} \\
 j_{\mbox{\tiny 125}} & \hspace{-2mm} s_{\mbox{\tiny 12}} & \hspace{-2mm} s_{\mbox{\tiny 15}}
\end{array} \right\}
\end{array}
\nn
\eeqa
\beqa
A_2=
\begin{array}{c}
 \psfrag{j1}{$j_{\mbox{\tiny 235}}$}
\psfrag{j2}{$j_{\mbox{\tiny 234}}$}
\psfrag{j3}{$j_{\mbox{\tiny 123}}$}
\psfrag{j4}{$j_{\mbox{\tiny 124}}$}
\psfrag{j5}{$j_{\mbox{\tiny 125}}$}
\psfrag{j6}{$j_{\mbox{\tiny 245}}$}
\psfrag{s25}{$s_{\mbox{\tiny 25}}$}
\psfrag{s23}{$s_{\mbox{\tiny 23}}$}
\psfrag{t23}{$t_{\mbox{\tiny 23}}$}
\psfrag{s12}{$s_{\mbox{\tiny 12}}$}
\psfrag{-}{$_-$}
\psfrag{+}{$_+$}
\includegraphics[scale=0.45]{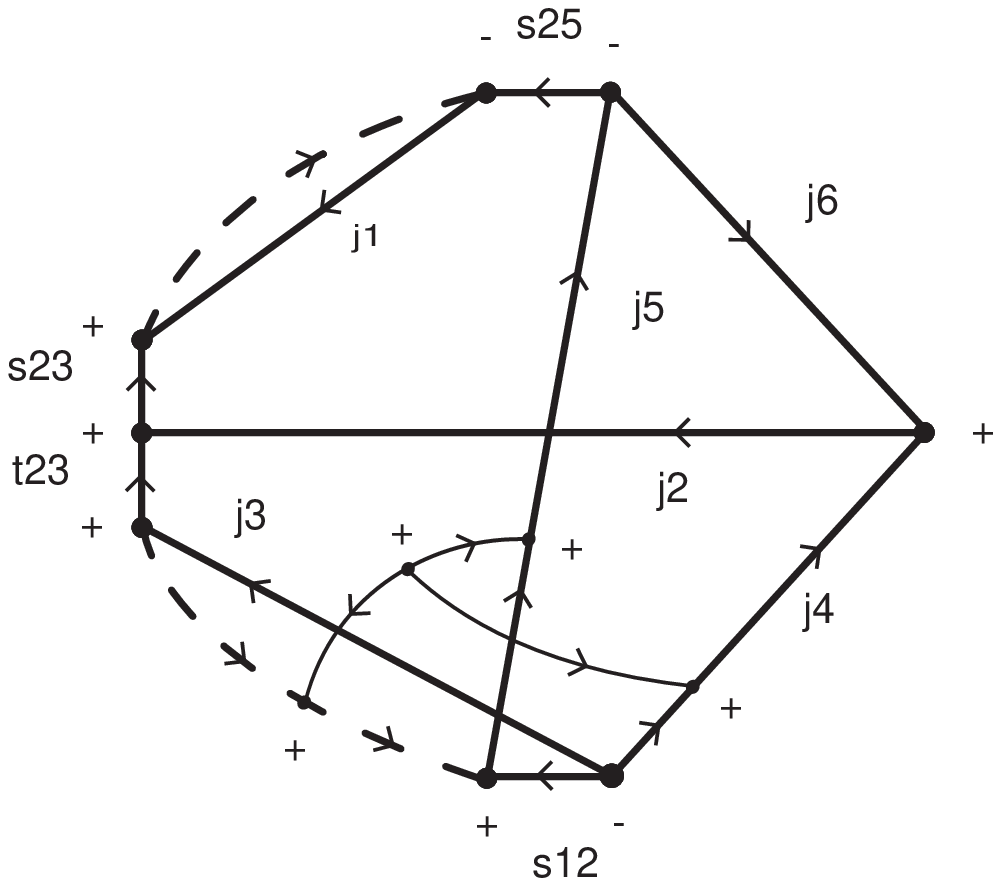}
\end{array}
\hspace{-4mm}
&=&
\hspace{-4mm}
\begin{array}{l}
\sum_{m,r}   (\dim s_{\mbox{\tiny 23}})^{-1}   \dim m \ \dim r   \   \delta_{s_{\mbox{\tiny 23}},s_{\mbox{\tiny 25}}   }
(-1)^{( 3s_{\mbox{\tiny 23}} + 3j_{\mbox{\tiny 245}}+ j_{\mbox{\tiny 234}} )}
\nn \\ \times
(-1)^{( m+2r + 2s_{\mbox{\tiny 12}} + 2j_{\mbox{\tiny 123}} + j_{\mbox{\tiny 235}} + 3t_{\mbox{\tiny 23}}   + 3j_{\mbox{\tiny 125}}                      + j_{\mbox{\tiny 124}}    +     1 )}
 \\  \times
\left\{ \begin{array}{ccc}
 j_{\mbox{\tiny 125}} & \hspace{-2mm} j_{\mbox{\tiny 124}} & \hspace{-2mm} t_{\mbox{\tiny 23}} \\
 j_{\mbox{\tiny 234}} & \hspace{-2mm} s_{\mbox{\tiny 23}} & \hspace{-2mm} j_{\mbox{\tiny 245}}
\end{array} \right\}
\left\{ \begin{array}{ccc}
 j_{\mbox{\tiny 124}} & \hspace{-2mm} j_{\mbox{\tiny 125}} & \hspace{-2mm} t_{\mbox{\tiny 23}} \\
 1 & \hspace{-2mm} r & \hspace{-2mm} j_{\mbox{\tiny 123}}
\end{array} \right\}
\left\{ \begin{array}{ccc}
1 &  1 &  1 \\
 m &  r &  t_{\mbox{\tiny 23}}
\end{array} \right\}
\nn \\  \times
\left\{ \begin{array}{ccc}
 j_{\mbox{\tiny 125}} & \hspace{-2mm}m & \hspace{-2mm} j_{\mbox{\tiny 124}} \\
1& \hspace{-2mm} j_{\mbox{\tiny 124}} & \hspace{-2mm} r
\end{array} \right\}
\left\{ \begin{array}{ccc}
 1& \hspace{-2mm}t_{\mbox{\tiny 23}} & \hspace{-2mm} m \\
 j_{\mbox{\tiny 123}} & \hspace{-2mm} \frac{1}{2} & \hspace{-2mm} \frac{1}{2}
\end{array} \right\}
\left\{ \begin{array}{ccc}
 \frac{1}{2} & \hspace{-2mm} j_{\mbox{\tiny 125}} & \hspace{-2mm} s_{\mbox{\tiny 12}} \\
 j_{\mbox{\tiny 124}} & \hspace{-2mm} j_{\mbox{\tiny 123}} & \hspace{-2mm} m
\end{array} \right\}
\end{array}
\nn
\eeqa

\beqa
A_3 =
\begin{array}{c}
\psfrag{j123}{$j_{\mbox{\tiny 123}}$}
\psfrag{j134}{$j_{\mbox{\tiny 134}}$}
\psfrag{j135}{$j_{\mbox{\tiny 135}}$}
\psfrag{j345}{$j_{\mbox{\tiny 345}}$}
\psfrag{j235}{$j_{\mbox{\tiny 235}}$}
\psfrag{j234}{$j_{\mbox{\tiny 234}}$}
\psfrag{s23}{$s_{\mbox{\tiny 23}}$}
\psfrag{t23}{$t_{\mbox{\tiny 23}}$}
\psfrag{s34}{$s_{\mbox{\tiny 34}}$}
\psfrag{s35}{$s_{\mbox{\tiny 35}}$}
\psfrag{-}{$_-$}
\psfrag{+}{$_+$}
\includegraphics[scale=0.45]{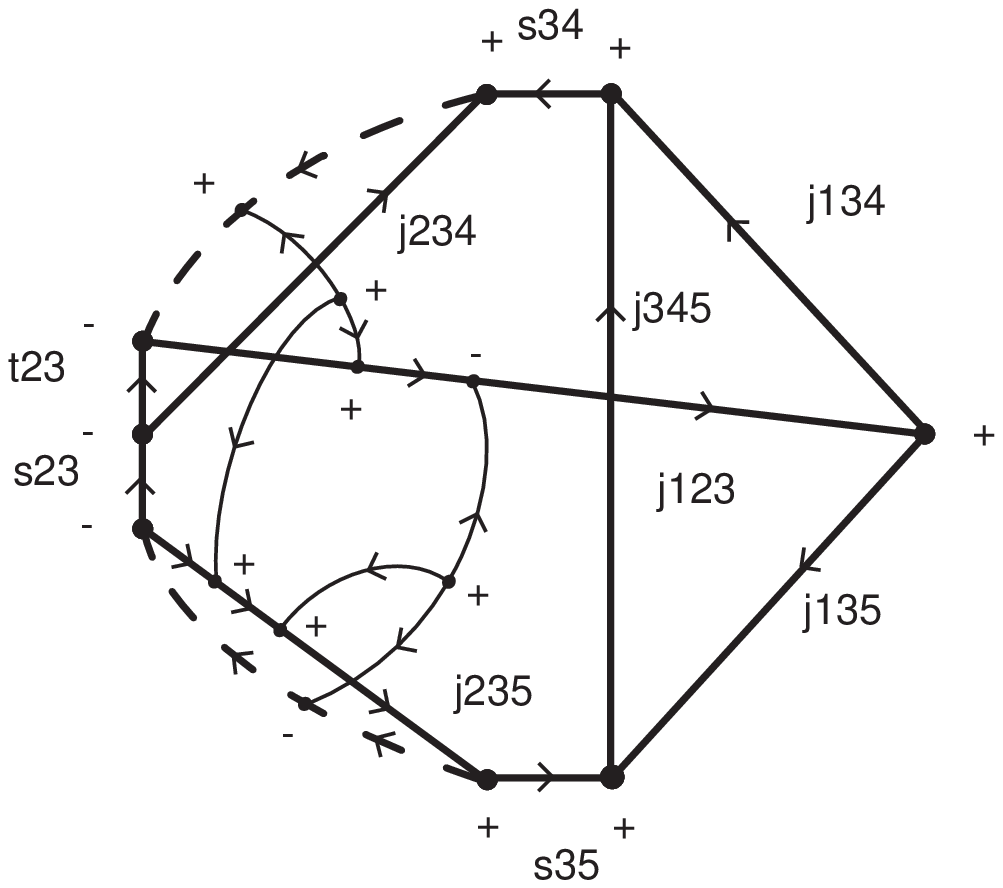}
\end{array}
 \hspace{-4mm}
&=&
 \hspace{-4mm}
\begin{array}{l}
\sum_{n,p,q,v} \dim n \ \dim p \ \dim q \ \dim v \ (-1)^{( j_{\mbox{\tiny 135}}+j_{\mbox{\tiny 345}}+ s_{\mbox{\tiny 34}} )}
\\ \times
(-1)^{(3n+q+v+ j_{\mbox{\tiny 123}} + 2j_{\mbox{\tiny 234}} + 2t_{\mbox{\tiny 23}} + s_{\mbox{\tiny 23}}  +j_{\mbox{\tiny 123}} +s_{\mbox{\tiny 35}}              +2j_{\mbox{\tiny 134}}+ s_{\mbox{\tiny 23}}
)}
\\ \times
\left\{ \begin{array}{ccc}
 j_{\mbox{\tiny 123}} & \hspace{-2mm} \frac{1}{2} & \hspace{-2mm} t_{\mbox{\tiny 23}} \\
 j_{\mbox{\tiny 234}} & \hspace{-2mm} s_{\mbox{\tiny 23}} & \hspace{-2mm} v
\end{array} \right\}
\left\{ \begin{array}{ccc}
 j_{\mbox{\tiny 345}} & \hspace{-2mm} j_{\mbox{\tiny 135}} & \hspace{-2mm} s_{\mbox{\tiny 35}} \\
 j_{\mbox{\tiny 123}} & \hspace{-2mm} s_{\mbox{\tiny 34}} & \hspace{-2mm} j_{\mbox{\tiny 134}}
\end{array} \right\}
\left\{ \begin{array}{ccc}
 s_{\mbox{\tiny 35}} & \hspace{-2mm} j_{\mbox{\tiny 123}} & \hspace{-2mm} s_{\mbox{\tiny 34}} \\
 j_{\mbox{\tiny 123}} & \hspace{-2mm} n & \hspace{-2mm} 1
\end{array} \right\}
\\ \times
\left\{ \begin{array}{ccc}
 1 & \hspace{-2mm} s_{\mbox{\tiny 35}} &  n \\
 j_{\mbox{\tiny 235}} & \hspace{-2mm} \frac{1}{2} &  \frac{1}{2}
\end{array} \right\}
\left\{ \begin{array}{ccc}
 1 & \hspace{-2mm} s_{\mbox{\tiny 35}} &  q \\
 j_{\mbox{\tiny 235}} & \hspace{-2mm} \frac{1}{2} &  \frac{1}{2}
\end{array} \right\}
\left\{ \begin{array}{ccc}
 s_{\mbox{\tiny 34}} & \hspace{-2mm} j_{\mbox{\tiny 123}} &  n \\
 j_{\mbox{\tiny 123}} & \hspace{-2mm} p &  1
\end{array} \right\}
\\ \times
\left\{ \begin{array}{ccc}
 \frac{1}{2} &  n & \hspace{-2mm} j_{\mbox{\tiny 235}} \\
1 &  j_{\mbox{\tiny 235}} & \hspace{-2mm} q
\end{array} \right\}
\left\{ \begin{array}{ccc}
 j_{\mbox{\tiny 123}} & \hspace{-2mm} p &  n \\
 1 & \hspace{-2mm} s_{\mbox{\tiny 23}} &   v
\end{array} \right\}
\left\{ \begin{array}{ccc}
 1 &  1 &  1 \\
 v &  s_{\mbox{\tiny 34}} &  p
\end{array} \right\}
\\ \times
\left\{ \begin{array}{ccc}
 \frac{1}{2} &  s_{\mbox{\tiny 23}} & \hspace{-2mm} j_{\mbox{\tiny 235}} \\
 1 &  j_{\mbox{\tiny 235}} & \hspace{-2mm} n
\end{array} \right\}
\left\{ \begin{array}{ccc}
 1 & \hspace{-2mm} \frac{1}{2} &  \frac{1}{2} \\
 j_{\mbox{\tiny 234}} & \hspace{-2mm} s_{\mbox{\tiny 34}} &  v
\end{array} \right\}
\end{array} \nn
\eeqa

\beqa
A_4=
\begin{array}{c}
 \psfrag{j145}{$j_{\mbox{\tiny 145}}$}
\psfrag{j245}{$j_{\mbox{\tiny 245}}$}
\psfrag{j345}{$j_{\mbox{\tiny 345}}$}
\psfrag{j234}{$j_{\mbox{\tiny 234}}$}
\psfrag{j134}{$j_{\mbox{\tiny 134}}$}
\psfrag{j12}{$j_{\mbox{\tiny 124}}$}
\psfrag{s14}{$s_{\mbox{\tiny 14}}$}
\psfrag{s45}{$s_{\mbox{\tiny 45}}$}
\psfrag{t45}{$t_{\mbox{\tiny 45}}$}
\psfrag{s34}{$s_{\mbox{\tiny 34}}$}
\psfrag{-}{$_-$}
\psfrag{+}{$_+$}
\includegraphics[scale=0.45]{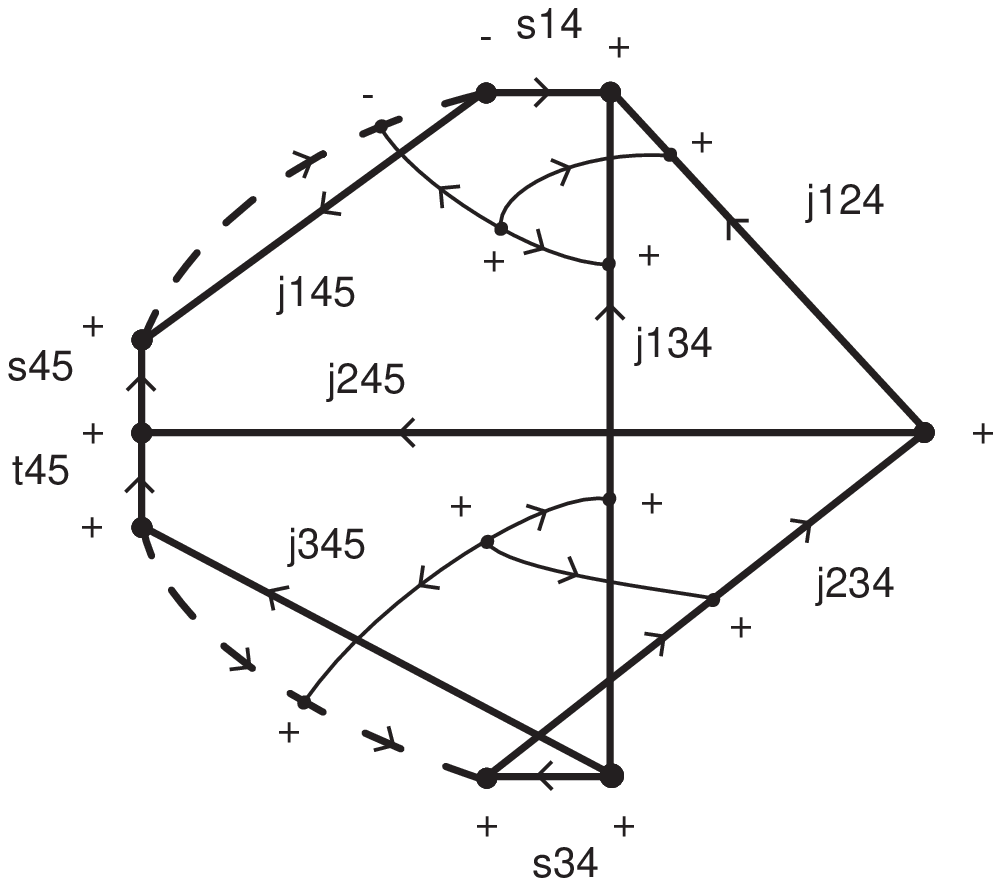}
\end{array}
\hspace{-4mm}
&=&
\hspace{-4mm}
\begin{array}{l}
\sum_{a,b,c} \dim a \; \dim b \; \dim c
\nn \\  \times
(-1)^{ ( 3a+3b+3c +3       s_{\mbox{\tiny 34}}  + j_{\mbox{\tiny 234}}+ j_{\mbox{\tiny 145}}+ j_{\mbox{\tiny 345}} +\frac{3}{2} +3s_{\mbox{\tiny 45}}       + 3 j_{\mbox{\tiny 134}}                          )      }
\nn \\  \times
\left\{ \begin{array}{ccc}
 j_{\mbox{\tiny 134}}  &  \hspace{-2mm} j_{\mbox{\tiny 234}} \hspace{-2mm}  & t_{\mbox{\tiny 45}}  \\
 j_{\mbox{\tiny 245}}  &\hspace{-2mm}   s_{\mbox{\tiny 45}}  & \hspace{-2mm} j_{\mbox{\tiny 124}}
\end{array} \right\}
\left\{ \begin{array}{ccc}
 j_{\mbox{\tiny 134}}  & \hspace{-2mm}  j_{\mbox{\tiny 234}}  &\hspace{-2mm} t_{\mbox{\tiny 45}}  \\
 c &\hspace{-2mm}  1  &\hspace{-2mm}  j_{\mbox{\tiny 134}}
\end{array} \right\}
\left\{ \begin{array}{ccc}
 j_{\mbox{\tiny 134}}  &\hspace{-2mm}  j_{\mbox{\tiny 124}}  &\hspace{-2mm} s_{\mbox{\tiny 45}}  \\
 b  &\hspace{-2mm}  1  &\hspace{-2mm}  j_{\mbox{\tiny 134}}
\end{array} \right\}
\\  \times
\left\{ \begin{array}{ccc}
 1  &  1  & 1  \\
 a &  t_{\mbox{\tiny 45}}  &  c
\end{array} \right\}
\left\{ \begin{array}{ccc}
 1  &  1 & 1  \\
 s_{\mbox{\tiny 14}}  &  b  &  s_{\mbox{\tiny 45}}
\end{array} \right\}
\left\{ \begin{array}{ccc}
 a  &  j_{\mbox{\tiny 134}}  &\hspace{-2mm} j_{\mbox{\tiny 234}}  \\
 1 &  j_{\mbox{\tiny 234}}  &\hspace{-2mm}  c
\end{array} \right\}
\\  \times
\left\{ \begin{array}{ccc}
 s_{\mbox{\tiny 14}}  &\hspace{-2mm}  j_{\mbox{\tiny 134}}  &\hspace{-2mm} j_{\mbox{\tiny 124}}  \\
 j_{\mbox{\tiny 124}}  &\hspace{-2mm}  1  &\hspace{-2mm}  b
\end{array} \right\}
\left\{ \begin{array}{ccc}
 1  &\hspace{-2mm}  s_{\mbox{\tiny 14}}  &\hspace{-2mm} s_{\mbox{\tiny 45}}  \\
 j_{\mbox{\tiny 145}}  &\hspace{-2mm} \frac{1}{2}  &\hspace{-2mm}  \frac{1}{2}
\end{array} \right\}
\left\{ \begin{array}{ccc}
 j_{\mbox{\tiny 234}}  &\hspace{-2mm}  \frac{1}{2}  &\hspace{-2mm} s_{\mbox{\tiny 34}}  \\
 j_{\mbox{\tiny 345}}  &\hspace{-2mm}  j_{\mbox{\tiny 134}}  &\hspace{-2mm}  a
\end{array} \right\}
\end{array}
\nn
\eeqa
\beqa
A_5= \hspace{-1mm}
\begin{array}{c}
\psfrag{j145}{$j_{\mbox{\tiny 145}}$}
\psfrag{j345}{$j_{\mbox{\tiny 345}}$}
\psfrag{j245}{$j_{\mbox{\tiny 245}}$}
\psfrag{j235}{$j_{\mbox{\tiny 235}}$}
\psfrag{j125}{$j_{\mbox{\tiny 125}}$}
\psfrag{j135}{$j_{\mbox{\tiny 135}}$}
\psfrag{s15}{$s_{\mbox{\tiny 15}}$}
\psfrag{t15}{$t_{\mbox{\tiny 15}}$}
\psfrag{s45}{$s_{\mbox{\tiny 45}}$}
\psfrag{t45}{$t_{\mbox{\tiny 45}}$}
\psfrag{s25}{$s_{\mbox{\tiny 25}}$}
\psfrag{s35}{$s_{\mbox{\tiny 35}}$}
\psfrag{-}{$_-$}
\psfrag{+}{$_+$}
\includegraphics[scale=0.50]{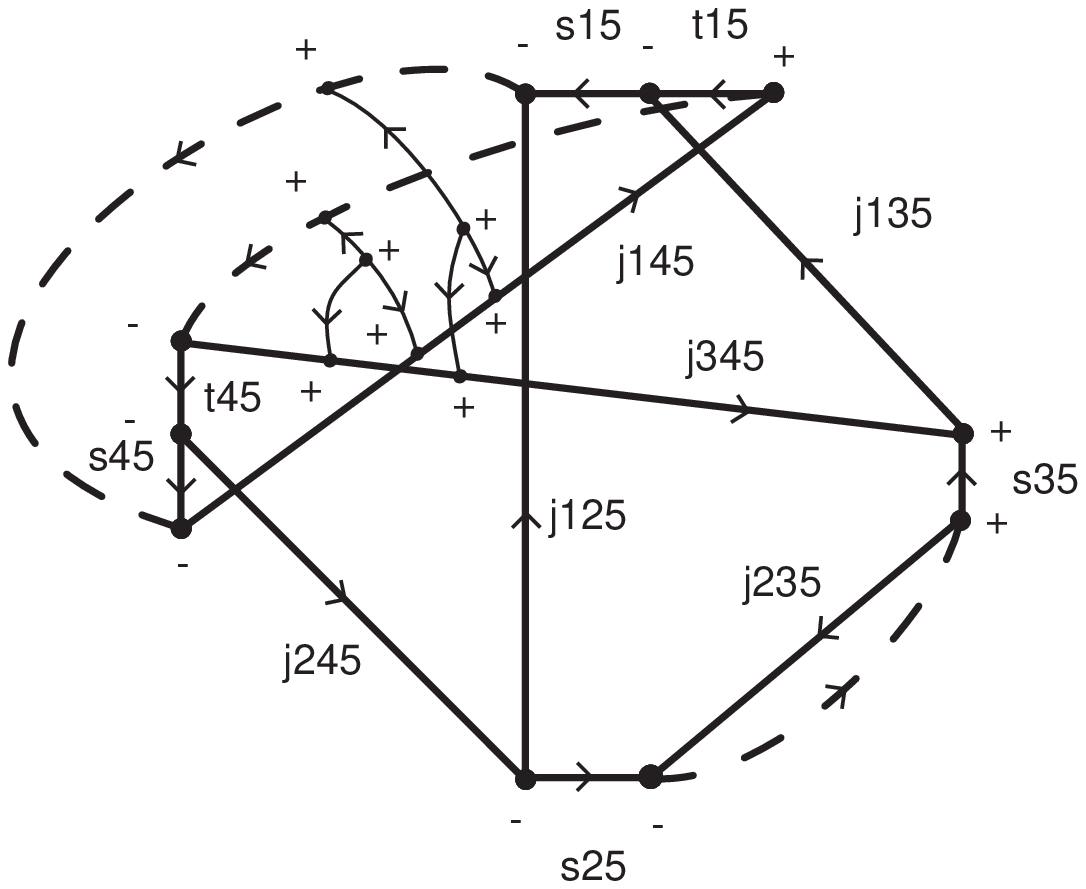}
\end{array}
\hspace{-4mm}
&=&
\hspace{-4mm}
\begin{array}{l}
\sum_{a,b,c,d,e,f,g,h,i }\dim a\dim b\;\dim c\;\dim d\;\dim e\;
\\ \times
\dim f \; \dim g \;\dim h \;\dim i \;   \delta_{s_{\mbox{\tiny 23}},s_{\mbox{\tiny 25}}   }  (  \dim s_{\mbox{\tiny 35}}   )^{-1}
\\ \times
(-1)^{( 3c+ 2d+ 2f+ 2g +3h   +  3i +1  2j_{\mbox{\tiny 125}}  + 3j_{\mbox{\tiny 245}}+ 3s_{\mbox{\tiny 25}} )} \\ \times
(-1)^{(
3j_{\mbox{\tiny 145}}+ j_{\mbox{\tiny 135}}  +
3j_{\mbox{\tiny 345}}   + 3s_{\mbox{\tiny 15}} + 3j_{\mbox{\tiny 45}}+ j_{\mbox{\tiny 235}} + s_{\mbox{\tiny 35}}    )    }
\\ \times
\left\{ \begin{array}{ccc}
 j_{\mbox{\tiny 125}} & \hspace{-2mm} j_{\mbox{\tiny 245}} & \hspace{-2mm} s_{\mbox{\tiny 25}} \\
 j_{\mbox{\tiny 345}} & \hspace{-2mm} j_{\mbox{\tiny 135}} & \hspace{-2mm} i
\end{array} \right\}
\left\{ \begin{array}{ccc}
i & \hspace{-2mm} j_{\mbox{\tiny 345}} & \hspace{-2mm} j_{\mbox{\tiny 24}} \\
 j_{\mbox{\tiny 345}} & \hspace{-2mm} g & \hspace{-2mm} 1
\end{array} \right\}
\left\{ \begin{array}{ccc}
 1 & \hspace{-2mm} i &  g \\
 j_{\mbox{\tiny 145}} & \hspace{-2mm} h &  d
\end{array} \right\}
\\ \times
\left\{ \begin{array}{ccc}
 j_{\mbox{\tiny 125}} & \hspace{-2mm} d & \hspace{-2mm} c \\
 j_{\mbox{\tiny 145}} & \hspace{-2mm} j_{\mbox{\tiny 135}} & \hspace{-2mm} i
\end{array} \right\}
\left\{ \begin{array}{ccc}
 h & \hspace{-2mm} g &j_{\mbox{\tiny 145}} \\
 j_{\mbox{\tiny 145}} & \hspace{-2mm} 1 &  e
\end{array} \right\}
\left\{ \begin{array}{ccc}
 1 &  1 &  1 \\
 e &  d &  h
\end{array} \right\}
\\ \times
\left\{ \begin{array}{ccc}
 a & \hspace{-2mm} e & \hspace{-2mm} 1 \\
 j_{\mbox{\tiny 145}} & \hspace{-2mm} j_{\mbox{\tiny 145}} & \hspace{-2mm} g
\end{array} \right\}
\left\{ \begin{array}{ccc}
 1 &  1&  1 \\
 a &  e &  f
\end{array} \right\}
\left\{ \begin{array}{ccc}
 e & 1 &  a \\
 \frac{1}{2} &  \frac{1}{2}  &  \frac{1}{2}
\end{array} \right\}
\\ \times
\left\{ \begin{array}{ccc}
 d & 1 &  e \\
 \frac{1}{2}  & \frac{1}{2} &  \frac{1}{2}
\end{array} \right\}
\left\{ \begin{array}{ccc}
 j_{\mbox{\tiny 125}} & \hspace{-2mm} \frac{1}{2}  &  s_{\mbox{\tiny 15}} \\
 \frac{1}{2}  & \hspace{-2mm} c &  d
\end{array} \right\}
\left\{ \begin{array}{ccc}
 j_{\mbox{\tiny 135}} & \hspace{-2mm} s_{\mbox{\tiny 15}} & \hspace{-2mm} t_{\mbox{\tiny 15}} \\
 \frac{1}{2}  & \hspace{-2mm} j_{\mbox{\tiny 145}} & \hspace{-2mm} c
\end{array} \right\}
\\ \times
\left\{ \begin{array}{ccc}
 \frac{1}{2}  &  j_{\mbox{\tiny 345}} & \hspace{-2mm} t_{\mbox{\tiny 45}} \\
 b &  \frac{1}{2}  & \hspace{-2mm} a
\end{array} \right\}
\left\{ \begin{array}{ccc}
 t_{\mbox{\tiny 45}} & \hspace{-2mm} j_{\mbox{\tiny 245}} & \hspace{-2mm} s_{\mbox{\tiny 45}} \\
 \frac{1}{2}  & \hspace{-2mm} j_{\mbox{\tiny 145}} & \hspace{-2mm} b
\end{array} \right\}
\end{array}
\nn
\eeqa

This completes the evaluation of $A^0_{\Gamma(2,3)}$.

\subsection{Gauge variant observable: Fermion propagator}

For the propagator we calculate one term of the expectation value of $\mathcal{O}^{\;\,\, A}_{f \;\, B}(2,3) = \psi_2^{A} \,\, \bar{\psi}_{3 B}$ in a particular gauge.
For this observable, the term in the Feynman expansion \eqref{expansion} that we consider is
\beqa
\mathcal{O}_{b} &=& ...+  (i\alpha)^{9} \left( \int_{\G} d \mu (\overline{\psi}_I, \hspace{1mm} \psi_I) \right) \hspace{1mm}
\psi_2^{A} \,\, \bar{\psi}_{3 B} \nn \\
&& \times
\overline{\psi}_{1C}   D_{12 \, D}^{\ \  C} \psi_{2}^D \
\overline{\psi}_{2E}   D_{21 \, F}^{\ \  E} \psi_{1}^F \
\overline{\psi}_{1G}   D_{15 \, H}^{\ \  G} \psi_{5}^H \
\overline{\psi}_{5I}   D_{51 \, J}^{\ \  I} \psi_{1}^J \
\overline{\psi}_{2K}   D_{23 \, L}^{\ \  K} \psi_{3}^L \
\nn \\
&& \times
\overline{\psi}_{3M}   D_{34 \, N}^{\ \  M} \psi_{4}^N \
\overline{\psi}_{4P}   D_{43 \, Q}^{\ \  P} \psi_{3}^Q \
\overline{\psi}_{4R}   D_{45 \, S}^{\ \  R} \psi_{5}^S \
\overline{\psi}_{5T}   D_{54 \, U}^{\ \  T} \psi_{4}^U \
 + ...,
\eeqa
including the factor of 2 coming from the identical contribution of the $S_{IJ}^2$ and $S_{JI}^2$ terms.

Applying the graphical method gives the admissible graph $\Gamma'(2,3)$ corresponding to the above term in the Feynman expansion
 $$
\psfrag{1}{$1$}
\psfrag{2}{$2$}
\psfrag{3}{$3$}
\psfrag{4}{$4$}
\psfrag{5}{$5$}
\begin{array}{c}
\includegraphics[scale=0.35]{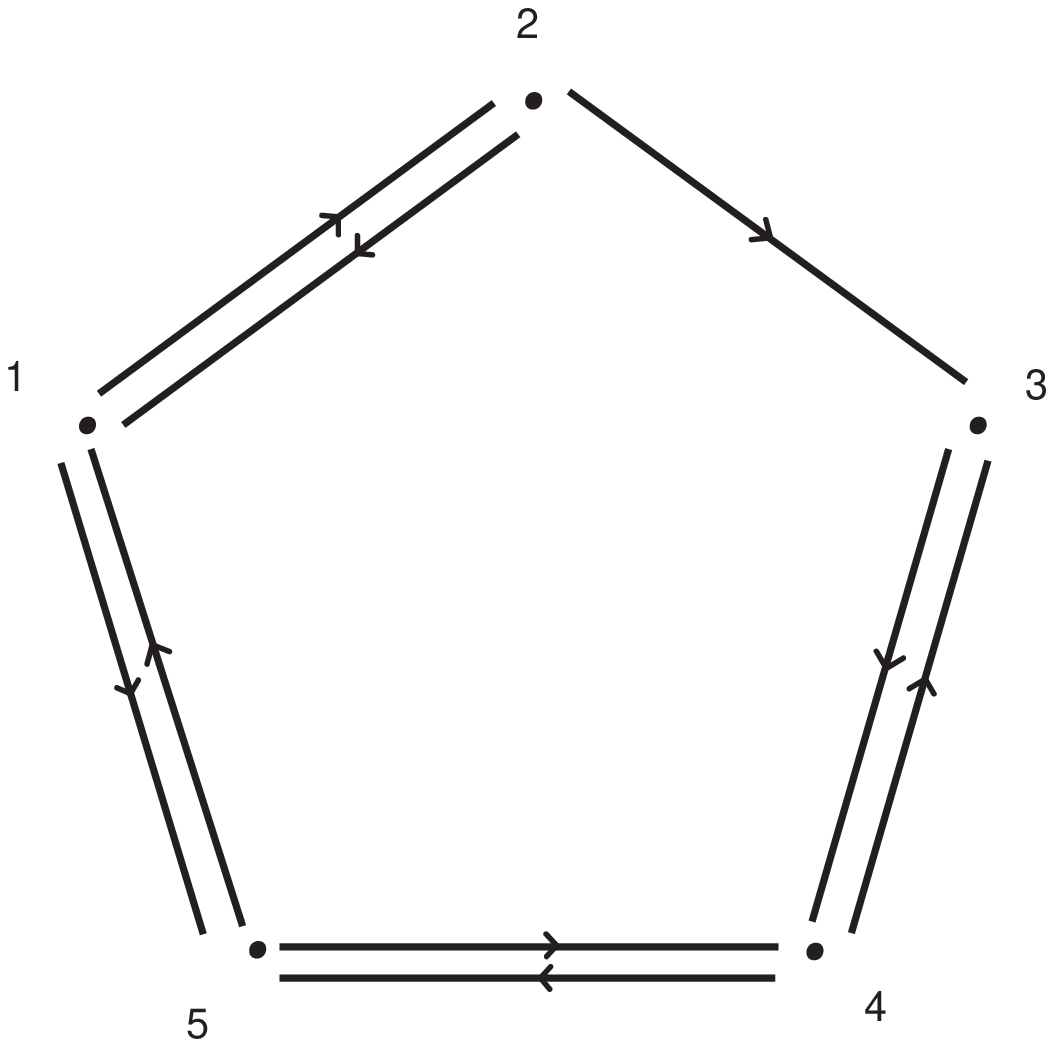}
\end{array}
 \ \ \ \ \ \ \ \ \Rightarrow \ \ \ \ \ \ \ \ \
\psfrag{1}{$1$}
\psfrag{2}{$2$}
\psfrag{3}{$3$}
\psfrag{4}{$4$}
\psfrag{5}{$5$}
\psfrag{p}{$\mathcal{P}'(2,3)$}
\psfrag{l}{$\mathcal{L}$}
\begin{array}{c}
\includegraphics[scale=0.35]{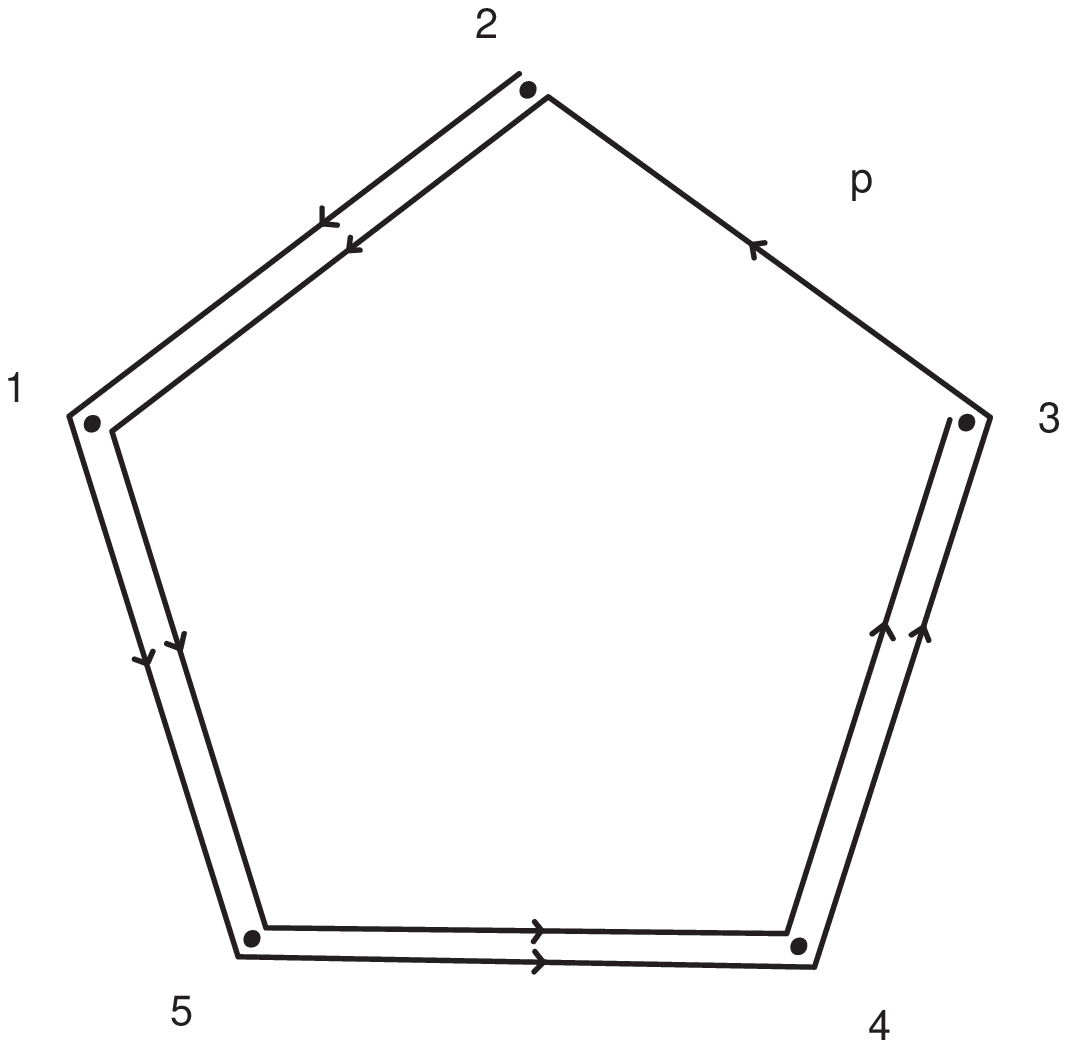}
\end{array}
$$
This gives a single path $\mathcal{P}'(2,3)$ of length $E_{\Gamma'(2,3) }= E_{\mathcal{P}'(2,3)} = 9$ passing through the vertices 2154321543.
Performing the integration and contracting the indices as before gives the sign factor $\epsilon_{\Gamma(2,3)}=1$ since there is no closed loop. The amplitude $ I_{\Gamma'(2,3)}$ of the graph $\Gamma'(2,3) $ is given by
\beq
\label{s3 example traces of D gauge variant}
I^{\,\,\;\;\;\;\;\;\;\;A}_{\Gamma'(2,3) \ B}
 =  (i \alpha)^9   \left( D_{21}D_{15}D_{54}D_{43}D_{32}D_{21}D_{15}D_{54}D_{43}\right)^A_{\ B} .
\eeq

This completes the fermionic integration, we can now consider the gravitational integration. As above, we will consider the $c=0$ configuration $I^{0}_{\Gamma'(2,3)}$, which is obtained from the above expression by replacing the Dirac matrices $D_{IJ}$ with the quantities $\Sigma_{IJ}U_{IJ}$.

We then choose the maximal gauge fixing tree $T$ on $\Delta^*$ given by the set of edges $T=\{21,15,54,43\}$.

We can now apply the graphical methods of section \ref{quantum theory} to obtain the generating functional and then the amplitude $A^0_{\Gamma'(2,3)}$ of the graph $\Gamma'(2,3)$ for the tree $T$. We obtain
\beqa
\label{s3 gauge variant amplitude}
A^0_{\Gamma'(2,3)}
&=&
 \left(\frac{\alpha}{4}\right)^{9}
  \prod_{I<J<K }
   \sum_{j_{IJK}}
 \dim j_{IJK}
(-1)^{2(j_{145} + j_{125} + j_{235} + j_{234} + j_{134}) + 1}
  \\
& & \times
\sum_{s_{23}} \dim s_{\mbox{\tiny 23}}
\nn \\
& & \times
  \Theta^3(j_{\mbox{\tiny 123}})
  \Theta^4(j_{\mbox{\tiny 125}})
  \Theta^2(j_{\mbox{\tiny 134}})
  \Theta^4(j_{\mbox{\tiny 145}})
  \Theta^3(j_{\mbox{\tiny 234}})
  \Theta^2(j_{\mbox{\tiny 345}})
 \nn \\
& & \times
 A(j_{IJK}, s_{23}    ),
\nn
\eeqa
where the amplitude $ A(j_{IJK}, s_{23}    ) \in \C$ is obtained by evaluating the
spin network diagram displayed below.
\beq
A(j_{IJK}, s_{23} )  =
\hspace{-30mm}
\begin{array}{c}
\psfrag{A}{$A$}
\psfrag{B}{$B$}
\psfrag{j123}{$j_{\mbox{\tiny 123}}$}
\psfrag{j124}{$j_{\mbox{\tiny 124}}$}
\psfrag{j125}{$j_{\mbox{\tiny 125}}$}
\psfrag{j134}{$j_{\mbox{\tiny 134}}$}
\psfrag{j135}{$j_{\mbox{\tiny 135}}$}
\psfrag{j145}{$j_{\mbox{\tiny 145}}$}
\psfrag{j234}{$j_{\mbox{\tiny 234}}$}
\psfrag{j235}{$j_{\mbox{\tiny 235}}$}
\psfrag{j245}{$j_{\mbox{\tiny 245}}$}
\psfrag{j345}{$j_{\mbox{\tiny 345}}$}
\psfrag{s23}{$s_{\mbox{\tiny 23}}$}
\psfrag{+}{$_+$}
\psfrag{-}{$_-$}
\psfrag{p}{ }
\includegraphics[scale=0.45]{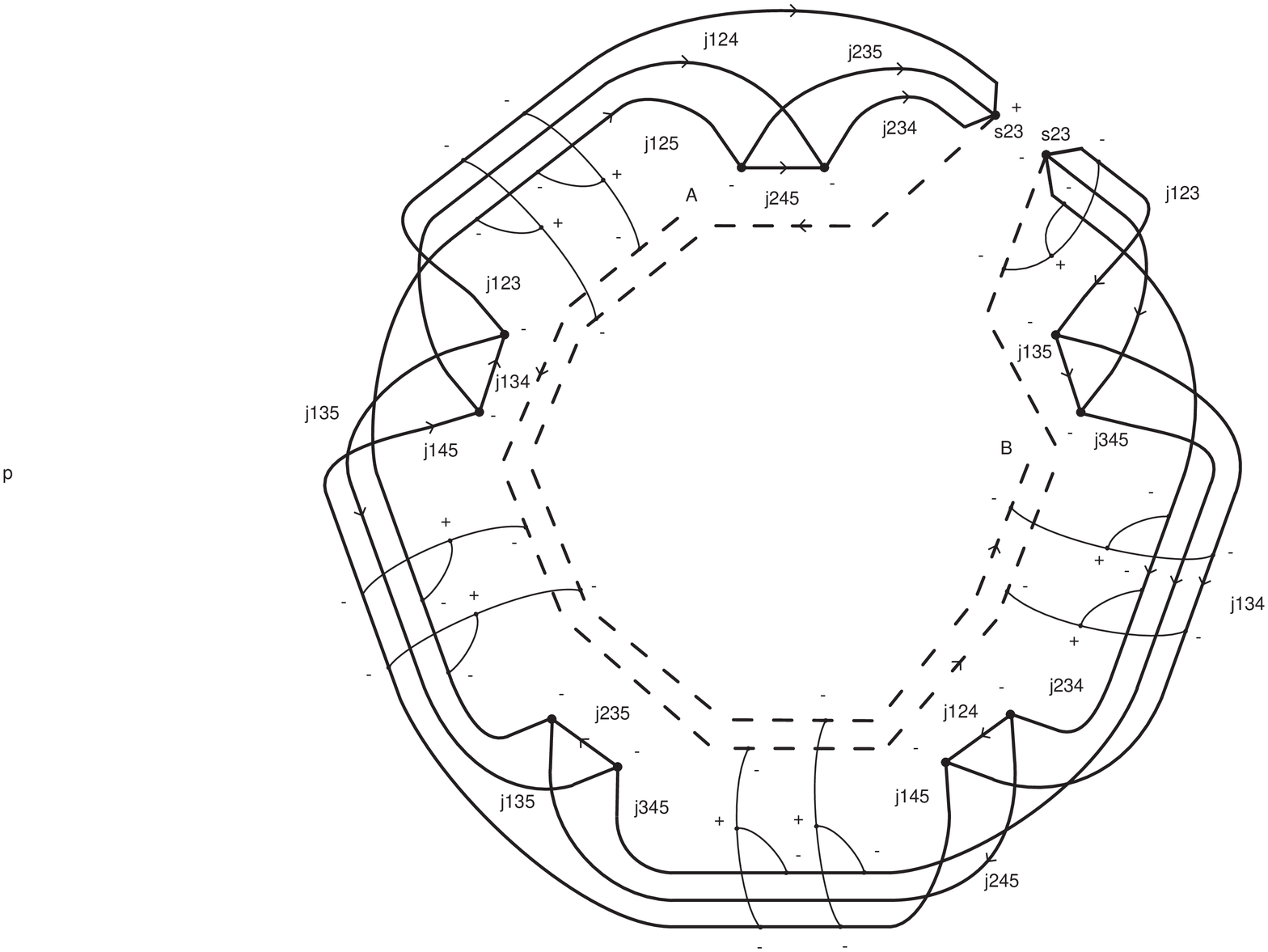}
\end{array}
\eeq
By Schur's Lemma, this will be proportional to the identity map $\delta^A_B$ in the fundamental representation multiplied by half the value of the closed spin network evaluation obtained by connecting the two open lines in the $j = 1/2$ representation. The closed diagram can then be recoupled and given as a combination of $6j$ symbols.


\section{Conclusions}

In this paper, we have shown how to compute expectation values of  fermionic observables in spinfoam models of 3d quantum gravity. We have identified and properly gauge fixed the symmetries of the model,  given a general procedure to compute gauge invariant and gauge dependent observables and illustrated the techniques on an example.

It is important to stress the achievements of this work with respect to the original proposal \cite{fairbairn-2007-39}. We have 1) extended the model to massless fermionic fields, 2) implemented a new procedure to compute the Berezin integrals, 3) introduced observables, 4) recoupled the vertex amplitudes to express them as combinations of $6j$ symbols.


Many issues remain open. First, unlike the point particle case, fields break the topological invariance of three dimensional gravity. The model is therefore only an approximation of the continuum theory, even if the spinfoam explicitly derives from a classical action. This leads to two questions. First, do the asymptotics ($j \rightarrow \infty$) of the vertex amplitudes reproduce (a simple function of) the classical Regge action coupled to the Feynman diagrams issued from the fermionic integration? Second, and most importantly: what is the behaviour of the model under refinement
of the triangulation? Since the finiteness of our Feynman diagram expansion is due to the cut-off on the number of degrees of freedom induced by the triangulation, this issue is of primary interest. This issue was considered in \cite{Dowdall:2009ds} where a group field theory approach to the model was constructed, in order to implement a sum over triangulations.
Finally, one would need to analyse the well known fermion doubling problem appearing in
lattice-like theories incorporating fermions. Does the spinfoam generate doublers? Could we
modify the action to suppress the extra species?

We close with the perspectives. Working with a three-dimensional non-topological model has taught us techniques that could be useful in the full four-dimensional theory coupled to matter. This leads us to the most interesting outlook which would be a generalisation of the ideas and techniques developed here to four spacetime dimensions. At first, one might expect the coupling of fermions to the EPRL or FK models \cite{Engle:2007wy,freidel-2007,Kaminski:2009cc} of four dimensional quantum gravity to be problematic. The reason is that the kinetic term of the standard Dirac action in a $d$-dimensional curved spacetime couples to the $d-1$ exterior power of the $d$-bein. Hence, in even dimensions, fermions are coupled to Palatini gravity via an odd power of the gravitational field not convertible into any power of the $d-2$-form B field of BF theory. However, if one uses the Capovilla, Dell, Jacobson, Mason chiral action \cite{Capovilla:1991qb} of fermions coupled to two-form gravity, the fermions indeed couple to the B field and one could imagine such an action as an interesting starting point for a four-dimensional construction.

\subsection*{Acknowledgements}

We are grateful to Laurent Lellouch for discussions on the fermion two-point function in lattice QCD and to John W. Barrett for key remarks on an early version of the draft. WF acknowledges support from the Royal Commission of the Exhibition of 1851, and the Emmy Noether grant ME 3425/1-1 of the German Research Foundation (DFG).


\appendix

\section{Basic facts on the representation theory of $\SU(2)$}

\subsection{Notation and conventions}

Our conventions are essentially those of references \cite{AMD,varshalovich}. The unitary irreducible representations of $\SU(2)$, $\pi^j : \SU(2) \rightarrow \Aut(V_j)$, are labelled by a spin $j \in \mathbb{N}/2$. The spin $j$ representation acts in the finite dimensional complex vector space $V_j \cong \C^{2j +1}$. In fact, $V_j$ can be endowed with a metric structure inherited from the pairing map $\epsilon^j : V_j \rightarrow V_j^*$, where $V_j^*$ is the vector space of linear forms on $V_j$. Choosing a particular basis $(e_{\alpha}^j)_{\alpha = -j , ..., j}$ of $V_j$, the components of the paring map, the Wigner $1j$ symbol, are given by
$$
\epsilon_{\alpha \beta}^j = \epsilon^{j \,\alpha \beta} = (-1)^{j - \beta} \delta_{\alpha, - \beta}.
$$
This map satisfies the properties
$$
\epsilon_{\beta \alpha}^j = (-1)^{2j} \epsilon_{\alpha \beta}^j  = (-1)^{j + \beta} \delta_{\beta, -\alpha}, \;\;\;\; \mbox{and} \;\;\;\; \epsilon_{\alpha \beta}^j \epsilon^{j \,\beta \gamma} = (-1)^{2j} \delta_{\alpha}^{\gamma}.
$$
The associated inner product is defined by $(v,w)_j = \epsilon^j(v)(w) = v^{\alpha} w^{\beta} \epsilon_{\alpha \beta}^j := v^{\alpha} w_{\alpha}$, for $v,w$ in $V_j$.

For all triple of unitary, irreducible representations $\pi^i, \pi^j, \pi^k$ of $\SU(2)$, let
\beq
C^{k}_{\; ij}: V_i \otimes V_j \rightarrow V_k, \hspace{3mm} \mbox{and} \hspace{3mm} C^{ij}_{\;\, k} : V_k \rightarrow V_i \otimes V_j,
\eeq
denote the Clebsh-Gordan intertwining operators whose coefficients
\beq
C^{ij}_{\;\, k}(e^k_{\gamma}) = \sum_{\alpha,\beta} \left( \begin{array}{cc} \alpha & \beta \\
                          i & j  \end{array} \right| \left. \begin{array}{c} k \\ \gamma \end{array} \right)  e^i_{\alpha} \otimes e^j_{\beta}, \hspace{3mm} \mbox{and} \hspace{3mm}
C^{k}_{\; ij}( e^i_{\alpha} \otimes e^j_{\beta}) = \sum_{\gamma} \left( \begin{array}{c} \gamma \\ k
 \end{array} \right| \left. \begin{array}{cc} i & j \\ \alpha & \beta \end{array} \right)  e^k_{\gamma} ,
\eeq
are well known from the quantum mechanics of angular momentum. These intertwiners are unique up to normalisation because the space of three-valent intertwiners is a vector space of dimension one.
The complete reducibility of the tensor product of representations
$$
V_i \otimes V_j \cong \bigoplus_{k = | i - j |}^{i+j} V_k,
$$
is then expressed as $\pi^i \otimes \pi^j = \sum_k C^{k}_{\; ij} \, \pi^k \, C^{ij}_{\;\, k}$. Relative to the chosen bases, this yields
\beq
\pi^{i \, \alpha}_{\;\;\; \beta} \, \pi^{j \, \gamma}_{\;\;\;\, \delta} = \sum_{k,\epsilon, \zeta} \left( \begin{array}{cc} \alpha & \gamma \\
                          i & j  \end{array} \right| \left. \begin{array}{c} k \\ \zeta \end{array} \right) \, \pi^{k \, \zeta}_{\;\;\;\; \epsilon} \, \left( \begin{array}{c} \epsilon \\ k
 \end{array} \right| \left. \begin{array}{cc} i & j \\ \beta & \delta \end{array} \right).
\eeq

The Clebsh-Gordan coefficients satisfy rather awkward symmetry properties . One can consider a rescaled version of these symbols that are more symmetric. These objects are called $3j$ symbols. They are evaluations of the $3j$ maps which are normalised three-valent intertwiners. Let $\iota_{ijk} : V_i \otimes V_j \otimes V_k \rightarrow \C$ be a $3j$ intertwining map and $u,v$ and $w$ be arbitrary vectors in $V_i,V_k$ and $V_k$ respectively. Clebsh-Gordan symbols and $3j$ symbols are related via the following evaluation
\beq
\iota_{ijk} (u \otimes v \otimes w) = \frac{\epsilon(i,j,k)}{\sqrt{\dim k}} \, (w , C^{k}_{\; ij}( u \otimes v)),
\eeq
where $\epsilon(i,j,k)$ is a sign usually chosen to be $(-1)^{-i + j - k}$, where $-i+j-k \in \mathbb{Z}$. The evaluation of the above equation on the vectors $u=e^i_{\alpha}$, $v=e^j_{\beta}$ and $w=e^k_{\gamma}$ reads
\beq
\left( \begin{array}{ccc}
 i & j & k \\ \alpha & \beta & \gamma \end{array} \right)  =  \frac{(-1)^{-i + j - k}}{\sqrt{\dim k}} \epsilon^k_{\gamma \delta} \left( \begin{array}{c} \delta \\ k
 \end{array} \right| \left. \begin{array}{cc} i & j \\ \alpha & \beta \end{array} \right) = \frac{(-1)^{-i + j + \gamma}}{\sqrt{\dim k}}  \left( \begin{array}{c} - \gamma \\ k
 \end{array} \right| \left. \begin{array}{cc} i & j \\ \alpha & \beta \end{array} \right),
\eeq
where the left hand side is a $3j$ symbol.
The above relation can be put into a more compact form by using the symplectic metric to raise and lower indices. Our conventions are such that indices are raised from the right and lowered from the left, that is, picturing only one column of the symbol,
$$
\left( \begin{array}{c}
 i \\ \alpha \end{array} \right)  = \epsilon^i_{\alpha \beta} \left( \begin{array}{c}
 \beta \\ i \end{array} \right), \;\;\;\; \mbox{and} \;\;\;\; \left( \begin{array}{c}
 \alpha \\ i \end{array} \right) = \left( \begin{array}{c}
 i  \\ \beta \end{array} \right) \epsilon^{i \,\beta \alpha}.
$$
Note that this implies the following property
\beq
\label{seesaw1}
\left( \begin{array}{c}  i \\ \alpha \end{array} \right) \left( \begin{array}{c}
 \alpha \\ i \end{array} \right) = (-1)^{2i} \left( \begin{array}{c}
 \alpha \\ i \end{array} \right) \left( \begin{array}{c}  i \\ \alpha \end{array} \right).
\eeq
From these rules, one can construct mixed $3j$ symbols, that is, $3j$ symbols associated to all possible three-fold tensor products involving three vector spaces together with their duals.
Furthermore, using the symmetry under permutations of the $3j$ symbol
\beq
\label{cyclicity}
\left( \begin{array}{ccc}
 i & j & k \\ \alpha & \beta & \gamma \end{array} \right)  = (-1)^{i+j+k} \left( \begin{array}{ccc}
 i & k & j \\ \alpha & \gamma & \beta \end{array} \right) 
 ,
 \eeq
which implies that the symbol is even under even permutations of the columns, the relation between $3j$ symbols and Clebsch-Gordan coefficients can be put in the form
\beq
\left( \begin{array}{ccc}
 k & i & j \\  \gamma & \alpha & \beta \end{array} \right) = \frac{(-1)^{-i + j - k}}{\sqrt{\dim k}}  \left( \begin{array}{c} k \\ \gamma
 \end{array} \right| \left. \begin{array}{cc} i & j \\ \alpha & \beta \end{array} \right).
\eeq

\subsection{Group integrals}

Integrals of tensor product of representations can be expressed in terms of $3j$ symbols. The first step is to realise that
$$
\int_{\SU(2)} dg \, \pi^k(g) = \delta^k_0.
$$
From the above expression and the complete reducibility, one can compute all integrals involved in this paper. The first in the bivalent integration. The calculation proceeds as follows.
\beqa
\int_{\SU(2)} dg \, \pi^i(g)^{\alpha}_{\;\; \beta} \, \pi^j(g)^{\gamma}_{\;\; \delta} &=& \sum_{k,\epsilon, \zeta} \left( \begin{array}{cc} \alpha & \gamma \\
                          i & j  \end{array} \right| \left. \begin{array}{c} k \\ \zeta \end{array} \right) \left( \begin{array}{c} \epsilon \\ k
 \end{array} \right| \left. \begin{array}{cc} i & j \\ \beta & \delta \end{array} \right) \int_{\SU(2)} dg  \, \pi^k(g)^{\zeta}_{\;\; \epsilon} \nn \\
 &=& \left( \begin{array}{cc} \alpha & \gamma \\
                          i & j  \end{array} \right| \left. \begin{array}{c} 0 \\ 0 \end{array} \right)  \, \left( \begin{array}{c} 0 \\ 0
 \end{array} \right| \left. \begin{array}{cc} i & j \\ \beta & \delta \end{array} \right) \nn \\
&=& \frac{\delta^{ij}}{ \dim j} \, \epsilon^{j \,\alpha \gamma}  \epsilon_{\beta \delta}^j,
\eeqa
where we have used that
$$
\left( \begin{array}{cc} \alpha & \beta \\
                          i & j  \end{array} \right| \left. \begin{array}{c} 0 \\ 0 \end{array} \right) = \frac{\delta^{ij}}{\sqrt{\dim j}} (-1)^{j-\alpha} \delta_{\beta,-\alpha} = (-1)^{2j} \frac{\delta^{ij}}{\sqrt{\dim j}} \epsilon^{j \,\alpha \beta},
$$
in the last step. From this result it is immediate to prove the orthogonality of characters \eqref{character}.

Using the above, one can compute the three-valent integral. The calculation is displayed below.
\beqa
\label{3valent integral}
\int_{\SU(2)} dg \, \pi^i(g)^{\alpha}_{\;\; \beta} \, \pi^j(g)^{\gamma}_{\;\; \delta} \pi^k(g)^{\epsilon}_{\;\; \zeta} &=& \sum_{l,\eta,\theta} \left( \begin{array}{cc} \alpha & \gamma \\
                          i & j  \end{array} \right| \left. \begin{array}{c} l \\ \theta \end{array} \right)\left( \begin{array}{c} \eta \\ l
 \end{array} \right| \left. \begin{array}{cc} i & j \\ \beta & \delta \end{array} \right)  \int_{\SU(2)} dg \,\pi^l(g)^{\theta}_{\;\; \eta} \pi^k(g)^{\epsilon}_{\;\; \zeta}  \nn \\
 &=& \frac{1}{\dim k} \sum_{\eta,\theta} \left( \begin{array}{cc} \alpha & \gamma \\
                        i & j  \end{array} \right| \left. \begin{array}{c} k \\ \theta \end{array} \right)
                          \epsilon^{k \,\theta \epsilon}  \epsilon_{\eta \zeta}^k \left( \begin{array}{c} \eta \\ k
 \end{array} \right| \left. \begin{array}{cc} i & j \\ \beta & \delta \end{array} \right) \nn \\
 &=& \left( \begin{array}{ccc}
\alpha & \gamma & \epsilon \\ i & j & k  \end{array} \right)  \left( \begin{array}{ccc}
 i & j & k \\ \beta & \delta & \zeta \end{array} \right).
\eeqa
To calculate the same integral, but with one group element inverted, on simply uses the fact that the dual pairing is in fact an isomorphism of representations, that is, a bijective intertwiner. This translates as
\beq
\label{conjugation}
\pi^i(g^{-1})^{\alpha}_{\;\; \beta} = (-1)^{2j} \epsilon^{j \,\alpha \gamma} \pi^i(g)^{\delta}_{\;\; \gamma} \epsilon^j_{\delta \beta}.
\eeq
From the above, we immediately obtain that
\beq
\int_{\SU(2)} dg \, \pi^i(g)^{\alpha}_{\;\; \beta} \, \pi^j(g)^{\gamma}_{\;\; \delta} \pi^k(g^{-1})^{\epsilon}_{\;\; \zeta} = (-1)^{2k} \left( \begin{array}{ccc}
\alpha & \gamma & k \\ i & j & \zeta  \end{array} \right)  \left( \begin{array}{ccc}
 i & j & \epsilon \\ \beta & \delta &  k \end{array} \right). \nn
\eeq
Because of the invariance of the Haar measure under inversion $g \mapsto g^{-1}$, we have, with the two above formulae, covered all possible cases of orientations that can occur in the model.

We can next proceed analogously for the four-valent case. We obtain the following result
\beqa
&& \int_{\SU(2)} dg \, \pi^i(g)^{\alpha}_{\;\; \beta} \, \pi^j(g)^{\gamma}_{\;\; \delta} \pi^k(g)^{\epsilon}_{\;\; \zeta}  \pi^l(g)^{\theta}_{\;\; \eta} = \sum_{m,\kappa,\iota} \sum_{n,\mu, \lambda} \left( \begin{array}{cc} \alpha & \gamma \\
                          i & j  \end{array} \right| \left. \begin{array}{c} m \\ \iota \end{array} \right)\left( \begin{array}{c} \kappa \\ m
 \end{array} \right| \left. \begin{array}{cc} i & j \\ \beta & \delta \end{array} \right)  \nn \\
 && \times \left( \begin{array}{cc} \epsilon & \theta \\
                          k & l  \end{array} \right| \left. \begin{array}{c} n \\ \mu \end{array} \right)\left( \begin{array}{c} \lambda \\ n
 \end{array} \right| \left. \begin{array}{cc} k & l \\ \zeta & \eta \end{array} \right) \int_{\SU(2)} dg \,\pi^m(g)^{\iota}_{\;\; \kappa} \pi^n(g)^{\mu}_{\;\; \lambda}  \nn \\
 &=& \frac{1}{\dim m} \sum_{m,\kappa,\iota,\mu,\lambda}  \left( \begin{array}{cc} \alpha & \gamma \\
                          i & j  \end{array} \right| \left. \begin{array}{c} m \\ \iota \end{array} \right) \epsilon^{m \,\iota \mu}  \left( \begin{array}{cc} \epsilon & \theta \\
                          k & l  \end{array} \right| \left. \begin{array}{c} m \\ \mu \end{array} \right)  \left( \begin{array}{c} \kappa \\ m
 \end{array} \right| \left. \begin{array}{cc} i & j \\ \beta & \delta \end{array} \right)  \epsilon_{\kappa\lambda}^m \left( \begin{array}{c} \lambda \\ m
 \end{array} \right| \left. \begin{array}{cc} k & l \\ \zeta & \eta \end{array} \right) \nn \\
 &=& \sum_{m}  \left( \begin{array}{cccc}
\alpha & \gamma & \epsilon & \theta \\ i & j & k & l \end{array} \right)_{m}  \left( \begin{array}{cccc}
 i & j & k & l \\ \beta & \delta & \zeta & \eta \end{array} \right)_{m},
\eeqa
where the obtained $4j$ symbol is defined as follows
\beq
\left( \begin{array}{cccc}
 i & j & k & l \\ \alpha & \beta & \gamma & \delta  \end{array} \right)_{m} = \sum_{\kappa}\sqrt{\dim m} 
 \left( \begin{array}{ccc}
i & j & \kappa \\ \alpha & \beta & m \\   \end{array} \right)  \left( \begin{array}{ccc}
m & k & l  \\  \kappa & \gamma & \delta \end{array} \right).
\eeq
Using the conjugation \eqref{conjugation}, we can also calculate the following integral
\beq
\int_{\SU(2)} dg \, \pi^i(g)^{\alpha}_{\;\; \beta} \, \pi^j(g)^{\gamma}_{\;\; \delta} \pi^k(g)^{\epsilon}_{\;\; \zeta}  \pi^l(g^{-1})^{\theta}_{\;\; \eta} = (-1)^{2l} \sum_{m}  \left( \begin{array}{cccc}
\alpha & \gamma & \epsilon & l \\ i & j & k & \eta \end{array} \right)_{m}  \left( \begin{array}{cccc}
 i & j & k & \theta \\ \beta & \delta & \zeta & l \end{array} \right)_{m}, \nn
\eeq
and also
\beq
\int_{\SU(2)} dg \, \pi^i(g)^{\alpha}_{\;\; \beta} \, \pi^j(g)^{\gamma}_{\;\; \delta} \pi^k(g^{-1})^{\epsilon}_{\;\; \zeta}  \pi^l(g^{-1})^{\theta}_{\;\; \eta} = (-1)^{2(k +l)} \sum_{m}  \left( \begin{array}{cccc}
\alpha & \gamma & k & l \\ i & j & \zeta & \eta \end{array} \right)_{m}  \left( \begin{array}{cccc}
 i & j & \epsilon & \theta \\ \beta & \delta & k & l \end{array} \right)_{m}. \nn
\eeq
This covers all possible four-valent orientations configurations.

Finally, we compute the five-valent integration as follows.
\beqa
&& \int_{\SU(2)} dg \, \pi^i(g)^{\alpha}_{\;\; \beta} \, \pi^j(g)^{\gamma}_{\;\; \delta} \pi^k(g)^{\epsilon}_{\;\; \zeta}  \pi^l(g)^{\theta}_{\;\; \eta} \pi^m(g)^{\iota}_{\;\; \kappa}
= \int_{\SU(2)} dg \, \sum_{n,\mu, \lambda} \sum_{o, \nu, \xi} \left( \begin{array}{cc} \alpha & \gamma \\
                          i & j  \end{array} \right| \left. \begin{array}{c} n \\ \lambda \end{array} \right) \nn \\
                          && \times \left( \begin{array}{c} \mu \\ n
 \end{array} \right| \left. \begin{array}{cc} i & j \\ \beta & \delta \end{array} \right)
 \pi^n(g)^{\lambda}_{\;\; \mu} \pi^k(g)^{\epsilon}_{\;\; \zeta} \left( \begin{array}{cc} \theta & \iota \\
                          l & m  \end{array} \right| \left. \begin{array}{c} o \\ \nu \end{array} \right) \pi^o(g)^{\nu}_{\;\; \xi}\left( \begin{array}{c} \xi \\ o
 \end{array} \right| \left. \begin{array}{cc} l & m \\ \eta & \kappa \end{array} \right)   \nn \\
 &=&  \sum_{n,\mu, \lambda} \sum_{o, \nu, \xi} \sum_{p,\rho, \sigma} \left( \begin{array}{cc} \alpha & \gamma \\
                          i & j  \end{array} \right| \left. \begin{array}{c} n \\ \lambda \end{array} \right)  \left( \begin{array}{c} \mu \\ n
 \end{array} \right| \left. \begin{array}{cc} i & j \\ \beta & \delta \end{array} \right)
  \left( \begin{array}{cc} \lambda & \epsilon \\
                          n & k  \end{array} \right| \left. \begin{array}{c} p \\ \rho \end{array} \right) \nn \\ && \times \left( \begin{array}{c} \sigma \\ p
 \end{array} \right| \left. \begin{array}{cc} n & k \\ \mu & \zeta \end{array} \right)
 \left( \begin{array}{cc} \theta & \iota \\
                          l & m  \end{array} \right| \left. \begin{array}{c} o \\ \nu \end{array} \right) \left( \begin{array}{c} \xi \\ o
 \end{array} \right| \left. \begin{array}{cc} l & m \\ \eta & \kappa \end{array} \right) \int_{\SU(2)} dg \, \pi^p(g)^{\rho}_{\;\; \sigma} \pi^o(g)^{\nu}_{\;\; \xi}   \nn \\
&=&  \sum_{n,o}  \left( \begin{array}{ccccc}
\alpha & \gamma & \epsilon & \theta & \iota \\ i & j & k & l & m \end{array} \right)_{n,o}  \left( \begin{array}{ccccc}
 i & j & k & l & m \\ \beta & \delta & \zeta & \eta & \kappa \end{array} \right)_{n,o},
\eeqa
where the $5j$ symbol is given by the following expression
\beq
\left( \begin{array}{ccccc}
 i & j & k & l & m \\ \beta & \delta & \zeta & \eta & \kappa \end{array} \right)_{n,o} = \sum_{\mu,\sigma} \sqrt{\dim n \dim o}\left( \begin{array}{ccc} i & j & \mu \\ \beta & \delta & n \end{array} \right)  \left( \begin{array}{ccc}  n & k & \sigma \\ \mu & \zeta & o \end{array} \right) \left( \begin{array}{ccc} o &  l & m \\ \sigma & \eta & \kappa \end{array} \right).
\eeq
Again, from the above, it is immediate to obtain that
\beqa
&&
\int_{\SU(2)} dg \, \pi^i(g)^{\alpha}_{\;\; \beta} \, \pi^j(g)^{\gamma}_{\;\; \delta} \pi^k(g)^{\epsilon}_{\;\; \zeta}  \pi^l(g)^{\theta}_{\;\; \eta} \pi^m(g^{-1})^{\iota}_{\;\; \kappa} \nn \\
&&= (-1)^{2m} \sum_{n,o}  \left( \begin{array}{ccccc}
\alpha & \gamma & \epsilon & \theta & m \\ i & j & k & l & \kappa\end{array} \right)_{n,o}  \left( \begin{array}{ccccc}
 i & j & k & l & \iota \\ \beta & \delta & \zeta & \eta & m \end{array} \right)_{n,o}, \nn
\eeqa
and that
\beqa
&&
\int_{\SU(2)} dg \, \pi^i(g)^{\alpha}_{\;\; \beta} \, \pi^j(g)^{\gamma}_{\;\; \delta} \pi^k(g)^{\epsilon}_{\;\; \zeta}  \pi^l(g^{-1})^{\theta}_{\;\; \eta} \pi^m(g^{-1})^{\iota}_{\;\; \kappa} \nn \\
&&= (-1)^{2(l +m)} \sum_{n,o}  \left( \begin{array}{ccccc}
\alpha & \gamma & \epsilon & l & m \\ i & j & k & \eta & \kappa \end{array} \right)_{n,o}  \left( \begin{array}{ccccc}
 i & j & k & \theta & \iota \\ \beta & \delta & \zeta & l & m \end{array} \right)_{n,o}. \nn
\eeqa
This covers all possible group integrals that can appear in the model.

\subsection{Diagrammatics and Recoupling identities}

The spin network vertex amplitudes obtained in this paper can be re-expressed as products of $6j$-symbols using recoupling identities. These identities are most conveniently express diagrammatically. A $3j$ symbol is represented as an ordered vertex with three lines. The clockwise (resp. anti-clockwise) ordering of the lines at a vertex is denoted with a $+$ (resp. $-$) symbol. If required, arrows will be depicted on the lines with the convention that ingoing lines at a vertex correspond to lowered magnetic components in the symbol. This implies that
\beq
\left( \begin{array}{ccc}
 i & j & k \\ \alpha & \beta & \gamma \end{array} \right) =
 \begin{array}{c}
\psfrag{i}{$i$}
\psfrag{j}{$j$}
\psfrag{k}{$k$}
\psfrag{a}{$ $}
\psfrag{b}{$ $}
\psfrag{c}{$ $}
\psfrag{+}{$_+$}
\includegraphics[scale=0.25]{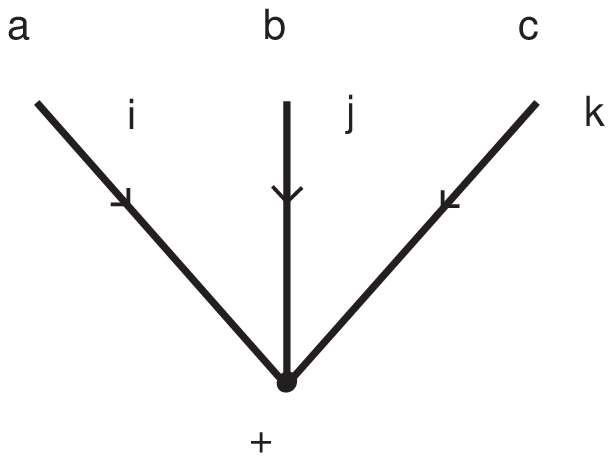}
\end{array}
 . \nn
\eeq
and
\beq
\left( \begin{array}{ccc}
\alpha & \beta & \gamma \\  i & j & k \end{array} \right) =
\begin{array}{c}
\psfrag{i}{$i$}
\psfrag{j}{$j$}
\psfrag{k}{$k$}
\psfrag{a}{$ $}
\psfrag{b}{$ $}
\psfrag{c}{$ $}
\psfrag{+}{$_+$}
\includegraphics[scale=0.25]{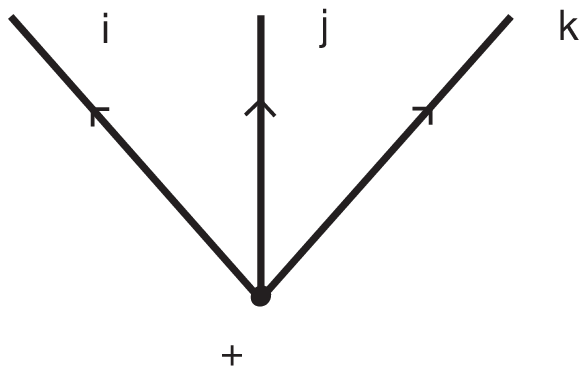}
\end{array}
. \nn
\eeq

The $3j$ symbols are normalised
\beq
\left( \begin{array}{ccc}
i & j & k \\
\alpha & \beta & \gamma
\end{array} \right)
\left( \begin{array}{ccc}
 \alpha & \beta & \gamma \\
 i & j & k
 \end{array} \right)
 =  Y(i,j,k),
\eeq
where $Y(i,j,k) = 1$ if $i,k$ and $l$ satisfy the triangle inequalities and yields zero otherwise. Diagrammatically, this means that the theta network is equal to (zero or) one
\beq
\label{theta}
\begin{array}{c}
\psfrag{a}{$i$}
\psfrag{b}{$j $}
\psfrag{c}{$k $}
\psfrag{+}{$_+$}
\psfrag{-}{$_-$}
\includegraphics[scale=0.4]{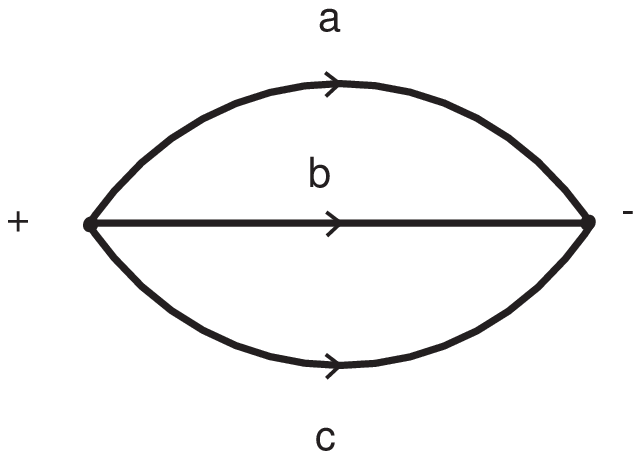}
\end{array} =
\begin{array}{c}
\psfrag{a}{$i$}
\psfrag{b}{$j $}
\psfrag{c}{$k $}
\psfrag{+}{$_+$}
\psfrag{-}{$_-$}
\includegraphics[scale=0.4]{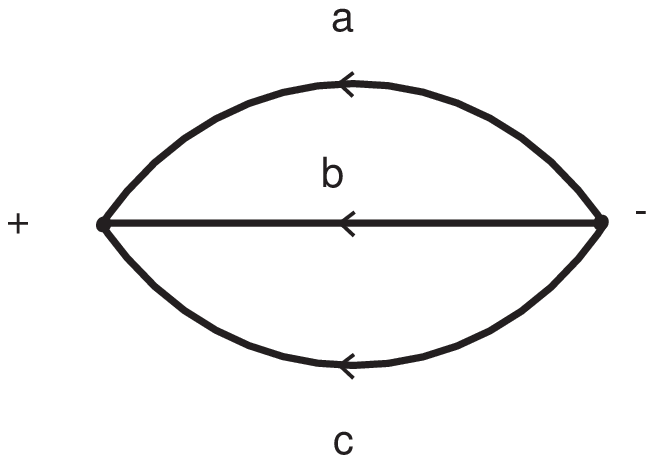}
\end{array} =
Y(i,k,l).
\eeq
Note that changing the orientation of an arrow, say of spin $k$, in the above diagrams produces a factor $(-1)^{2k}$. The corresponding intertwiner has norm (squared) given by $(-1)^{2k}Y(i,j,k)$, in accordance with the results found in the group integral section.

Changing the ordering at a vertex is equivalent to a permutation of the arguments of the $3j$ thus we have the diagrammatic identity
\beq
 \begin{array}{c}
\psfrag{i}{$i$}
\psfrag{j}{$j$}
\psfrag{k}{$k$}
\psfrag{a}{$ $}
\psfrag{b}{$ $}
\psfrag{c}{$ $}
\psfrag{+}{$_+$}
\includegraphics[scale=0.25]{ingoing}
\end{array}=
(-1)^{i+j+k}
 \begin{array}{c}
\psfrag{i}{$i$}
\psfrag{j}{$j$}
\psfrag{k}{$k$}
\psfrag{a}{$ $}
\psfrag{b}{$ $}
\psfrag{c}{$ $}
\psfrag{+}{$_-$}
\includegraphics[scale=0.25]{ingoing}
\end{array}
\eeq
This emphasises the need to have an ordering label at the vertices.

The $6j$ symbol is defined by the following four-fold contraction of $3j$ symbols
\beq
\left\{ \begin{array}{ccc}
j_1 & j_2 & j_3 \\ j_4 & j_5 & j_6 \end{array} \right\} =
\left( \begin{array}{ccc}
\alpha_1 & \alpha_2 & \alpha_3 \\ j_1 & j_2 & j_3 \end{array} \right)
 \left( \begin{array}{ccc}
 j_1 & \alpha_5 & j_6 \\ \alpha_1 & j_5 & \alpha_6 \end{array} \right)
 \left( \begin{array}{ccc}
 \alpha_6 & j_4 & j_2 \\ j_6 & \alpha_4 & \alpha_2 \end{array} \right)
 \left( \begin{array}{ccc}
 \alpha_4 & j_5 & j_3 \\ j_4 & \alpha_5 & \alpha_3 \end{array} \right). \nn
\eeq
Accordingly, the $6j$ symbol is equal to the following tetrahedral spin network evaluation
\beq
\left\{ \begin{array}{ccc}
j_1 & j_2 & j_3 \\ j_4 & j_5 & j_6 \end{array} \right\} =     \begin{array}{c}
\psfrag{a}{$j_1$}
\psfrag{b}{$j_3$}
\psfrag{c}{$j_2$}
\psfrag{d}{$j_4$}
\psfrag{e}{$j_6$}
\psfrag{f}{$j_5$}
\psfrag{+}{$_+$}
\includegraphics[scale=0.6]{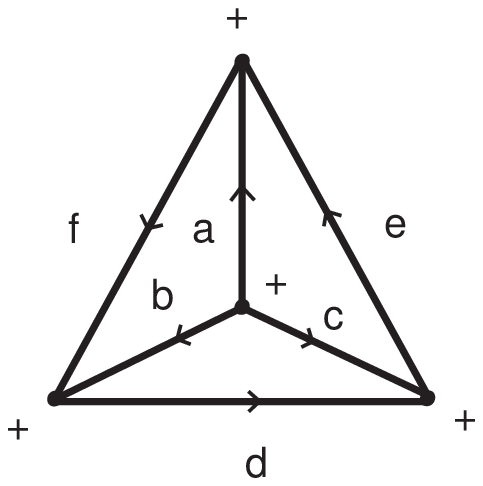}
\end{array}
\eeq
Note that reversing an arrow on a spin $j$ line of the diagram produces a factor $(-1)^{2j}$ because of the `see saw' property \eqref{seesaw1} of the inner product. This emphasises the need to have the arrows in the diagram. Thanks to the symmetries of the $3j$ symbol, the $6j$ symbol network is invariant under any permutation of the vertices.

We are now ready to introduce the recoupling properties used throughout the paper.
The first that we use is the second orthogonality of the $3j$ symbols
\beq
\left( \begin{array}{ccc}
i & \gamma & \delta \\ \alpha & k & l  \end{array} \right)  \left( \begin{array}{ccc}
 k & l & \beta \\  \gamma & \delta & j \end{array} \right) = \frac{(-1)^{2j}}{\dim j} \, \delta^i_j \, \delta^{\beta}_{\alpha} \, Y(i,k,l).
\eeq
This property can be understood as Schur's lemma. This is most transparent in the diagrammatic version of the above identity
\beq
\label{Schur}
\begin{array}{c}
\psfrag{i}{$i$}
\psfrag{j}{$j $}
\psfrag{k}{$l $}
\psfrag{l}{$k $}
\psfrag{+}{$_+$}
\psfrag{-}{$_-$}
\includegraphics[scale=0.35]{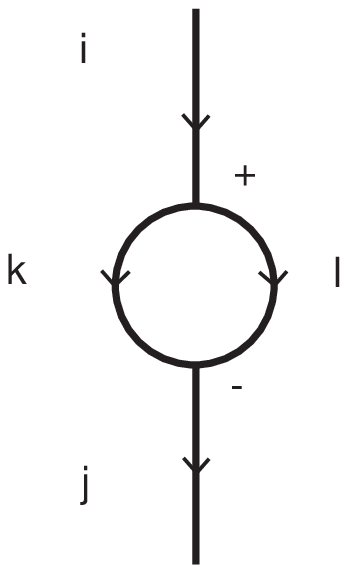}
\end{array} =   \frac{(-1)^{2j}}{\dim j} \delta^i_{j}
\begin{array}{c}
\psfrag{i}{$i $}
\includegraphics[scale=0.35]{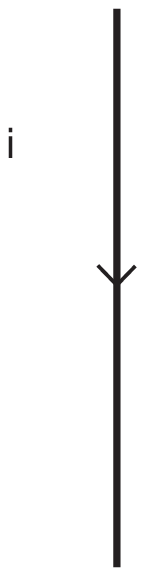}
\end{array} Y(i,k,l)
\eeq

We also made use of the recoupling theorem which provides the change of basis formula between two different recoupling schemes of a $4$-valent intertwiner
\beq
\left( \begin{array}{ccc}
\beta & i & m \\ j & \alpha & \epsilon  \end{array} \right)
\left( \begin{array}{ccc}
 \epsilon & \gamma &     l             \\
 m        & k      & \delta      \end{array} \right)
  = \sum_n C_n
 \left( \begin{array}{ccc}
i & l & \theta  \\
\alpha & \delta & n  \end{array} \right)
  \left( \begin{array}{ccc}
 n & \beta & \gamma  \\
 \theta & j & k \end{array} \right),
\eeq
where
$$
C_n = \dim n \, (-1)^{i+k+m+n}  \left\{ \begin{array}{ccc}
i & j & m\\ k & l & n \end{array} \right\}.
$$
In graphical language, the recoupling theorem yields
\beqa
\label{recoupling1}
\begin{array}{c}
\psfrag{a}{$i$}
\psfrag{b}{$j$}
\psfrag{c}{$m$}
\psfrag{d}{$k$}
\psfrag{e}{$l$}
\psfrag{+}{$_+$}
\psfrag{-}{$_-$}
\includegraphics[scale=0.4]{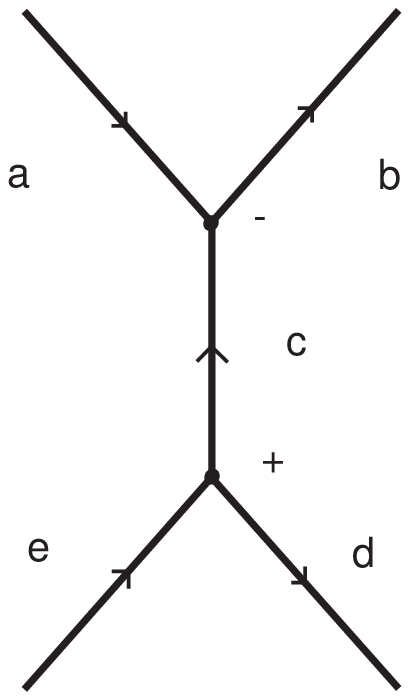}
\end{array}
&=& \sum_n \dim n
\begin{array}{c}
\psfrag{a}{$i$}
\psfrag{b}{$m$}
\psfrag{c}{$j$}
\psfrag{d}{$k$}
\psfrag{e}{$n$}
\psfrag{f}{$l$}
\psfrag{+}{$_+$}
\psfrag{-}{$_-$}
\includegraphics[scale=0.6]{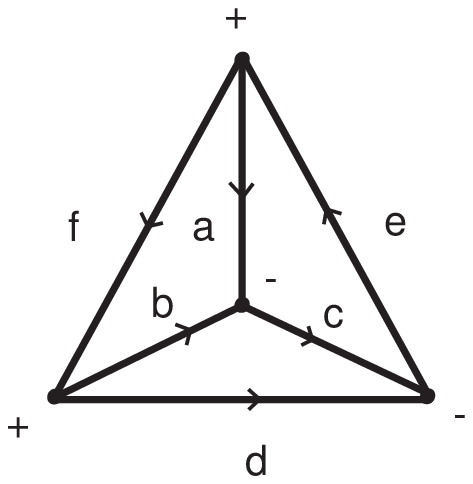}
\end{array}
\begin{array}{c}
\psfrag{a}{$i$}
\psfrag{b}{$j$}
\psfrag{d}{$k$}
\psfrag{e}{$l$}
\psfrag{c}{$n$}
\psfrag{+}{$_+$}
\psfrag{-}{$_-$}
\includegraphics[scale=0.4]{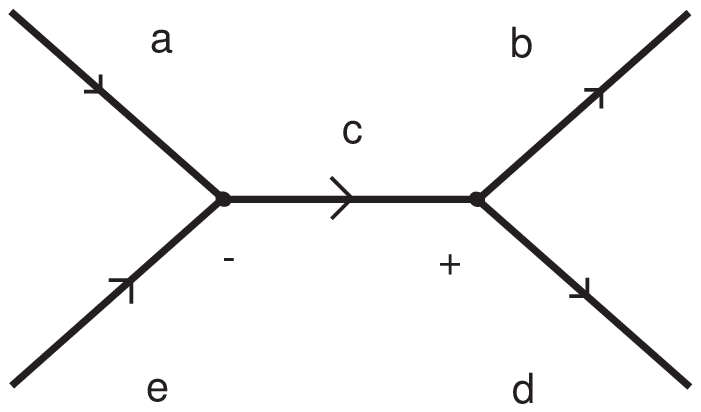}
\end{array}
\nn \\
&=& \sum_n \dim n (-1)^{i+k+m+n}
\begin{array}{c}
\psfrag{a}{$i$}
\psfrag{b}{$m$}
\psfrag{c}{$j$}
\psfrag{d}{$k$}
\psfrag{e}{$n$}
\psfrag{f}{$l$}
\psfrag{+}{$_+$}
\psfrag{-}{$_-$}
\includegraphics[scale=0.6]{tetnet2}
\end{array}
\begin{array}{c}
\psfrag{a}{$i$}
\psfrag{b}{$j$}
\psfrag{d}{$k$}
\psfrag{e}{$l$}
\psfrag{c}{$n$}
\psfrag{+}{$_+$}
\psfrag{-}{$_-$}
\includegraphics[scale=0.4]{recoupling2a}
\end{array}
\eeqa

A consequence of the recoupling identity is the following property
\beq
\left( \begin{array}{ccc}
i & \delta & n \\ \alpha & l & \epsilon  \end{array} \right)  \left( \begin{array}{ccc}
 l & j & \theta \\ \delta & \beta & m \end{array} \right)
 \left( \begin{array}{ccc}
m & k & \epsilon \\ \theta & \gamma & n  \end{array} \right) =
\left\{ \begin{array}{ccc}
i & k & j \\ m & l & n \end{array} \right\}
\left( \begin{array}{ccc}
 i & j & k \\ \alpha & \beta & \gamma \end{array} \right).
\eeq
Diagrammatically, this yields the fusion move displayed below
    \beq
    \label{fusion move}
    \begin{array}{c}
\psfrag{a}{$i$}
\psfrag{b}{$k$}
\psfrag{c}{$j$}
\psfrag{d}{$m$}
\psfrag{e}{$l$}
\psfrag{f}{$n$}
\psfrag{+}{$_+$}
\psfrag{-}{$_-$}
\includegraphics[scale=0.4]{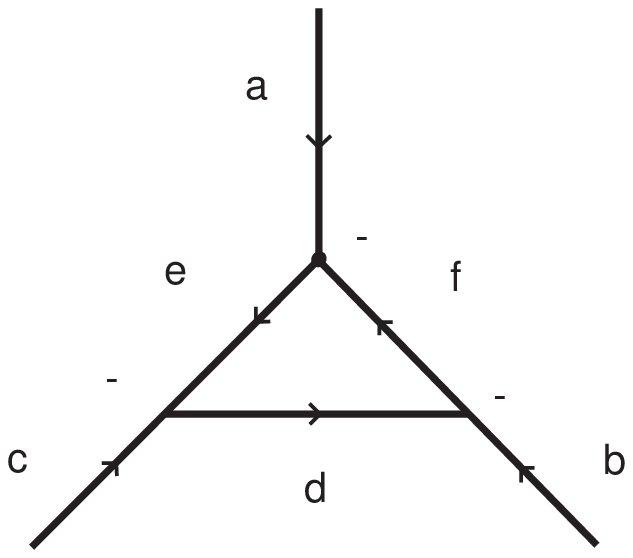}
\end{array}
    =
\begin{array}{c}
\psfrag{a}{$i$}
\psfrag{b}{$j$}
\psfrag{c}{$k$}
\psfrag{d}{$m$}
\psfrag{e}{$n$}
\psfrag{f}{$l$}
\psfrag{+}{$_+$}
\psfrag{-}{$_-$}
\includegraphics[scale=0.65]{tetnet2}
\end{array}
\begin{array}{c}
\psfrag{a}{$i$}
\psfrag{b}{$k$}
\psfrag{c}{$j$}
\psfrag{+}{$_+$}
\psfrag{-}{$_-$}
\includegraphics[scale=0.4]{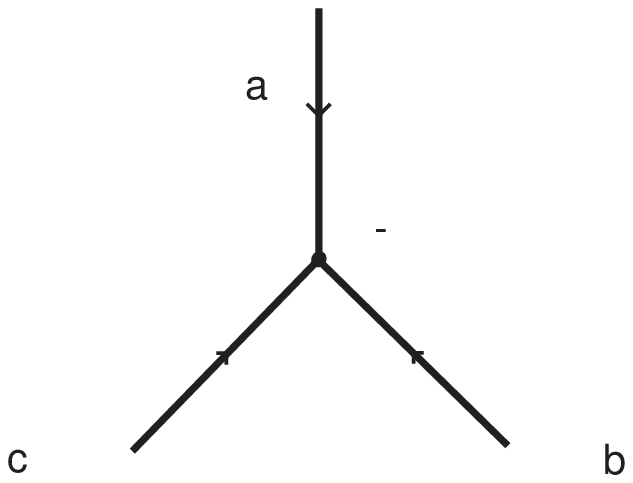}
\end{array}
    \eeq


\end{document}